\documentclass[12pt]{article}
\usepackage{a4wide,epsfig,psfrag,amsmath,amssymb,cite,scalefnt}
\usepackage[dvipsnames]{xcolor}
\usepackage{amsmath,braket}
\usepackage{placeins}
\usepackage{breqn}
\usepackage{slashed}

\graphicspath{ {figs/} }

\newcommand{\ms}[1]{\color{red} #1}

\allowdisplaybreaks
\newcommand{\qv}{\vec q \,}

\newcommand{\myintB}{\int\frac{{\rm d}^d k_1}{(2\pi)^d}\int\frac{{\rm d}^d k_2}{(2\pi)^d} \,}

\newcommand{\ep}{\varepsilon}

\begin{document}

\begin{flushright}

  P3H-20-73, TTP20-040
\end{flushright}

\begin{center}

  \vspace*{5em}

  {\bf\Large\boldmath On the relation between the $\overline{\mathrm{MS}}$ and
    the kinetic mass of heavy quarks}\\[15mm]

  \setlength {\baselineskip}{0.2in}
  {\large  Matteo Fael, Kay Sch\"onwald and Matthias Steinhauser}\\[5mm]
  {\it Institut f\"ur Theoretische Teilchenphysik, 
    Karlsruhe Institute of Technology (KIT),\\
    76128 Karlsruhe, Germany.}\\[3cm] 

\end{center} 

\centerline{\bf Abstract}

\bigskip

We compute the relation between the pole mass and the kinetic mass of a heavy
quark to three loops. Using the known relation between the pole and the
$\overline{\rm MS}$ mass we obtain precise conversion relations between the
$\overline{\rm MS}$ and kinetic masses.  The kinetic mass is defined via the
moments of the spectral function for the scattering involving a heavy quark
close to threshold.  This requires the computation of the imaginary part of a
forward scattering amplitude up to three-loop order. We discuss in detail the
expansion procedure and the reduction to master integrals. For the latter
analytic results are provided. We apply our result both to charm and bottom
quark masses.  In the latter case we compute and include finite charm quark
mass effects.  Furthermore, we determine the large-$\beta_0$ result for the
conversion formula at four-loop order. For the bottom quark we estimate the
uncertainty in the conversion between the $\overline{\rm MS}$ and kinetic
masses to about 15~MeV which is an improvement by a factor two to three as
compared to the two-loop formula. The improved precision is crucial for the
extraction of the Cabibbo-Kobayashi-Maskawa matrix element $|V_{cb}|$ at
Belle~II.


\thispagestyle{empty}

\newpage


\section{\label{sec::intro}Heavy quark mass definitions}

Quark masses enter the QCD Lagrangian density as free parameters and as such
they have to be renormalized once higher order corrections are
considered. There are two distinguished renormalization schemes, the pole (or
on-shell) and the (modified) minimal subtraction scheme.  
The pole mass scheme (OS)
has the advantage that it is based on a physical definition since it
requires that, order-by-order in perturbation theory, the inverse heavy-quark
propagator has a zero at the position of the pole mass. On the other hand, the
minimal subtraction scheme only subtracts the divergent parts of the quantum
corrections to the quark two-point function and combines them with the bare
mass to arrive at the renormalized ${\rm MS}$ (or $\overline{\rm MS}$) quark
mass.

In high-energy reactions it is appropriate to use the $\overline{\rm MS}$
mass, which does not suffer from intrinsic uncertainties. However, typical
energy scales in $B$ meson decays are smaller than the bottom quark mass which
is the reason that the $\overline{\rm MS}$ mass is less appropriate in such
situations.  The pole mass, on the other hand, suffers from renormalon
ambiguities, which manifest themselves through an ill-behaved perturbative
series. This can already be seen in the relation between the on-shell and
$\overline{\rm MS}$ mass which suffers from large higher order
corrections~\cite{Beneke:1994sw,Bigi:1994em,Beneke:1994rs}.  For example, the
four-loop term in the mass relation amounts to about 100~MeV for bottom
quarks~\cite{Marquard:2015qpa,Marquard:2016dcn}, which is much larger than the
current uncertainty of the $\overline{\rm MS}$ mass as, e.g., extracted from
lattice calculations or low-moment sum rules (see, e.g.,
Refs.~\cite{Chetyrkin:2017lif}).

In order to combine the advantages of the two canonical mass schemes,
so-called threshold masses as, e.g., the potential subtracted
(PS)~\cite{Beneke:1998rk}, 1S~\cite{Hoang:1998hm,Hoang:1998ng,Hoang:1999zc},
renormalon subtracted (RS)~\cite{Pineda:2001zq} or
MRS~\cite{Hoang:2008yj,Hoang:2017suc} mass have been developed. They are of
short-distance nature and are free of renormalon ambiguities, as the
$\overline{\rm MS}$ mass.  On the other hand, the threshold masses are
suitable parameters to be used for cross sections near threshold, decay rates
and heavy quark bound state properties.  In this article, we concentrate on
the kinetic mass~\cite{Bigi:1996si,Czarnecki:1997sz}, that is defined via the
so-called Small-Velocity (SV) sum rules and involves the relation between the
masses of heavy quark and heavy mesons up to the kinetic energy term.

The various mass definitions can be converted into each other within
perturbation theory. Such a conversion is frequently needed in practical calculations
as well as in the extraction of mass values from experiments.  In order to
achieve high precision, it is mandatory to know the conversion relations
between the different mass schemes as precisely as possible.  For the most
commonly used ones, their relation to the pole mass is known to
next-to-next-to-next-to-leading order (N$^3$LO)~\cite{Marquard:2016dcn}. 

In this work, we concentrate on the kinetic mass and on the methods used in
the calculation of the relation between the $\overline{\mathrm{MS}}$ and the
kinetic mass to $O(\alpha_s^3)$ presented in~\cite{Fael:2020iea}.  In
particular, we describe our approach based on \textit{expansion by
  regions}~\cite{Beneke:1997zp,Smirnov:2012gma} to compute the SV sum
rules~\cite{Bigi:1994ga,Bigi:1996si} that define the kinetic mass to higher
orders in $\alpha_s$.  We give an account of the reduction to master integrals
and summarize our strategies for their analytic computation at three loops.
Moreover, we improve our work in~\cite{Fael:2020iea} by including finite charm
mass effects in the mass relation for the bottom quark. We also present the
large-$\beta_0$ contribution to the conversion formula at four-loop order.
Our results, available as ancillary files attached to this paper and also
implemented in~\texttt{RunDec} and~\texttt{CRunDec}~\cite{Herren:2017osy},
allow us to carefully assess for bottom and charm quark the theoretical
uncertainties in the mass scheme conversions.  For the bottom quark, such
uncertainty is reduced by a factor two to three compared to the two-loop
estimates in Ref.~\cite{Gambino:2011cq}.  Our results are crucial for future
extractions of $|V_{cb}|$ from $B \to X_c \ell \bar \nu_\ell$ decays at Belle
II, in particular to better constrain global fits of the branching ratios and
the moments of inclusive semileptonic $B$ decays.

The  paper is structured as follows: in the next
section we motivate the definition of the kinetic mass and derive the
relevant formulae that are needed for the practical calculation.  In
Section~\ref{sec::calc} we provide several technical details on the
calculation and in Section~\ref{sec::charm} we describe the calculation of the
charm quark mass effects to the bottom mass relation. We
discuss our analytic three-loop results in Section~\ref{sec::ana} and
 the four-loop large-$\beta_0$ terms in
Section~\ref{sec::BLM}. The numerical effects of the new correction
terms are discussed in Section~\ref{sec::num}, with special emphasis on
the charm quark effects. Finally we summarize our findings in
Section~\ref{sec::concl}.  

In Appendix A we report the analytic expressions up to $O(\alpha_s^3)$ of the
HQET parameters $\overline{\Lambda}, \mu_\pi, \rho_D$ and $r_E$ computed in
perturbative QCD.  In Appendix B we discuss in detail the calculation of the
most difficult master integral while in Appendix C we provide analytic
results of auxiliary integrals which were useful in the course of our
calculation.


\section{Why the kinetic mass?}

In this section, we first summarize the motivations behind the kinetic mass
scheme and the main findings of Refs.~\cite{Bigi:1994ga,Bigi:1996si}.
Afterwards we introduce the rigorous definition of the relation between the OS
and the kinetic mass in terms of SV rum rules, and present our method for its
calculation to higher orders in $\alpha_s$ based on the expansion by regions and
in a fully covariant formalism.

The basis of the precise theoretical prediction for inclusive $B\to X_c \ell \bar\nu_\ell$ decays
is the Heavy-Quark Expansion (HQE), which allows us to predict various observables, 
as the total semileptonic rate as well as the moments of differential distributions 
(lepton energy, hadronic energy, hadronic invariant mass, etc.), as a double expansion 
in $\alpha_s(m_b)$ and $\Lambda_{\mathrm{QCD}}/m_b$.
The starting point is the optical theorem which 
relates the decay rate to the forward matrix element of a scattering amplitude:
\begin{equation}
    \Gamma = \frac{1}{m_B} \mathrm{Im} \,
    \int {\rm d}^4x \,
    \bra{B(p_b)}
    T\{ \mathcal{H}_{\mathrm{eff}}^\dagger (x) \mathcal{H}_{\mathrm{eff}}(0) \} 
    \ket{B(p_b)}\,.
\end{equation}
The time-ordered product can be written in terms of an operator product expansion that allows us to write
\begin{equation}
   \int {\rm d}^4x \,
   T\{ \mathcal{H}_{\mathrm{eff}}^\dagger (x) \mathcal{H}_{\mathrm{eff}} (0) \}  
   =
   \sum_{n,i}
   \frac{1}{m_b^n} \mathcal{C}_{n,i}\, \mathcal{O}_{n+3,i}\,,
\end{equation}
where $\mathcal{O}_{{n},i}$ is a set (labeled by $i$) of operators of
dimension $n$,
and $\mathcal{C}_{n,i}$ are the Wilson coefficients calculable in perturbative QCD.
Taking the forward matrix element of this expression, we obtain the decay rate in terms 
of the Wilson coefficients and hadronic matrix elements of the operators
$\mathcal{O}_{{n},i}$ 
encoding the non-perturbative input into the decay rate.
The general structure of the expansion for an observable ${\rm d}\Gamma$ is
\begin{equation}
    {\rm d}\Gamma =
    {\rm d}\Gamma_0 
    + {\rm d}\Gamma_{\mu_\pi}  \frac{\mu_\pi^2}{m_b^2} 
    + {\rm d}\Gamma_{\mu_G}  \frac{\mu_G^2}{m_b^2} 
    + O\left( \frac{\Lambda_{\mathrm{QCD}}^3}{m_b^3} \right)\,,
\end{equation}
where the coefficients ${\rm d}\Gamma_i$ are functions of $m_c/m_b$
and have an expansion in $\alpha_s(m_b)$, while $\mu_\pi$ and $\mu_G$
are dimension-full parameters of order $\Lambda_{\mathrm{QCD}}$.
Since the bottom quark vector current $\bar b \gamma^\mu b$ is
conserved, there are for $n=0$ only perturbative corrections, i.e.,
${\rm d}\Gamma_0$ corresponds to the decay of a free $b$ quark.  Note
that there are no linear $1/m_b$ terms in the HQE as was shown in
Refs.~\cite{Chay:1990da,Bigi:1993fe,Manohar:1993qn,Mannel:1994kv}.
The first two non-perturbative contributions, denoted by $\mu_\pi$ and
$\mu_G$, emerge at order $\Lambda_{\mathrm{QCD}}^2/m_b^2$ and can be
written in terms of two matrix elements:
\begin{align}
    -2M_B \mu_\pi^2 &= \bra{H_\infty (v)} \bar{h}_v \, (iD_\perp)^2 \, h_v  \ket{H_\infty (v)}\,, \notag \\
    2M_B \mu_G^2 &= \bra{H_\infty (v)} \bar{h}_v \, \sigma \cdot G \, h_v  \ket{H_\infty (v)}\,,
    \label{eqn:mupimugmat}
\end{align}
with $\sigma \cdot G \equiv (iD_\perp^\mu) (iD_\perp^\nu)
(-i\sigma_{\mu \nu})$, $\sigma^{\mu \nu} =
\frac{i}{2}[\gamma_\mu,\gamma_\nu]$ and $iD^\mu = iv^\mu \, (iv \cdot
D) + D_\perp^\mu$.  In Eq.~\eqref{eqn:mupimugmat} $h_v(x)$ is the $b$
quark field in the heavy quark effective theory, $v=p_B/M_B$ is the
velocity of the $B$ meson and $\ket{H_\infty(v)}$ is the pseudoscalar
or vector meson's state in the infinite mass
limit.\footnote{Note that the heavy quark expansion for semileptonic
  $B$ decays is often written in terms of operators with the field
  $b_v(x) = e^{im_b v \cdot x} b(x)$ and the meson state
  $\ket{B(p_B)}$ in full QCD, which differ from $h_v$ and
  $\ket{H_\infty(v)}$ by higher power corrections in $1/m_b$, as for
  instance $b_v(x) = (1 + i \slashed D_\perp/2m_b+\dots) h_v(x)$.  In
  the following we will consider terms only up to $1/m_b^2$, so such
  difference can be ignored.}  The parameter $\mu_\pi^2$ corresponds
to the kinetic energy of the heavy quark inside the heavy meson, while
$\mu_G^2$ is its chromo-magnetic moment.

The HQE has a strong dependence on the mass of the heavy quark
$m_b$. Therefore, in order to obtain precise predictions for decay
rates, the quark mass has to be carefully chosen.  This choice is
closely intertwined with the size of the QCD corrections to the decay
rates.  As already mentioned above, perturbative calculations 
which use the pole mass scheme are affected by the renormalon
ambiguity and thus show a bad behaviour of the perturbative
series. Indeed if the semileptonic width $\Gamma_{\mathrm{sl}}$ is
expressed in terms of the pole mass of the $b$ quark, the expression
for $\Gamma_{\mathrm{sl}}$ contains a factorially divergent series in
powers of $\alpha_s$\cite{Beneke:1994sw,Bigi:1994em,Beneke:1994rs}:
\begin{equation}
  \Gamma_{\mathrm{sl}} \sim 
    \sum_k k! \left( \frac{\beta_0}{2} \frac{\alpha_s}{\pi} \right)^k.
\end{equation}

However, also in the $\overline{\mathrm{MS}}$ mass scheme, the $\alpha_s$
corrections to the $\Gamma_{\mathrm{sl}}$ have a bad convergence. Indeed,
removing the infra-red (IR) renormalons by using a short distance mass
definition does not guarantee yet that we have a fast convergent
perturbative series.  Semileptonic decays of a heavy quark are in fact also
affected by large corrections of the type $(n \alpha_s)^k$, with $n=5$, which
arise from the conversion of the overall factor $m_b^5$
from the pole scheme to the $\overline{\mathrm{MS}}$ scheme.
Note that the $(n \alpha_s)^k$ enhanced terms are not related to the running of
$\alpha_s$ and are present even for a vanishing $\beta$ function.

A further argument against the use of the $\overline{\mathrm{MS}}$
bottom quark mass at a scale $\mu = m_b$ for inclusive decays is that the maximal
energy of the final hadronic system is limited by $m_b-m_c$. Moreover, since
the two leptons in $B \to X_c \ell \nu$ carry away a significant fraction of
the energy, the mass scale $\mu$ to be used is even smaller. However, at such
low scale the logarithmic running of the $\overline{\mathrm{MS}}$ mass is
considered unphysical.

The kinetic scheme for the mass of a heavy quark, $m_Q^{\mathrm{kin}}$, was introduced in~\cite{Bigi:1994ga,Bigi:1996si} 
to resum in the semileptonic rate the $(n \alpha_s)^k$-enhanced terms via a suitable short-distance definition. It relies on a set of QCD sum rules which 
hold in the so-called small velocity limit, i.e.\ in the limit where the three-momentum components of the final state $X_c=D,D^*$ are much smaller than $m_b$ and $m_c$ in the rest frame of the $B$ meson. 
The SV sum rules are relations between the physical differential rate and the parameters 
$\mu_\pi$ and $\mu_G$, as well as $\overline \Lambda$, the binding-energy of a heavy hadron.
They are obtained by considering moments of the hadronic energy spectrum
\begin{equation}
    I_n (\vec q \,^2) = 
    \int_{|\qv|}^{q_0^{\mathrm{max}}} {\rm d} q_0\, \omega^n \, \frac{{\rm d}^2\Gamma_{\mathrm{sl}}}{{\rm d}q_0 {\rm d}\qv \,^2}\,,
    \label{eqn:svrules}
\end{equation}
where $q = (q_0, \qv)$ is the momentum of the di-lepton system in the rest frame of the $B$ meson 
and $q_0^{\mathrm{max}}= M_B-\sqrt{M^2_D+\qv^2}$.
The moments $I_n$ are evaluated at fixed values of $\vec q \,^2$.
The variable $\omega$ is the excitation energy of the $X_c$ state, i.e.\ the difference between the energy 
of the hadronic system and the minimum energy necessary to produce a $D$ meson with a spacial component $\qv$:
\begin{equation}
  \omega = M_B-q_0-\sqrt{M^2_D+\qv^2}= q_0^{\mathrm{max}}-q_0\,.
  \label{eq::omega_decay}
\end{equation}
The factor $\omega^n$ in~\eqref{eqn:svrules} eliminates for $n>0$ the
``elastic peak'' corresponding to the elastic process $B\to D \ell \nu$, so
the integral is saturated only by the inelastic contributions.  Moreover, all
moments are finite. Indeed, the case $n=1$ gives the expectation value of
the excitation energy, which is bounded by the decay kinematics. Therefore
the differential rate cannot scale worse than $1/\omega$ in the
$\omega \to 0$ limit.

We are interested in the leading term of the SV sum rules in an expansion in $|\qv|\ll m_c \sim m_b$, 
i.e.\ the small velocity limit, and in $\Lambda_{\mathrm{QCD}} \ll m_c \sim m_b$, the heavy quark expansion.
The first and the second sum rules are obtained inserting in~\eqref{eqn:svrules} the decay rate of a free quark at tree level:
\begin{equation}
  \frac{{\rm d}^2\Gamma_{\mathrm{sl}}^{\mathrm{free}}}{{\rm d}q_0 {\rm d} \vec q \, ^2} =
    \frac{G_F^2 |V_{cb}|^2}{8 \pi^3} \frac{|\qv|}{\sqrt{m_c^2+\qv^2}}
    \left[
     m_b \left( q_0^2  - \frac{\qv^2}{3} \right) + q_0 \, \qv^2 - q_0^3
   \right] \delta(q_0 - \widetilde{q_0}^{\mathrm{max}})\,,
    \label{eqn:gammaslquark}
\end{equation}
where $\widetilde{q_0}^{\mathrm{max}} = m_b - \sqrt{m_c^2+\qv^2}$ is the maximum energy of the 
leptonic system in the free quark approximation. We then expand the heavy meson masses appearing 
in $\omega$ in terms of heavy quark masses,
\begin{equation}
    M_H = m_Q + \overline{\Lambda} + \frac{\mu_\pi^2 + d_H \mu_G^2}{2m_Q} 
    + O \left( \Lambda_{\mathrm{QCD}}^3 \right),
\end{equation}
with $H=B^{(*)},D^{(*)}$, $Q=b,c$ and $d_H =3$ ($d_H=-1$) for a pseudoscalar (vector) meson. 
Keeping the leading terms in $|\qv|$ and $\Lambda_{\mathrm{QCD}}$, one finds the first and the second sum rule~\cite{Bigi:1994ga}:
\begin{align}
    I_0(\vec q \,^2) &= |\qv| \frac{G_F^2 |V_{cb}|^2}{8 \pi^3}  (m_b-m_c)^2 
    + O(\qv^2, \Lambda_{\mathrm{QCD}})\,, \\
    I_1(\vec q \,^2) &= I_0 \frac{\qv^2}{2 m_c^2} \overline{\Lambda} 
    + O(\qv^3, \Lambda_{\mathrm{QCD}}^2)\,.
\end{align}
The third sum rule is obtained by employing in~\eqref{eqn:svrules} the differential rate computed 
up to $O(1/m_b^2)$ in the HQE (see e.g.~\cite{Manohar:2000dt}), which yields
\begin{equation}
  I_2(\qv) = I_0 \frac{\qv^2}{3m_c^2} \mu_\pi^2 + O(\qv^3,\Lambda_{\mathrm{QCD}}^3)\,.
\end{equation}
Even if these sum rules are obtained with a $V$-$A$ weak current, the leading term of the ratios $I_n/I_0$ is actually independent on the specific current mediating the $b \to c$
transition. This is a consequence of the heavy quark
symmetries~\cite{Shifman:1987rj,Isgur:1988gb,Isgur:1989ed} in the infinite
mass limit.

Let us now discuss how the sum rules are modified once radiative corrections
are included.  At tree level only the peak at the end point of the partonic
spectrum -- the $\delta$ function in~\eqref{eqn:gammaslquark} -- contributes.
Radiative corrections add a perturbative tail corresponding to the additional emission of gluons in the final state.  
In this case, it is mandatory to introduce
a Wilsonian cutoff~$\mu$ in order to separate gluons with energy smaller than
$\mu$, that should be treated as soft and belonging to the non-perturbative
regime, and hard gluons that can be described in perturbative QCD.  We must
therefore modify~\eqref{eqn:svrules} as follows
\begin{equation}
    I_n (\vec q \,^2) = 
    \int_{q_0^{\mathrm{max}} - \mu}^{q_0^{\mathrm{max}}} {\rm d} q_0\, \omega^n \, \frac{{\rm d}^2\Gamma_{\mathrm{sl}}}{{\rm d}q_0 {\rm d}\qv \,^2}
    +
    \int^{q_0^{\mathrm{max}} - \mu}_{|\qv|} {\rm d} q_0\, \omega^n \, \frac{{\rm d}^2\Gamma_{\mathrm{sl}}}{{\rm d}q_0 {\rm d}\qv \,^2}
    \,,
    \label{eqn:svrules2}
\end{equation}
and rewrite the sum rules for $\overline{\Lambda}$ and $\mu_\pi^2$ as
\begin{align}
    I_1(\vec q \,^2) &= I_0 \frac{\qv^2}{2 m_c^2} \overline{\Lambda}(\mu) 
    +
    \int^{q_0^{\mathrm{max}} - \mu}_{|\qv|} {\rm d} q_0\, \omega \, \frac{{\rm d}^2\Gamma_{\mathrm{sl}}}{{\rm d}q_0 {\rm d}\qv \,^2}  
    +
    O(\qv^4, \Lambda_{\mathrm{QCD}}^2)\,, \label{eqn:Lbarrad} \\[5pt]
    I_2(\qv) &= 
    I_0 \frac{\qv^2}{3m_c^2} \mu_\pi^2(\mu) 
    +
    \int^{q_0^{\mathrm{max}} - \mu}_{|\qv|} {\rm d} q_0\, \omega^2 \, \frac{{\rm d}^2\Gamma_{\mathrm{sl}}}{{\rm d}q_0 {\rm d}\qv \,^2} 
    + O(\qv^3,\Lambda_{\mathrm{QCD}}^3)\,. \label{eqn:mupirad}
\end{align}
The integrals on the right-hand sides correspond to the 
perturbative contribution with gluons of total energy greater
than $\mu$. For this reason the value of $\mu$ must be chosen large enough
to justify the applicability of perturbative QCD:
$\Lambda_{\mathrm{QCD}} \ll \mu \ll m_B$.

In the end, the SV sum rules provide an operative definition on how to extract
$\overline \Lambda$, $\mu_\pi^2$, etc., from the measurement of physical
spectra. Note that since the moments $I_n$ are independent on $\mu$,
Eqs.~\eqref{eqn:Lbarrad} and~\eqref{eqn:mupirad} show that
$\overline \Lambda(\mu)$ and $\mu_\pi(\mu)$ change under the variation of the
Wilsonian cutoff. Their running is not logarithmic but instead power-like.

At the same time, the SV sum rules give us an unambiguous procedure for the
definition of $m_Q$ via the relation between heavy quark and
heavy meson mass:
\begin{equation}
  m_Q(\mu) = M_{\bar H} - \overline{\Lambda}(\mu) - \frac{\mu_\pi^2(\mu)}{2 m_Q(\mu)} + \dots \, ,
   \label{eqn:mqmass}
\end{equation}
which shows that any conceivable short distance definition of the heavy quark
mass must necessarily include a cutoff $\mu$.  Note that there is no
$\mu_G$ term on the right-hand side of Eq.~(\ref{eqn:mqmass}) since it
cancels after averaging over $H$ and $H^*$ mesons: $M_{\bar H} \equiv (M_H+3 M_{H^*})/4$.
The quantities
$\overline \Lambda(\mu) $ and $\mu_\pi^2(\mu)$ can be obtained by taking the ratios
between SV sum rules, and evaluating them in the infinite heavy quark mass
limit and at zero recoil:
\begin{align}
  \overline{\Lambda}(\mu) = 
    \lim_{\vec v \to 0} \, \lim_{m_Q \to 0}\, 
    \frac{2}{\vec v \, ^2} \frac{
    \int_{q_0^{\mathrm{max}} - \mu}^{q_0^{\mathrm{max}}} 
    {\rm d} q_0\, \omega \, \frac{{\rm d}^2\Gamma_{\mathrm{sl}}}{{\rm d}q_0 {\rm d}\qv \,^2}
    }{
    \int_{q_0^{\mathrm{max}} - \mu}^{q_0^{\mathrm{max}}} 
    {\rm d} q_0\, \frac{{\rm d}^2\Gamma_{\mathrm{sl}}}{{\rm d}q_0 {\rm d}\qv \,^2}
    }\,, 
    \label{eqn:defLambdabar}\\
    \mu_\pi^2(\mu) = 
    \lim_{\vec v \to 0} \, \lim_{m_Q \to 0}\,
    \frac{3}{\vec v \, ^2} \frac{
    \int_{q_0^{\mathrm{max}} - \mu}^{q_0^{\mathrm{max}}} 
    {\rm d} q_0\, \omega^2 \, \frac{{\rm d}^2\Gamma_{\mathrm{sl}}}{{\rm d}q_0 {\rm d}\qv \,^2}
    }{
    \int_{q_0^{\mathrm{max}} - \mu}^{q_0^{\mathrm{max}}} 
    {\rm d} q_0\, \frac{{\rm d}^2\Gamma_{\mathrm{sl}}}{{\rm d}q_0 {\rm d}\qv \,^2}
    }\,,
    \label{eqn:defmupi}
\end{align}
where $|\vec v|  \ll 1$ is the velocity of the quark in the final state.

The SV sum rules give an insight on how to avoid the appearance of
large $(n \alpha_s)^k$ corrections in semileptonic rates, as those
affecting the $\overline{\mathrm{MS}}$ mass definition.  The authors
of Ref.~\cite{Bigi:1996si} employed the SV sum rules to show that the
dependence on the fifth power of the meson mass ($M_B^5$), that one
would naively expect for the total semileptonic width, is actually
substituted by the heavy quark mass (raised to the fifth power), which
becomes the relevant parameter of the process
\begin{equation}
  \Gamma_{\mathrm{sl}} \simeq \frac{G_F^2 |V_{cb}|^2}{192 \pi^3}
\Big(
M_B - \overline{\Lambda}
\Big)^{5}\,.
\end{equation}
There is a cancellation of the infrared contribution in the semileptonic width:
 $\Gamma_{\mathrm{sl}}$ is insensitive to long-distance effects responsible for the heavy meson binding energy.

So far our discussion focused on the SV sum rules for meson decays. Let us now turn our attention to perturbative QCD and how the SV sum rules can be employed to 
give a short distance definition of the heavy quark mass relevant for perturbative calculations. 
It was observed in~\cite{Bigi:1996si}, that the same kind of cancellation of infrared contribution to $\Gamma_{\mathrm{sl}}$ happens in perturbative QCD, granted that we substitute each term in Eq.~\eqref{eqn:mqmass} with its perturbative
version:
\begin{align}
  m_Q(\mu) &\rightarrow m_Q^{\mathrm{kin}}(\mu), & 
  M_{\bar H}      &\rightarrow m_Q^{\mathrm{OS}},\notag \\
  \overline{\Lambda}(\mu)  &\rightarrow 
  [\overline{\Lambda}(\mu)]_{\mathrm{pert}}, &
  {\mu_\pi^2(\mu)} &\rightarrow 
  [\mu_\pi^2(\mu)]_{\mathrm{pert}} \,.
\end{align}
The role of the scale-independent heavy meson mass is played in this
case by the pole mass $m_Q^{\rm OS}$, while the perturbative version
of $\overline{\Lambda}$ and $\mu_\pi^2$ are obtained utilizing the
same set of SV sum rules presented before, with the difference that
the rate has to be computed in perturbative QCD.  This provides us
with a scale-dependent short-distance mass definition for heavy
quarks, the ``kinetic mass'' $m_Q^{\mathrm{kin}}$~\cite{Bigi:1996si}
which is given by
\begin{equation}
  m_Q^{\mathrm{kin}}(\mu) = 
  m_Q^{\mathrm{OS}} -
  [\overline{\Lambda}(\mu)]_{\mathrm{pert}}
  -\frac{[\mu_\pi^2(\mu)]_{\mathrm{pert}}}{2 m_Q^{\mathrm{kin}}(\mu)} - \, \dots \,,
  \label{eq::mkin}
\end{equation}
where the ellipses stand for higher order $1/m_Q^{\rm kin}$ terms.
Note that in this definition the renormalon ambiguity present in the
on-shell mass cancel against the ones in $\overline{\Lambda}$ and
$\mu_\pi^2$.  The quantities $[\overline \Lambda(\mu)]_{\mathrm{pert}}$ and
$[\mu_\pi(\mu)]_{\mathrm{pert}}$ can be computed by considering the heavy
quark transition $Q \to Q'$ induced by a generic current
$J = \bar{Q}'\, \Gamma \, Q$ in the heavy quark ($m_{Q,Q'} \to \infty$) and SV
($\vec v = \vec q/m_Q'$) limits.  In the following we will
consider a generic scattering $J Q \to Q'$ of an external current $J$ on the
heavy quark $Q$.  As said before, the nature of the current $J$ is
irrelevant since the final result does not depend on it. Below we will
consider a scalar and a vector current. Moreover, for simplicity, we
consider the case $Q = Q'$.

Note that the relation between the kinetic mass and the $\overline{\rm MS}$
mass is obtained after inserting the $m^{\rm OS}$--$\overline{m}$ relation
into Eq.~(\ref{eq::mkin}). For our purpose we need this relation to three-loop
accuracy~\cite{Chetyrkin:1999qi,Marquard:2018rwx,Melnikov:2000qh,Marquard:2007uj,Fael:2020bgs}.

From now on for simplicity, let us identify the heavy quark $Q$ with the bottom quark $b$.
We denote the external momentum of the bottom by $p^\mu =(m_b,\vec{0} \,)$
with $p^2=m_b^2$, and we introduce $s=(p+q)^2$.
We can rewrite Eqs.~(\ref{eqn:defLambdabar}) and~(\ref{eqn:defmupi}) as
\begin{align}
  [\overline{\Lambda}(\mu)]_{\rm pert}
  &= 
      \lim_{\vec{v}\to0}\lim_{m_b\to\infty} 
      \frac{2}{\vec{v}\,^2} 
      \frac{
      \displaystyle
      \int_0^\mu \omega \, W(\omega,\vec{v}\,) \, {\rm d}\omega}
      {
      \displaystyle
      \int_0^\mu W(\omega,\vec{v}\,) {\rm d}\omega}
      \,,\notag\\[5pt]
      [\mu_\pi^2(\mu)]_{\rm pert} &= 
      \lim_{\vec{v}\to0}\lim_{m_b\to\infty} 
      \frac{3}{\vec{v}\,^2} 
      \frac{\displaystyle
      \int_0^\mu \omega^2 \, W(\omega,\vec{v}\,) \, {\rm d}\omega}
      {\displaystyle
      \int_0^\mu W(\omega,\vec{v}\,) {\rm d}\omega}
      \,.
      \label{eq::Lam_mupi}
\end{align}
where $W$ is the structure function, which is obtained from the 
imaginary part of the forward-scattering amplitude $T$
\begin{eqnarray}
  W(q_0,\vec{q}) &=& 2 \mbox{Im} \left[ T(q_0,\vec{q}\,) \right]
                     \,,
\end{eqnarray}
defined through
\begin{eqnarray}
  T(q_0,\vec{q}\,) &=& \frac{i}{2m_b} \int {\rm d}^4{x \,} e^{-iqx}\langle b|T J(x)J^\dagger(0)|b\rangle
  \,.
  \label{eq::Tkin}
\end{eqnarray}
For later convenience, we separate the energy and the three-momentum
components of the external momentum $q$.  

For the scattering process that we consider in the
following, we must define the excitation energy $\omega$,
i.e.\ the sum of all final state gluons' and quarks' energies,
as\footnote{Although we use the same letter as for the decay in
  Eq.~(\ref{eq::omega_decay}) there should be no confusion
  possible. From now on only the excitation energy in
  Eq.~(\ref{eq::omega_scatt}) is of relevance.}
\begin{eqnarray}
  \omega &\equiv & q_0-q_0^{\rm min} = 
  q_0 - \frac{m_b \vec v \,^2}{2} + O(\vec v \, ^4)\,,
  \label{eq::omega_scatt}
\end{eqnarray}
where 
\begin{eqnarray}
  q_0^{\rm min} &\equiv& \sqrt{\vec{q}\,^2+m_b^2} - m_b
  = \frac{m_b \vec v \,^2}{2} + { O}(\vec v \,^4)
                    \,
\end{eqnarray}
is the threshold value obtained from the condition $s=m_b^2$; for smaller
value of $s$ the structure function $W$ is zero.  From now on we
consider $W$ as a function of $\omega$ and $\vec{v}$. Its generic expression
can be written as
\begin{eqnarray}
  W(\omega,\vec{v}) 
  &=& 
  W_{\rm el}(\vec v \,) \, \delta(\omega) 
  + \frac{\vec{v}\,^2}{\omega} W_{\rm real}(\omega) \, \theta(\omega)
  + {O}\left(v^4, \omega^0 \right)\,,
      \label{eq::W}
\end{eqnarray}
where $W_{\rm el}$ describes the elastic $Jb\to b$ transition which receives
contributions from tree-level and virtual diagrams. $W_{\rm real}$ comes from
real emissions in the limit of small $\vec v \, ^2$ and $\omega$.  Both
contributions can be computed as a series in the strong coupling constant:
$W_i = \sum_n \alpha_s^n \, W_i^{(n)}$. The expansion starts at $n=0$ for
$W_{\rm el}$ (tree level) while for $W_{\rm real}$ it starts at $n=1$ which
leads to the following expression for $\overline{\Lambda}$ (and similarly for
$\mu_\pi^2$)
\begin{equation}
  [\overline{\Lambda}(\mu)]_{\rm pert} =
      \lim_{\vec{v}\to0}\lim_{m_b\to\infty} 
      \frac{2}{\vec{v}\,^2} 
      \frac{\displaystyle
        \sum_{n=1} \alpha_s^n
	\int_0^\mu \omega \,
         \frac{\vec{v}\,^2}{\omega}
       W_{\rm real}^{(n)}(\omega,\vec v) \, {\rm d}\omega}
      {
      \displaystyle
      \sum_{n=0} \alpha_s^n 
      W_{\rm el}^{(n)}
  } \,.
    \label{eq::Lam_2}
\end{equation}

From Eq.(\ref{eq::Lam_2}) it is clear that we expand $W$ at most up to order
$\vec{v}\,^2$ because higher orders are eliminated by the limit
$\vec{v}\to 0$.  Moreover, we retain only the leading $1/\omega$ term since
higher orders, which scale as $(\omega/m_b)^n$, are eliminated by the limit
$m_b\to + \infty$.  Due to the factors $\omega^k$ ($k=1$ for
$\overline{\Lambda}$ and $k=2$ for $\mu_\pi^2$) in the integrand of the
numerator it is furthermore clear that the $\delta$-function distribution in
Eq.~(\ref{eq::W}) is only present in the denominator.  As a consequence the
virtual corrections are needed to one order less than the real radiation
contributions.  Vice versa, we can discard real corrections at the denominator
since, after expansion in $\alpha_s$, they become of order $\vec v \,^4$ and
so eliminated by the $\vec v \to 0$ limit.

\begin{figure}[t]
    \centering
    \begin{minipage}[b]{0.3\textwidth}
     \includegraphics[width=\textwidth]{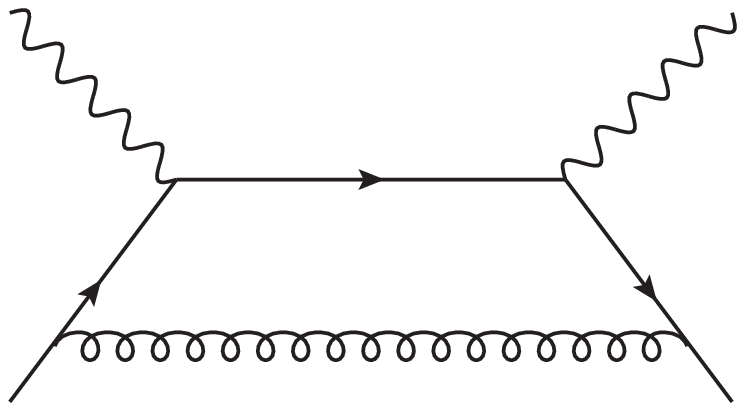}
    \end{minipage}
    \begin{minipage}[b]{0.3\textwidth}
     \includegraphics[width=\textwidth]{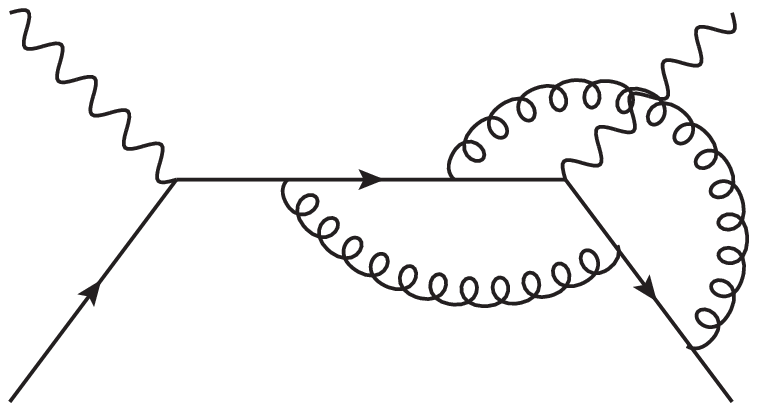}
    \end{minipage}
    \begin{minipage}[b]{0.3\textwidth}
     \includegraphics[width=\textwidth]{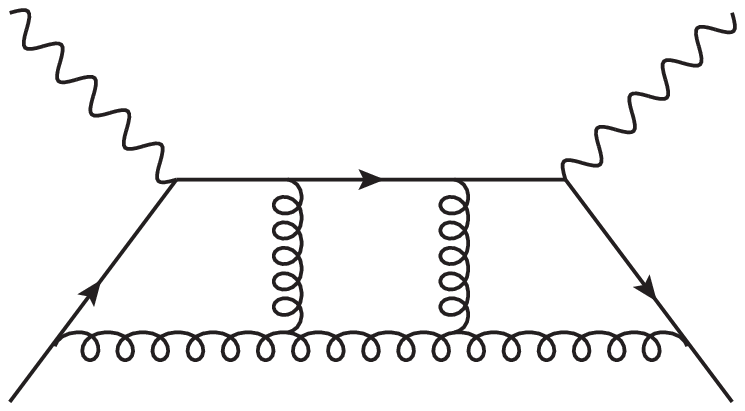}
    \end{minipage}
  \caption{\label{fig::sample_diags}Sample Feynman diagrams
    for the scattering process of an external current (wavy line)
    and a heavy quark (solid line). Gluons are represented by curly
    lines.}
\end{figure}

From Eqs.~(\ref{eq::mkin}) and~(\ref{eq::Lam_mupi}), we conclude that the
calculation of the kinetic mass up to order $\alpha_s^3$ reduces to the
computation of the function $W_{\rm real }(\omega)$ in Eq.~(\ref{eq::W}).
Two-loop virtual corrections to the heavy quark form factors are known
(cf. Section~\ref{sub::virt}). $W_{\rm real }(\omega)$ describes the
dipole radiation (cf.\ classical electrodynamics).  It is obtained from the
imaginary part of the forward scattering amplitude $T(q_0,\vec q)$ of a bottom
quark onto an external current $J$.  Examples of Feynman diagrams at one, two
and three loops are shown in Fig.~\ref{fig::sample_diags}.
%

Furthermore, for the practical calculation it is convenient to express the non-relativistic
quantities $\omega$ and $\vec{v}$ in terms of Lorenz invariants.
To this end we introduce 
\begin{eqnarray}
  y &\equiv& m_b^2 - s = - \omega \left( 2m_b \sqrt{ 1+\vec{v}\,^2 } + \omega \right)
  = - m_b \, \omega ( 2 + \vec{v}\,^2 ) + {O}(\omega^2,\vec{v}\,^4 )
  \label{eq:scaling1}
        \,, \\
  q^2 &\equiv& \left[ m_b \left( \sqrt{ 1+\vec{v}\,^2 } - 1 \right) + \omega \right]^2 - m_b^2 \vec{v}\,^2
  =  - m_b \, \vec{v}\,^2  ( m_b - \omega ) + {O}(\omega^2,\vec{v}\,^4 )
  \label{eq:scaling2}
  \,.
\end{eqnarray}
From these definitions, one can see that we can realize the
non-relativistic limits $\lim_{\vec{v}\to0}$ and $\lim_{m_b\to\infty}$
by expanding the amplitude $T$ around the threshold $s=(p+q)^2=m_b^2$ and a subsequent expansion in $q$.
In fact, we interpret $\lim_{m_b\to\infty}$ as an expansion in the quantity
\begin{eqnarray}
  y = m_b^2-s \le 0 \,,
\end{eqnarray}
which we realize with the help of {\it expansion by
  regions}~\cite{Beneke:1997zp,Smirnov:2012gma}.  The expansion $\vec{v}\to0$, on
the other hand, reduces to a naive Taylor expansion in $q$.  From the
definition of the kinetic mass and the relations in Eqs.~(\ref{eq:scaling1})
and~(\ref{eq:scaling2}) it is clear that we only have to consider terms up to
${O}(y^{-1})$ and ${O}(q^2)$.

Note that the two limits $\lim_{\vec{v}\to0}$ and $\lim_{m_b\to\infty}$ do not commute.
In case we apply first $\lim_{\vec{v}\to0}$ to $T$ there is no imaginary part.



\section{\label{sec::calc}Details of the calculation}

In this Section we provide technical details to our calculation and
discuss in particular the application of the method of
regions~\cite{Beneke:1997zp}, the reduction to master integrals and
the computation of the latter. 
We remark that in Ref.~\cite{Czarnecki:1997sz} no technical details 
for the calculation to $\mathcal{O}(\alpha_s^2)$ are provided.


\subsection{Method of regions}\label{sub::regions}
From Eqs.~(\ref{eq::mkin}) and~(\ref{eq::Lam_mupi}) we know that we have to
compute the imaginary part of $T(\omega, \vec v)$ in the limit
$m_b\to\infty$, which corresponds to an expansion around $y \to 0$. 
To this end, we apply the threshold expansion developed
in Ref.~\cite{Beneke:1997zp}, see also
Ref.~\cite{Smirnov:2012gma}. Ref.~\cite{Beneke:1997zp} considered the threshold expansion
of the heavy quark-photon vertex  and identified four different
scalings for the loop momenta: \textit{hard, soft, potential and
ultra-soft}. 
In our case, we only have to consider the threshold of one heavy
quark. Thus, the soft and potential regions lead to scaleless integrals, which
are set to zero within dimensional regularization.  We remain with two regions
(hard and ultra-soft) for each loop momentum $k_i$ ($i=1,2,3$) with the scalings
\begin{eqnarray}
  \mbox{hard (h):} && k_i \sim m_b\,,\nonumber\\
  \mbox{ultra-soft (us):} && k_i \sim y/m_b\,,
  \label{eq::h-us_scaling}
\end{eqnarray}
where $m_b$ is the heavy quark mass and $y=m_b^2-s$ (with $|y| \ll m_b^2$)
measures the distance to the
threshold. Note that in our case we have $y<0$.
When expanding the denominators we assume that both $p$ and $q$ scale as $m_b$.

At one-loop order, there are only two regions. At two loops, we have the regions
(uu), (uh) and (hh), and at three loops we have (uuu), (uuh), (uhh) and
(hhh). For each diagram, we have cross-checked the scaling of the loop momenta
using the program {\tt asy}~\cite{Pak:2010pt}. Note that the contributions
where all loop momenta are hard can be discarded since there are no imaginary
parts. The mixed regions are expected to cancel after renormalization and
decoupling of the heavy quark from the running of the strong coupling
constant. Nevertheless we performed an explicit calculation of the (uh), (uuh)
and (uhh) regions and used the cancellation as cross check.  The physical
result for the quark mass relation is solely provided by the purely ultra-soft
contributions.

A subtlety in connection with the expansion of the denominators arises at two
and three loops where either an individual loop momentum or a linear
combination of loop momenta can have a definite scaling.  Let us call
``naive regions'' those that can be obtained by assigning a definite scaling
to the loop momenta according to Eq.~(\ref{eq::h-us_scaling}), e.g.\ at two
loops (uu) $\equiv k_1, k_2 \sim y/m_b$, (hu) $\equiv k_1 \sim m_b$ and
$k_2 \sim y/m_b$, (hh) $\equiv k_1, k_2 \sim m_b$.

In case a linear combination of loop momenta flows through a gluon line,
it might happen that fewer regions are found than actually
exist.  No such problem appears with the (heavy) quark lines since they always
have a hard component.

\begin{figure}[htb]
  \centering
  \begin{minipage}[b]{0.3\textwidth}
     \includegraphics[width=\textwidth]{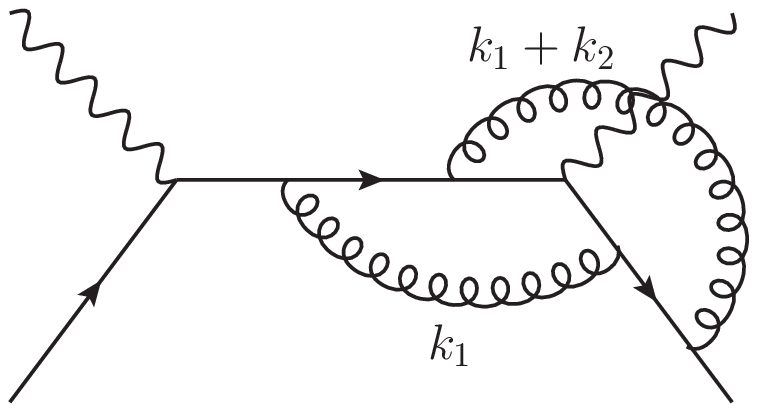}
     \centering (a)
  \end{minipage}\quad
  \begin{minipage}[b]{0.3\textwidth}
     \includegraphics[width=\textwidth]{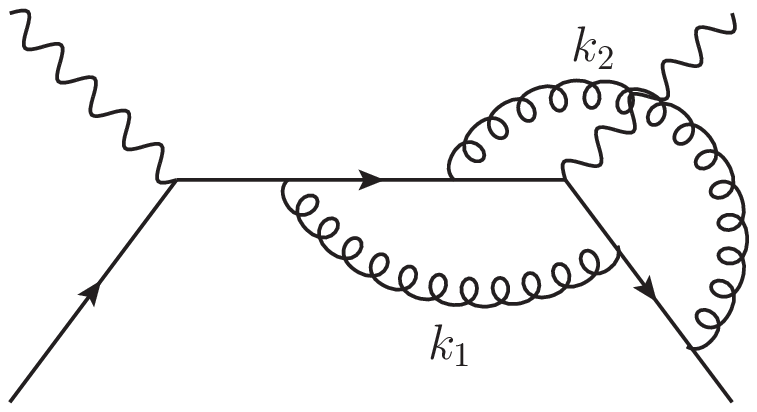}
     \centering (b)
  \end{minipage}
  \caption{Two possible momentum routing of a two-loop diagram.
    The naive regions in the first case (a) do not correspond to all regions 
    while the routing in (b) correctly reveals all regions.
  }
  \label{fig:scaling}
\end{figure}

Let us, e.g., consider the two-loop diagram in Fig.~\ref{fig:scaling} and let
us assume that $k_1+k_2$ flows through one of the gluon lines, as shown in
Fig.~\ref{fig:scaling}(a).  If both loop momenta are ultra-soft there is no
problem.  In case $k_1$ is hard and $k_2$ is ultra-soft,
the gluon line is always hard and there is no imaginary part.  Thus one has to
consider the case where $k_1+k_2$ is ultra-soft and both $k_1$ and $k_2$ are
hard.  On the contrary, if we consider the momentum routing shown in
Fig.~\ref{fig:scaling}(b), the gluon line can be ultra-soft $k_2 \sim y/m_b$,
while the other gluon can be hard $k_1 \sim m_b$. With this second routing we
see that the naive regions cover all possibilities.
Therefore for certain choices of momentum routing, 
the restriction of the scaling to individual loop momenta 
might miss some of the regions as it ignores potential ultra-soft scaling 
of linear combinations. 

To be sure that we considered all relevant regions, we proceeded as follows:
for each diagram we checked that the number of naive regions
and the scaling of individual loop momenta according to
Eq.~(\ref{eq::h-us_scaling}) agree with those found by {\tt
  asy}~\cite{Pak:2010pt}.  If we found fewer regions then we re-routed the
loop momenta through the gluon lines, applied the scaling rules and checked
again against {\tt asy}.

As mentioned before, we have to compute the expansions of the individual
diagrams up to $\mathcal{O}(y^{-1},q^2)$.  The expansion in $y$ is implemented
with the help of expansion by regions and we thus have a definite power
counting for the leading behaviour in $y$ for individual terms.  However, the
Taylor expansion in the momentum $q$ is effectively an expansion in the scalar
products
\begin{eqnarray}
  q^2 \quad \text{and} \quad p \cdot q = -\frac{1}{2} \left( y + q^2 \right) ~.
\end{eqnarray}
In order not to miss terms up to the desired order, we have to expand 
sufficiently deep in $q$.
Since the worst scaling in the ultra-soft region is $\sim y^{-1}$, we have to consider
two terms in the $q$ expansion at least.
The mixed regions at three-loop order show a behaviour $\sim y^{-3}$.
Here we have to consider up to six terms in the $q$ expansion,
which leads to high numerator and denominator powers. 


\subsection{\label{sub::sing}Singlet-type diagrams}

Let us in the following discuss the diagrams where one or both external
currents couple to a closed massive fermion loop,
which is connected to the external heavy quark line via gluons.
We refer to these contributions as ``singlet-type'' diagrams.
They occur for the first time at two loops.

The momentum $p$ is always routed through the heavy quark line.
As we have seen above, a diagram develops an imaginary part only in those
cases where the external heavy quark line is part of a ultra-soft
loop and carries the
external momentum $p+q$, which leads to a ``$-y$'' term in the denominator.

\begin{figure}[t]
  \centering
  \begin{minipage}[b]{0.23\textwidth}
     \includegraphics[width=\textwidth]{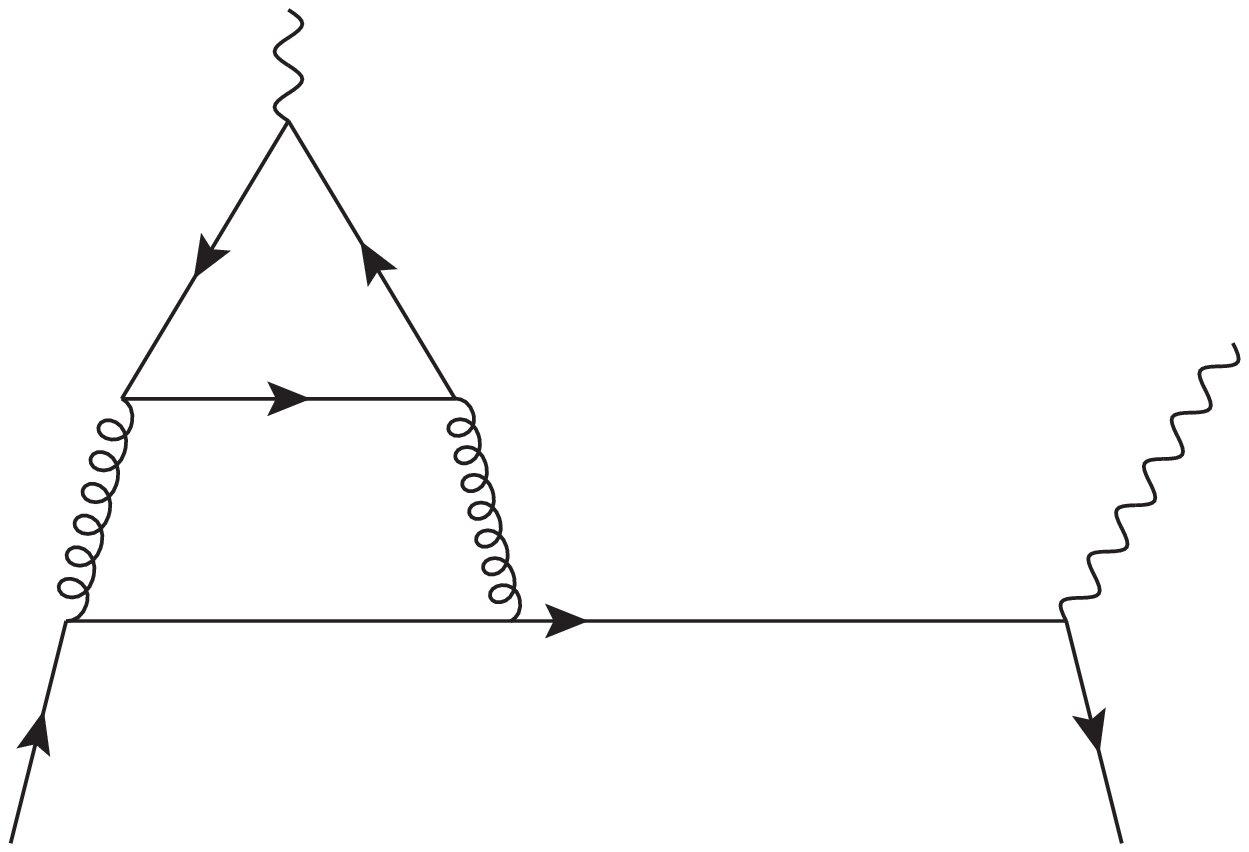}
     \centering (a)
  \end{minipage}\quad
  \begin{minipage}[b]{0.23\textwidth}
     \includegraphics[width=\textwidth]{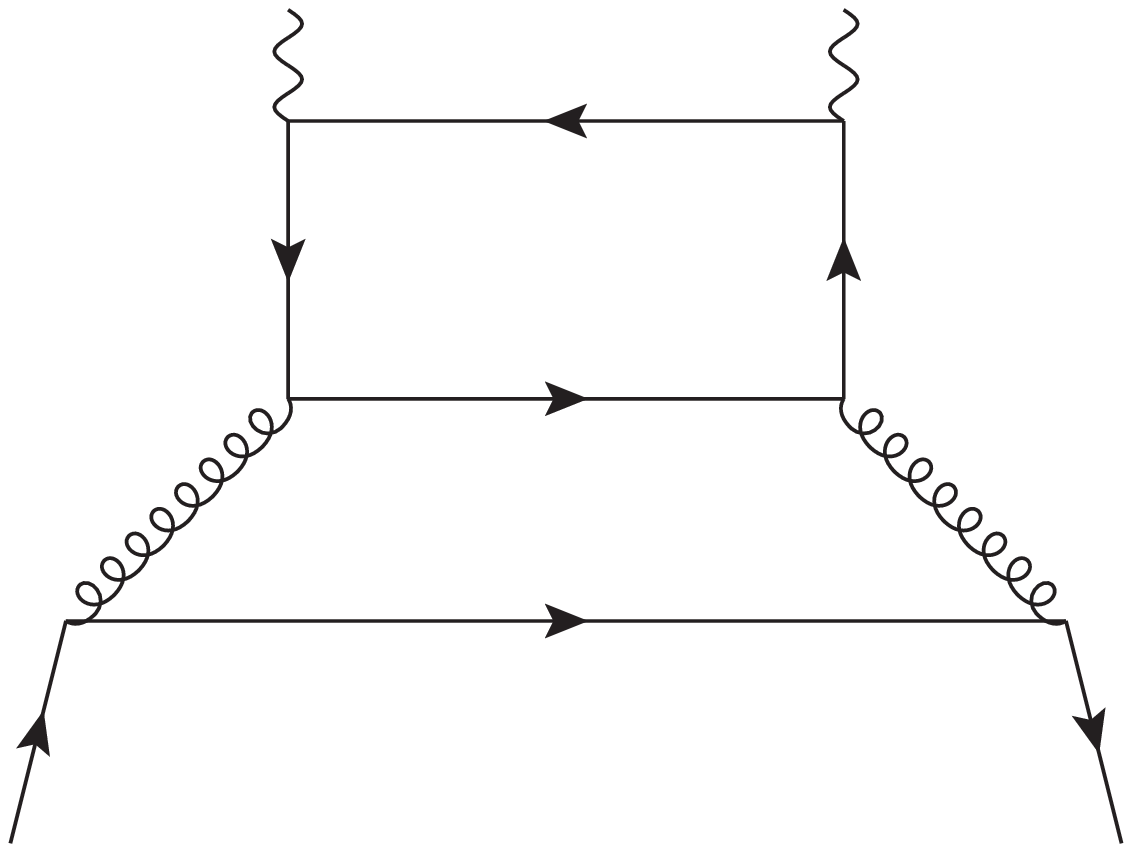}
     \centering (b)
  \end{minipage}\quad
  \begin{minipage}[b]{0.23\textwidth}
     \includegraphics[width=\textwidth]{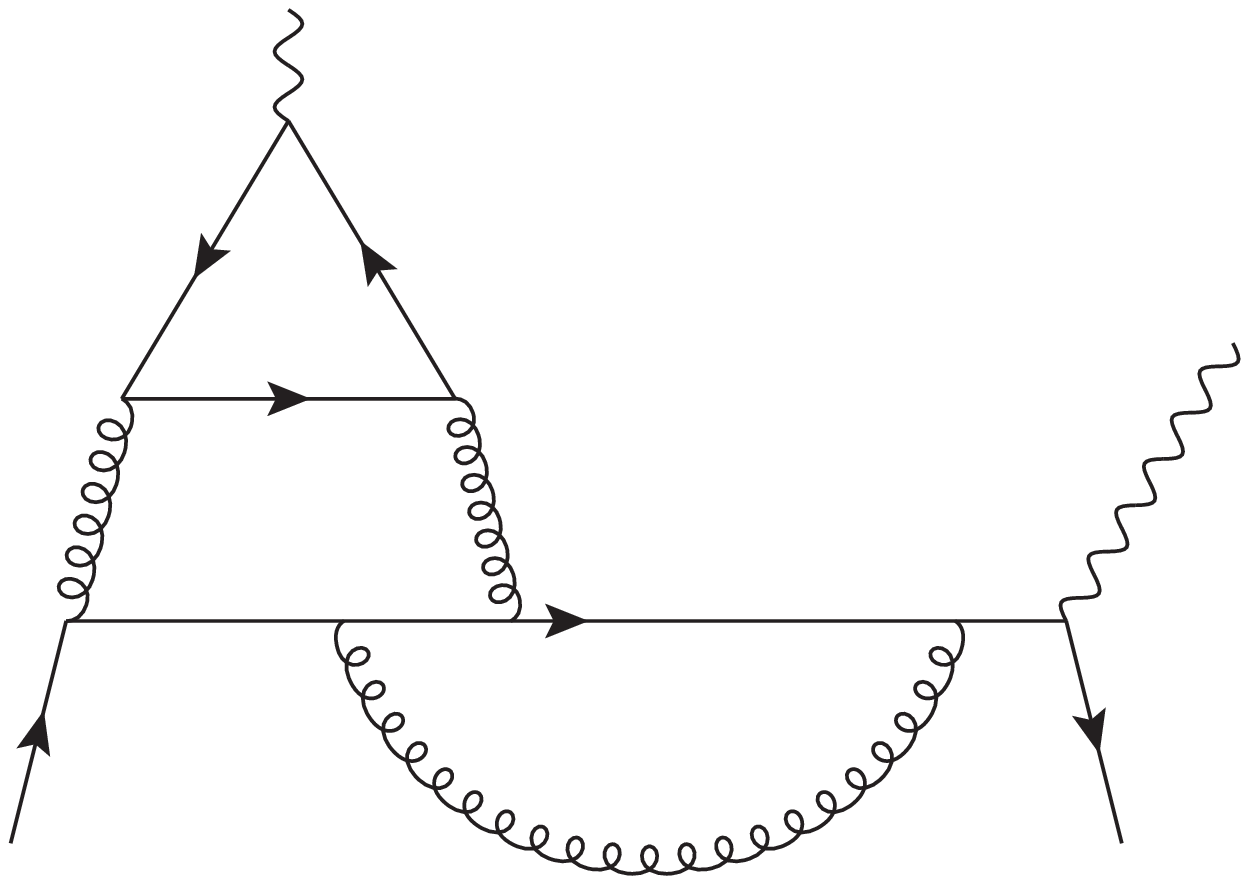}
     \centering (c)
  \end{minipage}\quad
  \begin{minipage}[b]{0.23\textwidth}
     \includegraphics[width=\textwidth]{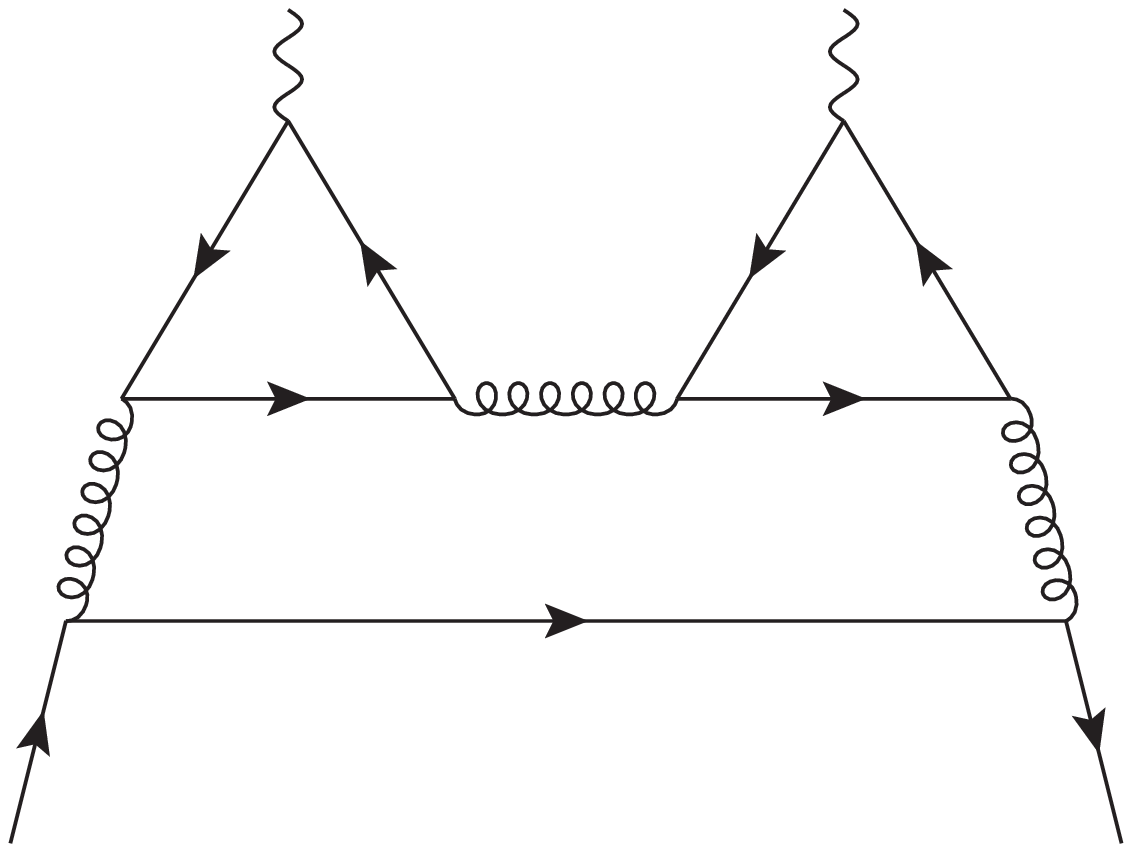}
     \centering (d)
  \end{minipage}
  \caption{\label{fig::diags}
    Sample singlet-type Feynman diagrams. 
    The external currents are drawn with wavy lines, 
    heavy quarks with solid lines and gluons are represented by curly lines.
    Diagrams (a), (c) and (d) are zero for an external vector current 
    but not for an external  scalar current.}
\end{figure}
At two loops ``singlet-type'' diagrams appear in two
versions:
\begin{itemize}
\item One external current couples to a quark triangle which is
  connected with two gluons to the external heavy quark line (see Fig.~\ref{fig::diags}(a)).
  The other external current is directly connected to the latter.

  Such contributions have no heavy quark line which is part of a ultra-soft
  loop and carries the momentum $p+q$.  In fact, the application of the method
  of regions together with the condition that at least one of the loops is
  ultra-soft, immediately leads to scaleless integrals.  For vector currents,
  such contributions are zero due to Furry's theorem.

\item In a second class of diagrams, the two external currents couple to
  a quark box which is connected with two gluons to the external heavy quark
  line (see Fig.~\ref{fig::diags}(b)). Again, no imaginary part can be developed 
  at the threshold $s = m_b^2$.

\end{itemize}

At three loops there are the same two classes of Feynman diagrams as at two
loops, supplemented by an additional gluon.  After applying the same arguments
it is easy to see that also here no contribution to the imaginary part of
$T(q_0,\vec{q}\,)$ in Eq.~(\ref{eq::Tkin}) can be constructed, with the exception
of diagrams like that one in Fig.~\ref{fig::diags}(c).
In these diagrams, one of the currents couples to a quark triangle that is
connected to the external heavy quark line  with two gluons.
An additional gluon couples only to the (external) heavy quark
forming the third loop.
In that case, the first two loops can be hard and the third loop
develops an imaginary part in analogy to the one-loop contribution.

Note that due to Furry's theorem, these kind of diagrams vanish for external
vector currents.  However, for the scalar currents they lead to non-zero
contributions. We have checked that they cancel against the virtual
corrections, which at two-loop order also have contributions from singlet
diagrams, see Section~\ref{sub::virt}. 

There is also a three-loop contribution where both currents are
connected to different closed fermion loops (Fig.~\ref{fig::diags}(d)) that
are connected to each other and to the external heavy quark line. Here the
loop momenta of the closed quark loops are hard, but the third loop momentum
can be ultra-soft and in principle produce an imaginary part.  However, 
an explicit calculation shows that these kind of diagrams scale $\sim y^0$
and therefore do not enter in the relation for the kinetic mass.


\subsection{Vector and scalar currents}

As external currents, we use for our calculation both
a vector and a scalar current which in coordinate space
are given by
\begin{eqnarray}
  J^\mu_V &=& \bar{b}(x) \gamma^\mu b(x)\,, \nonumber\\
  J_S &=& m_b \, \bar{b}(x) b(x)\,.
          \label{eq::J}
\end{eqnarray}
In the case of $J_S$ we introduce the factor $m$ such that
$J_S$ has vanishing anomalous dimension.
Note that $m$ enters the same renormalization procedure as the
mass parameter in the heavy quark propagators.

In spinor space the amplitude $T$ can be written as (ignoring Lorentz indices for
an external vector current) $T = p\!\!\!/\,\, \Sigma_V + m_b \Sigma_S$. We
multiply by $p\!\!\!/\,\, + m_b $ and take the trace. This leads to
\begin{eqnarray}
  \mbox{Tr}\left[ ( p\!\!\!/\,\, + m_b ) T \right] 
  &=&
      4 m_b^2 \left(\Sigma_V + \Sigma_S\right)\,.
\end{eqnarray}

In the case of $J_V$ the forward scattering amplitude $T$ 
becomes a tensor of rank~2 and can be parameterized through 
two structure functions $T_A$ and $T_B$, which we define as
\begin{eqnarray}
  T^{\mu\nu} &=& T_A \left( g^{\mu\nu} - \frac{q^\mu q^\nu}{q^2} \right) 
                 + T_B \left( \frac{p^\mu p^\nu}{p\cdot q} - \frac{p^\mu q^\mu
                 + p^\nu q^\mu}{q^2} + \frac{p\cdot q}{q^2} g^{\mu\nu}
                 \right)\,.
\end{eqnarray}
We have used the symmetry of the forward scattering amplitude
$T^{\mu\nu} = T^{\nu\mu}$ and the transversity
$q_\mu T^{\mu\nu} = q_\nu T^{\mu\nu} = 0$.

One can construct projectors on $T_A$ and $T_B$ which can be written
as linear combinations of the two structures
\begin{eqnarray}
    P_1^{\mu\nu} &=& \frac{p^\mu p^\nu}{m_b^2}\,,
    \\
    P_2^{\mu\nu} &=& g^{\mu\nu}\,.
\end{eqnarray}
We find
\begin{eqnarray}
    (d-2) P_A^{\mu\nu} &=& \frac{m_b^2 q^2 }{m_b^2 q^2 - (p\cdot q)^2}P_2^{\mu\nu} - \frac{m_b^2 q^2 \left((d-2) (p\cdot q)^2 + m_b^2 q^2 \right)}{\left((p\cdot q)^2-m_b^2 q^2 \right)^2} P_1^{\mu\nu} 
    \nonumber \\
    &\stackrel{q^2\to0}{\longrightarrow}& P_2^{\mu\nu} - P_1^{\mu\nu} + \frac{y}{2 m_b^2} \left( P_2^{\mu\nu} - d P_1^{\mu\nu} \right)\,,
                                          \label{eq::PA}
    \\
    (d-2) P_B^{\mu\nu} &=& \frac{(d-1) m_b^2 p\cdot q (q^2)^2 }{\left((p\cdot q)^2-m_b^2 q^2 \right)^2} P_1^{\mu\nu} + \frac{p\cdot q q^2 }{p\cdot q^2-m_b^2 q^2} P_2^{\mu\nu}  
    \nonumber \\
    &\stackrel{q^2\to0}{\longrightarrow}& \frac{2y}{m_b^2} \left( P_2^{\mu\nu} - (d-1) P_1^{\mu\nu} \right)\,.
       \label{eq::proj}
\end{eqnarray}
For both projectors the limits for $y \to 0$ and $q^2 \to 0$ exist.
Thus, in practice we can simply apply $P_{1}^{\mu\nu}$ and
$P_{2}^{\mu\nu}$ and construct the physical structure functions
afterwards by considering the proper linear combinations. It is
interesting to note that $P_{1}^{\mu\nu}$ and $P_{2}^{\mu\nu}$ applied
to $T^{\mu\nu}$ lead to a scaling $\sim 1/y$. For this reason the term
$P_2^{\mu\nu} - P_1^{\mu\nu}$ in Eq.~(\ref{eq::PA}) has to vanish and
both $P_A^{\mu\nu}$ and $P_B^{\mu\nu}$ (considered as linear
combinations of $P_{1}^{\mu\nu}$ and $P_{2}^{\mu\nu}$) have to scale
$\sim y$ in the limit $q^2 \to 0$.  As a consequence, we can apply
either $P_{1}^{\mu\nu}$ or $P_{2}^{\mu\nu}$ to compute the kinetic
mass.  The difference to the application of proper linear combinations
(i.e. $P_A$ and $P_B$) is a $d$- and $m$-dependent prefactor which
drops out in the definition of the quantities
$\overline{\Lambda}(\mu)$ and $\mu_\pi^2(\mu)$ from
Eq.~(\ref{eq::Lam_mupi}).


\subsection{Virtual corrections}\label{sub::virt}

Virtual corrections enter the denominator of Eq.~(\ref{eq::Lam_2}).  For
the two- and three-loop kinetic mass we need one- and two-loop virtual
corrections, respectively. Furthermore, we only need the static limit
($q^2=0$), which can be obtained, e.g.,
from~\cite{Lee:2018rgs,Blumlein:2018tmz}.  Note that in this limit the form
factors are infrared finite (as are the real radiation corrections which we
compute).

For the case of the vector current the effective vertex $\Gamma^V_{\mu}$
can be expressed in terms of two form factors contributing to the virtual
corrections
\begin{eqnarray}
  \Gamma^V_\mu &=& - i \left( 
                   F_1 \gamma_\mu + F_2 \frac{i}{2m_b} \sigma_{\mu\nu} q^\nu  \right),
\end{eqnarray}
with $\sigma_{\mu\nu} = \frac{i}{2}\left[ \gamma_\mu , \gamma_\nu \right]$.
After inserting $\Gamma^V_\mu$ into the tree level expression, we see that the
contribution of the virtual corrections is given by
\begin{eqnarray}
  V_i &=& \delta(y) \text{Tr} \left[ (\slashed{p} + m_b) {\Gamma^V_\mu}^* 
          ( \slashed{p} + \slashed{q} + m_b ) \Gamma^V_\nu \right] P_i^{\mu\nu} ,
    \label{eq:virtual1}
\end{eqnarray}
with $P_1^{\mu\nu}=p^{\mu} p^{\nu}/m_b^2$ and $P_2^{\mu\nu}=g_{\mu\nu}$.
The delta function $\delta(y)$ ensures that we have $s=m_b^2$.
We find
\begin{eqnarray}
  V_1 &=& \delta(y) \left[ -2 \left|F_1\right|^2 \left( q^2 - 4 m_b^2 \right) + \left|F_2\right|^2 \frac{q^4}{8m_b^4} \left( q^2 - 4 m_b^2  \right) \right]\,,
  \nonumber\\
  V_2 &=& \delta(y) \left[ \left|F_1\right|^2 \left( 2 (d-2) q^2 + 8 m_b^2 \right) + \left|F_2\right|^2 \frac{q^2}{2} \left( 8 - 4 d - \frac{q^2}{m_b^2} \right) \right]\,.
  \label{eq::V1V2}
\end{eqnarray}
From the definition of the kinematic mass we see that virtual
corrections always multiply lower-order real emissions (which
vanish for $q^2\to 0$). Therefore only the non-vanishing parts of
Eq.~(\ref{eq::V1V2}) in the limit $q^2 \to 0$
contribute, which is proportional to $|F_1|^2$. Note, however, that
$F_1$ has a vanishing static limit to all orders in perturbation
theory $F_1(q^2=0)=0$ and thus the kinetic mass does not receive
contributions from virtual corrections in the case of an external
vector current.

This is different for the scalar current. We define
\begin{eqnarray}
    \Gamma^S &=& -i F_S
                 \,.
\end{eqnarray}
which leads to
\begin{eqnarray}
    V_S &=& \delta(y) \left|F_S\right|^2 \left[ 8m_b^2 - 2 q^2  \right]
            \,.
\end{eqnarray}
Since $F_S(q^2=0) \neq 0$ we are left with a non-vanishing contribution.
For the three-loop correction to the $m^{\rm kin}$--$m^{\rm OS}$ relation we
need $F_S(q^2=0)$ up to two loops which is given
by (see, e.g., Refs.~\cite{Lee:2018rgs,Blumlein:2018tmz})
\begin{eqnarray}
  F_S(0) &=& \frac{\alpha_s}{4\pi} C_F
  \left( 
    -2 + 3 l_m
  \right)
  + \left(\frac{\alpha_s}{4\pi}\right)^2 C_F
  \biggl\{
    C_A \biggl[
      -6 \zeta _3
      +\frac{11 l_m^2}{2}
      +\frac{53 l_m}{6}
      +\pi ^2 \left(4l_2-\frac{4}{3}\right)
\nonumber \\ &&
      -\frac{123}{8}
    \biggr]
    +C_F \left[
      12 \zeta _3
      +\frac{9 l_m^2}{2}
      -\frac{9 l_m}{2}
      +\pi ^2 \left(5-8 l_2\right)
      +\frac{193}{8}\right]
\nonumber \\ &&
      +T_F\left[
        n_h \left(
          -2 l_m^2
          -\frac{2 l_m}{3}
          -\frac{8 \pi^2}{3}
          +\frac{51}{2}
        \right)
      +n_l \left(
        -2 l_m^2
        -\frac{2 l_m}{3}
        -\frac{4 \pi^2}{3}
        +\frac{11}{2}
      \right) 
    \right]
\nonumber \\ &&
    + T_F n_h \left(\frac{16 \pi ^2}{3}-32\right) 
  \biggr\} \,,
  \label{eq:virtalS}
\end{eqnarray}
with $l_m = \ln(\mu_s^2/m_b^2)$ and $l_2=\ln(2)$. $C_A=N_C$ and
$C_F=(N_C^2-1)/(2N_C)$ are SU($N_C$) colour factors, $T_F=1/2$, $n_l$ is the
number of massless quarks and $n_h=1$ is introduced for convenience for
closed loops of fermions with mass $m_b$. The last term in
Eq.~(\ref{eq:virtalS}) corresponds to the contributions from singlet-type
diagrams.  Note that our final result does not depend on the renormalization
scheme used for the external currents. In fact, the vector current does not
get renormalized and in the case of the scalar current we renormalize the mass
parameter $m_b$ introduced in Eq.~(\ref{eq::J}) in the
$\overline{\mathrm{MS}}$ scheme.


\subsection{Partial fraction decomposition}

The  starting point of our calculation are 
four-point functions with forward-scattering
kinematics. 
After we Taylor-expand in $q$, we
remain with only one external momentum, which is present
in the denominators. Thus, at most
2, 5 and 9 denominators can be linear independent
at one, two and three loops, respectively.
On the other hand, general four-point functions
contain up to 4, 7 and 10 lines and thus, in general,
a partial fraction decomposition is required, which
decomposes products of linear dependent propagators
into terms with only linear
independent factors.

At one- and two-loop order, it is straightforward to implement
the partial fraction decomposition manually. However, at three loops
many different cases appear and an automation of the procedure is
recommended. In our calculation we use the program \texttt{LIMIT} developed by
Florian Herren~\cite{Davies:2019xzc,Herren:2020ccq}.
The program is written in \texttt{Mathematica} and internally uses \texttt{LiteRed}~\cite{Lee:2012cn}.
Let us briefly summarize its mode of operation.

We start by grouping diagrams according to their denominator 
structure into preliminary families, which we supplement
with irreducible numerators in order to have complete families.
This is a necessary step for the reduction to master integrals which
is performed at a later stage.
Note that some of the denominators can still be linearly
dependent. Furthermore, at this step we do not apply any symmetry
transformation to minimize the number of different families.
The goal of the program is to find all relations due to
partial fraction decomposition. Afterwards the
resulting set of families is minimized.

In the first part the program goes through the list of denominators of
each family, selects those that are linearly dependent and produces
replacement rules that allow for partial fraction decomposition after
their iterative application.  This step has to be applied recursively
to ensure that all denominators are linearly independent.  Note that
partial fraction decomposition increases the number of families. In
our application we start at two loops with \{48,16\} in the
\{(uu),(uh)\} regions and we end up with \{90,23\} families with linearly
independent denominators.  At three loops, we have \{510,339,314\}
families in the \{(uuu),(uuh),(uhh)\} regions which result in
\{2650,906,531\} families after partial fraction decomposition.

Many of the resulting families are equivalent and can be mapped onto each
other.  The second part of the program finds these relations and provides
rules to map the scalar integrals into a minimal set of families.  The program
relies on \texttt{LiteRed} to find these rules.  In general the program has to
find two types of mappings.  The first type corresponds to mappings between
families that differ only by their momentum routing. These mappings are
obtained by computing the $U$ and $F$ polynomials for all families and using
the \texttt{LiteRed} command \texttt{FindExtSymmetries[]} to map all families
with the same polynomials to a representative one.  The second type
corresponds to mappings of families with a larger number of numerators onto
families with fewer numerators but more denominators.  All replacement rules
can be exported to \texttt{FORM}~\cite{Ruijl:2017dtg} statements.

In total we find \{2,2\} families in the \{(uu),(uh)\} regions at two loops
and at three-loop order \{14,4,3\} in the \{(uuu),(uuh),(uhh)\} regions, respectively.
Their definitions are given in the next subsection.


\subsection{Integral families and reduction to MIs}

After partial fraction decomposition and mapping of equivalent families to
each other, we are left with only a small number of families. In general they
have a number of irreducible numerators which are either formed by the scalar
product of the loop momenta with the external momenta $k_i\cdot q$, $k_i\cdot p$ 
or scalar products of loop momenta $k_i \cdot k_j$. They appear in particular in those
cases where both hard and ultra-soft regions are present since the integrals
factorize.  In principle one can apply a tensor decomposition to get rid of
such scalar products. However, we chose to include them into the
definition of the integral families. Thus, also for the cases
where the (two- and three-loop) integrations factorize we pass the 
corresponding scalar functions to the reduction programs \texttt{LiteRed}~\cite{Lee:2012cn} and
\texttt{Fire}~\cite{Smirnov:2019qkx}, which means that effectively the tensor reduction is performed
by these programs.
Note that in such cases all master integrals factorize into
a hard and ultra-soft part.

Since the expansions in the mixed regions have to be quite deep in order to
calculate the diagrams up to $\mathcal{O}(y^{-1},q^2)$, the indices of the
scalar integrals become large.  In the mixed regions, we had to reduce about
$10^6$ integrals with the absolute value of the indices reaching up to 12.  In
the ultra-soft region, we only had to reduce about $10^5$ scalar integrals
with indices reaching up to 6.  Nevertheless, reducing these integrals using
either \texttt{LiteRed} or \texttt{Fire} took roughly two days.  We observed
that in particular in the mixed regions \texttt{LiteRed} performed
better in those cases where high numerator and denominator powers had to be
reduced.

In the following we provide the definition of the integral families up to
three loops where the factors after the semi-colon correspond to numerators.
We do not show the ``$-i0$'' prescription which is present
in all denominator factors.  At two- and three-loop order we have both the
pure-ultra-soft and the mixed hard-ultra-soft regions.

At one-loop order we only have one family which is given by
\begin{center}
  \begin{tabular}{lll}
    fam1lu: & $-k_1^2, -2k_1\cdot p + y;$ & $-k_1\cdot q$ 
  \end{tabular}
\end{center}

At two loops we have two families in the (uu)-region:
\begin{center}
\begin{tabular}{lll}
  fam2luu1: & $-k_1^2, -(k_1-k_2)^2, -k_2^2, -2p\cdot k_1 + y, -2 p\cdot k_2 + y;$ & $- q\cdot k_1, - q\cdot k_2 $;  \\
  fam2luu2: & $-k_1^2, -(k_1-k_2)^2, -k_2^2, -2p\cdot k_1 + y, -2 p\cdot k_2;$ &  $- q\cdot k_1,-q\cdot k_2; $
\end{tabular}
\end{center}
and one family in the uh-region:
\begin{center}
\begin{tabular}{lll}
    fam2luh1: & $-k_1^2, -k_2^2, -(k_2+p)^2 + m_b^2, -2 p\cdot k_1+y;$& 
    $-k_1\cdot k_2, - k_1\cdot q, - k_2\cdot q$.
\end{tabular}
\end{center}
Here also the scalar product $k_1\cdot k_2$ is an irreducible numerator.
For the calculation with a massive charm quark (see Section~\ref{sec::charm})
we have in addition the following family:
\begin{center}
\begin{tabular}{lll}
    fam2luh2: & $-k_1^2, -k_2^2 + m_c^2, -2 p\cdot k_1+y;$& 
    $-k_1\cdot k_2,-k_2\cdot p , - k_1\cdot q, - k_2\cdot q$.
\end{tabular}
\end{center}

At three loops we have 14 families in the (uuu), two in the (uuh) and two in
the (uhh) regions for the calculation with a massless charm quark.
A massive charm quark requires two and one additional families in the (uuh)
and (uhh) regions, respectively. All definitions are given in Tab.~\ref{tab::fam3}.

\begin{table}[p]
  \begin{center}
    {\scalefont{0.75}
    \begin{tabular}{ll}
      \hline\\
      fam3luuu1:
      &$-k_1^2,-k_2^2,-k_3^2,-(k_1-k_3)^2,-(k_2-k_3)^2,-2 p\cdot k_1+y,-2 p\cdot k_2+y,-2 p\cdot k_3+y;$
      \\&$-k_1\cdot k_2,-q\cdot k_1,-q\cdot k_2,-q\cdot k_3$
      \\
      fam3luuu2: 
      &$-k_1^2,-k_2^2,-(k_1-k_2)^2,-(k_1-k_3)^2,-(k_2-k_3)^2,-2 p\cdot k_1+y,-2 p\cdot k_2+y,-2 p\cdot k_3+y;$
      \\&$-k_3^2,-q\cdot k_1,-q\cdot k_2,-q\cdot k_3$
      \\
      fam3luuu3:
      &$-k_1^2,-k_2^2,-k_3^2,-(k_1-k_2)^2,-(k_1-k_3)^2,-(k_2-k_3)^2,-2 p\cdot k_1+y,-2 p\cdot k_2+y;$
      \\&$-p\cdot k_3,-q\cdot k_1,-q\cdot k_2,-q\cdot k_3$
      \\
      fam3luuu4:
      &$-2 p\cdot k_2,-k_2^2,-(k_1-k_2)^2,-k_3^2,-(k_1-k_3)^2,-(k_1-k_2-k_3)^2,-2 p\cdot k_1+y,-2 p\cdot k_3+y;$
      \\&$-k_1^2,-q\cdot k_1,-q\cdot k_2,-q\cdot k_3$
      \\
      fam3luuu5: 
      &$-2 p\cdot k_1,-2 p\cdot k_2,-k_2^2,-(k_1-k_2)^2,-k_3^2,-(k_1-k_3)^2,-(k_1-k_2-k_3)^2,-2 p\cdot k_3+y;$
      \\&$-k_1^2,-q\cdot k_1,-q\cdot k_2,-q\cdot k_3$
      \\
      fam3luuu6:
      &$-k_1^2,-2 p\cdot k_2,-k_2^2,-(k_1-k_2)^2,-k_3^2,-(k_1-k_3)^2,-2 p\cdot k_1+y,-2 p\cdot k_3+y;$
      \\&$-k_2\cdot k_3,-q\cdot k_1,-q\cdot k_2,-q\cdot k_3$
      \\
      fam3luuu7:
      &$-2 p\cdot k_1,-k_1^2,-2 p\cdot k_2,-k_2^2,-(k_1-k_2)^2,-k_3^2,-(k_1-k_3)^2,-2 p\cdot k_3+y;$
      \\&$-k_2\cdot k_3,-q\cdot k_1,-q\cdot k_2,-q\cdot k_3$
      \\
      fam3luuu8:
      &$-2 p\cdot k_1,-k_1^2,-(k_1-k_2)^2,-k_3^2,-(k_1-k_3)^2,-(k_2-k_3)^2,-2 p\cdot k_2+y,-2 p\cdot k_3+y;$
      \\&$-k_2^2,-q\cdot k_1,-q\cdot k_2,-q\cdot k_3$
      \\
      fam3luuu9:
      &$-2 p\cdot k_1,-k_1^2,-2 p\cdot k_2,-(k_1-k_2)^2,-k_3^2,-(k_1-k_3)^2,-(k_2-k_3)^2,-2 p\cdot k_3+y;$
      \\&$-k_2^2,-q\cdot k_1,-q\cdot k_2,-q\cdot k_3$
      \\
      fam3luuu10:
      &$-k_1^2,-2 p\cdot k_2,-k_2^2,-(k_1-k_2)^2,-k_3^2,-(k_1+k_3)^2,-(k_2+k_3)^2,-2 p\cdot k_1+y;$
      \\&$-p\cdot k_3,-q\cdot k_1,-q\cdot k_2,-q\cdot k_3$
      \\
      fam3luuu11:
      &$-2 p\cdot k_1,-k_1^2,-k_2^2,-(k_1-k_2)^2,-2 p\cdot k_3,-k_3^2,-(k_2-k_3)^2,-2 p\cdot k_2+y;$
      \\&$-k_1\cdot k_3,-q\cdot k_1,-q\cdot k_2,-q\cdot k_3$
      \\
      fam3luuu12: 
      &$-2 p\cdot k_1,-k_1^2,-2 p\cdot k_2,-k_2^2,-(k_1-k_2)^2,-(k_1-k_3)^2,-(k_2-k_3)^2,-2 p\cdot k_3+y;$
      \\&$-k_3^2,-q\cdot k_1,-q\cdot k_2,-q\cdot k_3$
      \\
      fam3luuu13:
      &$-k_1^2,-k_2^2,-(k_1-k_3)^2,-(k_2-k_3)^2,-(k_1+k_2-k_3)^2,-2 p\cdot k_1+y,-2 p\cdot k_2+y,-2 p\cdot k_3+y;$
      \\&$-k_3^2,-q\cdot k_1,-q\cdot k_2,-q\cdot k_3$
      \\
      fam3luuu14:
      &$-2 p\cdot k_1,-k_1^2,-2 p\cdot k_2,-k_2^2,-(k_1-k_3)^2,-(k_2-k_3)^2,-(k_1+k_2-k_3)^2,-2 p\cdot k_3+y;$
      \\&$-k_3^2,-q\cdot k_1,-q\cdot k_2,-q\cdot k_3$
      \\
      \hline\\
      fam3luuh1:&$- k_1^2,-k_2^2,-(k_1-k_2)^2,-k_3^2,-(k_3+p)^2-m_b^2,-2 p\cdot k_1+y,-2 p\cdot k_2+y;$
      \\ &$-k_1\cdot k_3,-k_2\cdot k_3,-q\cdot k_1,-q\cdot k_2,-q\cdot k_3$
      \\
      fam3luuh2:&$- k_1^2,-k_2^2,-2 p\cdot k_1+2 p\cdot k_2,-(k_1+k_2)^2,-k_3^2,-(k_3+p)^2-m_b^2,-2 p\cdot k_2+y;$
      \\ &$-k_1\cdot k_3,-k_2\cdot k_3,-q\cdot k_1,-q\cdot k_2,-q\cdot k_3$
      \\
      fam3luuh3:&$- k_1^2,-k_2^2,-(k_1-k_2)^2,-k_3^2+m_c^2,-2 p\cdot k_1+y,-2 p\cdot k_2+y;$
      \\ &$-k_1\cdot k_3,-k_2\cdot k_3,-p\cdot k_3,-q\cdot k_1,-q\cdot k_2,-q\cdot k_3$
      \\
      fam3luuh4:&$- k_1^2,-k_2^2,-(k_1-k_2)^2,-k_3^2+m_c^2,-2 p\cdot k_1+y,-2 p\cdot k_2;$
      \\ &$-k_1\cdot k_3,-k_2\cdot k_3,-p\cdot k_3,-q\cdot k_1,-q\cdot k_2,-q\cdot k_3$
      \\
      \hline\\
      fam3luhh1:&$- k_1^2,-k_2^2,-2 p\cdot k_2+k_2^2,-k_3^2,-2 p\cdot k_3+k_3^2,-2 p\cdot k_2+2 p\cdot k_3+k_2^2+2 k_2\cdot k_3+k_3^2,-2 p\cdot k_1+y;$
      \\&$-k_1\cdot k_2,-k_1\cdot k_3,-q\cdot k_1,-q\cdot k_2,-q\cdot k_3$
      \\
      fam3luhh2:&$- k_1^2,-k_2^2,-2 p\cdot k_2+k_2^2,-k_3^2,-2 p\cdot k_3+k_3^2,-k_2^2-2 k_2\cdot k_3+k_3^2,-2 p\cdot k_1+y;$
      \\&$-k_1\cdot k_2,-k_1\cdot k_3,-q\cdot k_1,-q\cdot k_2,-q\cdot k_3$
      \\
      fam3luhh3:&$- k_1^2,-k_2^2,-2 p\cdot k_2+k_2^2,-k_3^2+m_c^2,-(k_2-k_3)^2+m_c^2,-2 p\cdot k_1+y;$
      \\&$-k_1\cdot k_2,-k_1\cdot k_3,-p\cdot k_3,-q\cdot k_1,-q\cdot k_2,-q\cdot k_3$
      \\
      \hline
\end{tabular}
}
\caption{\label{tab::fam3}Three-loop integral families in the (uuu), (uuh) and
  (uhh) regions. The factors after the semi-colon correspond to numerators.}
\end{center}
\end{table}


\subsection{Master integrals}

After reduction to master integrals and their subsequent minimization 
across all families, the amplitude can be expressed in terms of 
1, 3 and 20 ultra-soft master integrals at one-, two- and three-loop order,
respectively. Many of them can be computed introducing Feynman parameters
 and integrating step-by-step, even for general dimension $d=4-2\epsilon$.
In the following we denote them by the letter $I$. Master integrals
in the mixed regions are denoted by the letters $J$ and $K$.

\begin{figure}[t]
    \centering
    \begin{minipage}[b]{0.14\textwidth}
      \raisebox{-0.2em}{\includegraphics[width=\textwidth]{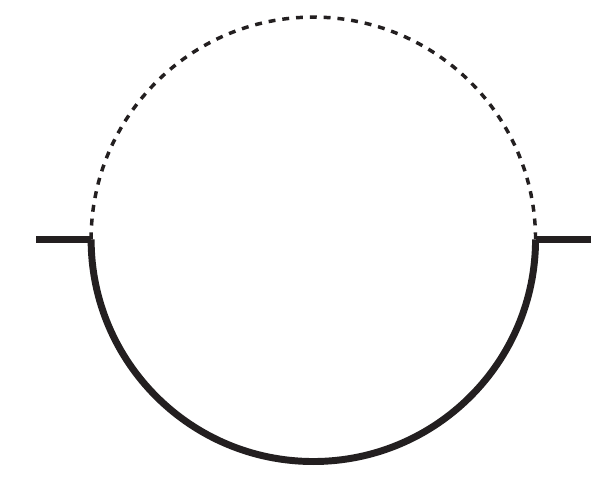}}
     
     \vspace*{0.19\textwidth}
     \centering $I^{1l}$
    \end{minipage}
    \begin{minipage}[b]{0.19\textwidth}
     \includegraphics[width=\textwidth]{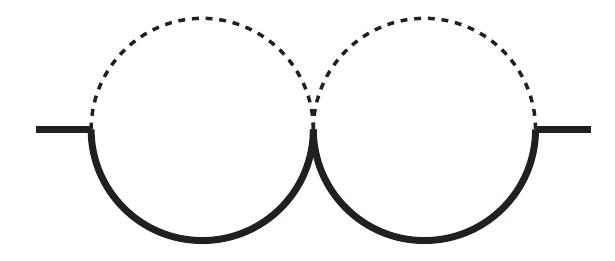}
     
     \vspace*{0.19\textwidth}
     \centering $I^{2l}_1$
    \end{minipage}
    \begin{minipage}[b]{0.19\textwidth}
     \includegraphics[width=\textwidth]{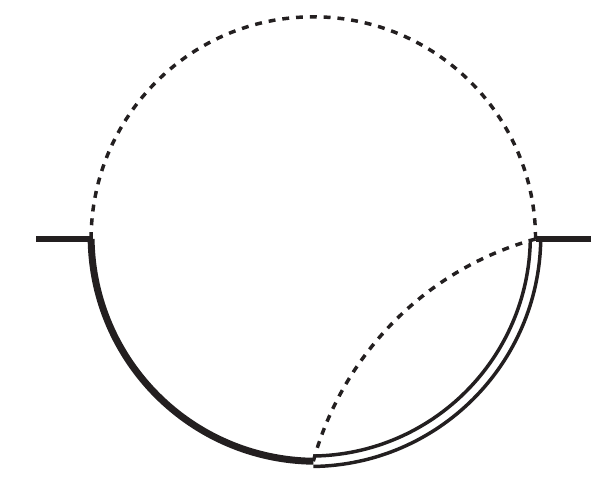}
     
     \centering $I^{2l}_2$
    \end{minipage}
    \begin{minipage}[b]{0.19\textwidth}
     \includegraphics[width=\textwidth]{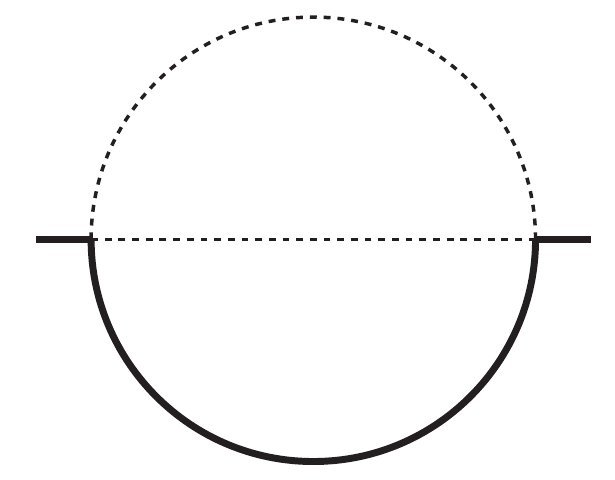}
     
     \centering $I^{2l}_3$
    \end{minipage}
    \caption{The one- and two-loop master integrals for the double ultra-soft region.
      Dashed lines represent massless propagators, while solid lines and
      double lines represent the linear massive ($2p\cdot k_i-y$) and massless
      ($2p\cdot k_i$) HQET-like propagators, {\ms respectively,} 
      with $p^2=m_b^2$ and the
      loop momentum $k_i$ ($i=1,2$).}
    \label{fig:mi2l}
\end{figure}

At one- and two-loop order the results for the master integrals
are given by
\begin{eqnarray}
    I^{1l} &=& 
    N y^{d-3} (m_b^2)^{1-d/2} \Gamma(d/2-1) \Gamma(3-d)
        \,,
    \nonumber\\ 
    J^{1l} &=& \int\frac{{\rm d}^dk_1}{(2\pi)^d} \frac{1}{-k_1^2 + m_b^2}
    = N (m_b^2)^{d/2-1} \Gamma(1-d/2)
    \,,
\end{eqnarray} 
and
\begin{eqnarray}
    I_1^{2l} &=& \left( I^{1l} \right)^2
    = N^2 y^{2d-6} (m_b^2)^{2-d} \Gamma^2\left(\tfrac{d}{2}-1\right) \Gamma^2\left(3-d\right) ,
    \nonumber\\[5pt]
    I_2^{2l} &=& 
    N^2 y^{2d-6} (m_b^2)^{2-d} \Gamma^2\left(\tfrac{d}{2}-1\right) \Gamma\left(3-d\right) \frac{\Gamma(2d-5)\Gamma(6-2d)}{\Gamma(d-2)}
        \,,
	\nonumber\\[5pt] 
    I_3^{2l} &=& 
    N^2 y^{2d-5} (m_b^2)^{2-d} \Gamma^2\left(\tfrac{d}{2}-1\right) \Gamma(5-2d)
        \,,
	\nonumber\\[5pt]
    J_1^{2l} &=& I^{1l} J^{1l}
    = N^2 y^{d-3} \Gamma\left( 1 - \tfrac{d}{2} \right) \Gamma\left( \tfrac{d}{2}-1 \right) \Gamma(3-d)
        \,,
\end{eqnarray} 
with
\begin{eqnarray}
  N &=& \frac{i}{(4\pi)^{d/2}}\,.
\end{eqnarray}
Note that the last integral originates from the (uh) region and factorizes
into a massive tadpole integral and the one-loop ultra-soft master integral.
Graphical representation of the ultra-soft one- and two-loop 
master integrals can be found in Fig.~\ref{fig:mi2l}.

Also at three loops, eleven (out of 20) master integrals can be expressed in term
of $\Gamma$ functions and are thus available to all orders in $\epsilon$.
They are given by 
\begin{align}
   I_1^{3l} &=
   N^3
   y^{3d-7} (m_b^2)^{3-3d/2} \,
   \Gamma^3(d/2-1)
   \Gamma(7-3d), \nonumber\\[5pt]
    I_2^{3l} &= N^3 
    y^{3d-8} (m_b^2)^{3-3d/2}\,
    \frac{
    \Gamma^3(d/2-1)
    \Gamma^2(3-d)
    \Gamma(8-3d)
    }
    {
    \Gamma(6-2d)
    },\nonumber\\[5pt]
    I_3^{3l} &= N^3 
    y^{3d-8} (m_b^2)^{3-3d/2}\, 
    \Gamma^3(d/2-1)
    \Gamma(3-d)
    \Gamma(5-2d),\nonumber\\[5pt]
    I_4^{3l} &= N^3 
    y^{3d-9} (m_b^2)^{4-3d/2}\,
    \frac{
    \Gamma^4(d/2-1)
    \Gamma^2(2-d/2)
    \Gamma(3d/2-4)
    \Gamma(9-3d)
    }
    {
    \Gamma^2(d-2)
    \Gamma(4-d)
    },\nonumber\\[5pt]
    I_6^{3l} &= N^3 
    y^{3d-9} (m_b^2)^{3-3d/2}\, 
    \Gamma^3(d/2-1)
    \Gamma^3(3-d),\nonumber\\[5pt]
    I_8^{3l}  &= N^3 
    y^{3d-8} (m_b^2)^{3-3d/2}\, 
    \frac{\Gamma^3(d/2-1)
    \Gamma(3-d)
    \Gamma(2d-5)
    \Gamma(8-3d)}
    {\Gamma(d-2)}
    ,\nonumber\\[5pt]
    I_9^{3l}  &= N^3 
    y^{3d-8} (m_b^2)^{3-3d/2}\, 
    \frac{
    \Gamma^3(d/2-1)
    \Gamma(5-2d)
    \Gamma(3d-7)
    \Gamma(8-3d)
    }{\Gamma(d-2)}
    ,\nonumber\\[5pt]
    I_{10}^{3l} &= N^3 
    y^{3d-8} (m_b^2)^{3-3d/2}\, 
    \frac{
    \Gamma^3(d/2-1)
    \Gamma(3-d)
    \Gamma(3d-7)
    \Gamma(8-3d)
    }{
    \Gamma(2d-4)}
    ,\nonumber\\[5pt]
    I_{13}^{3l} &= N^3 
    y^{3d-9} (m_b^2)^{3-3d/2}\, 
    \frac{
    \Gamma^3(d/2-1)
    \Gamma(3-d)
    \Gamma(2d-5)
    \Gamma(6-2d)
    \Gamma(3d-8)
    \Gamma(9-3d)
    }{
    \Gamma^2(d-2)}
    ,\nonumber\\[5pt]
    I_{15}^{3l} &= N^3 
    y^{3d-9} (m_b^2)^{3-3d/2}\, 
    \frac{
    \Gamma^3(d/2-1)
    \Gamma^2(3-d)
    \Gamma(2d-5)
    \Gamma(6-2d)
    }{
    \Gamma(d-2)}
    ,\nonumber\\[5pt]
    I_{17}^{3l} &= N^3 
    y^{3d-9} (m_b^2)^{3-3d/2}\, 
    \frac{
    \Gamma^3(d/2-1)
    \Gamma^2(3-d)
    \Gamma(3d-8)
    \Gamma(9-3d)
    }{
    \Gamma(d-2)}
                  \,,
    \label{eq::simpleMIs}
\end{align}
where the corresponding integral representation is easily obtained from the
pictures shown in Fig.~\ref{fig:MI3l}.
In Appendix~\ref{app:aux}, we provide auxiliary integrals useful
to obtain the results in Eq.~(\ref{eq::simpleMIs}).


\begin{figure}[t]
    \centering
    \begin{minipage}[b]{0.19\textwidth}
     \includegraphics[width=\textwidth]{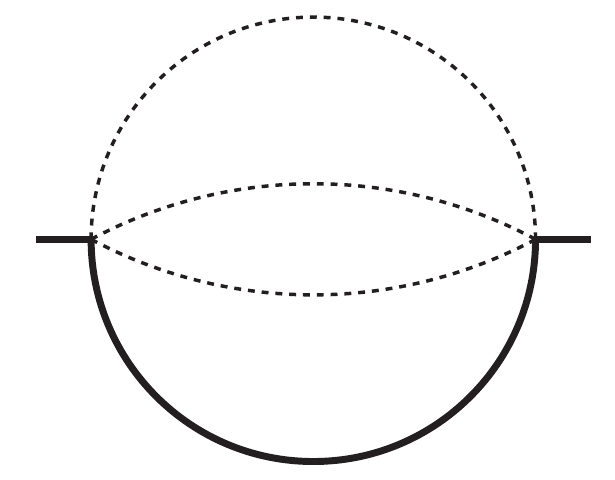}
     
     \centering $I^{3l}_1$
    \end{minipage}
    \begin{minipage}[b]{0.19\textwidth}
     \includegraphics[width=\textwidth]{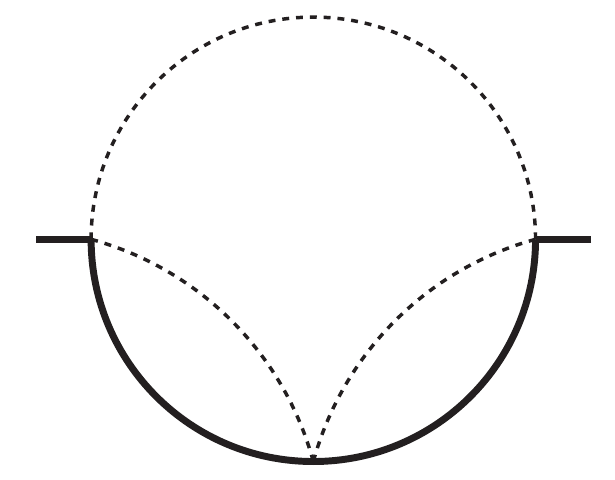}
     
     \centering $I^{3l}_2$
    \end{minipage}
    \begin{minipage}[b]{0.19\textwidth}
     \includegraphics[width=\textwidth]{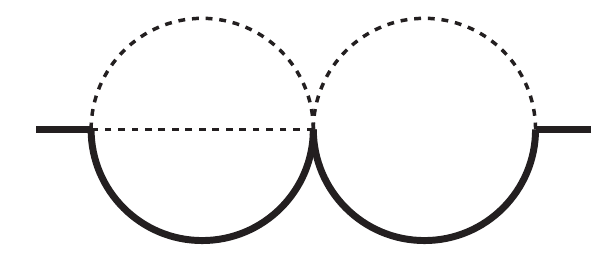}
     
     \vspace*{0.18\textwidth}
     \centering $I^{3l}_3$
    \end{minipage}
    \begin{minipage}[b]{0.19\textwidth}
     \includegraphics[width=\textwidth]{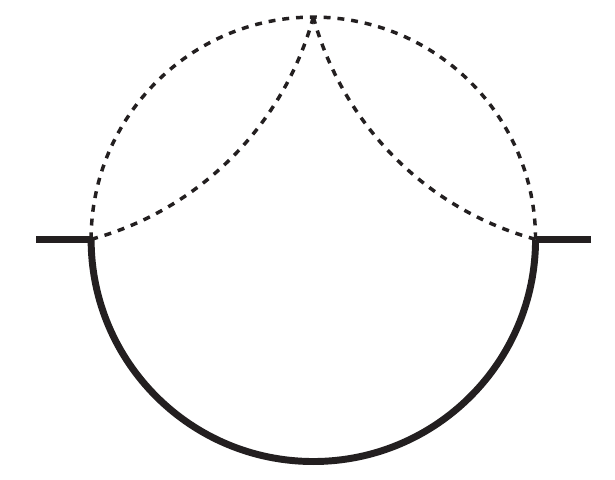}
     
     \centering $I^{3l}_4$
    \end{minipage}
    \begin{minipage}[b]{0.19\textwidth}
     \includegraphics[width=\textwidth]{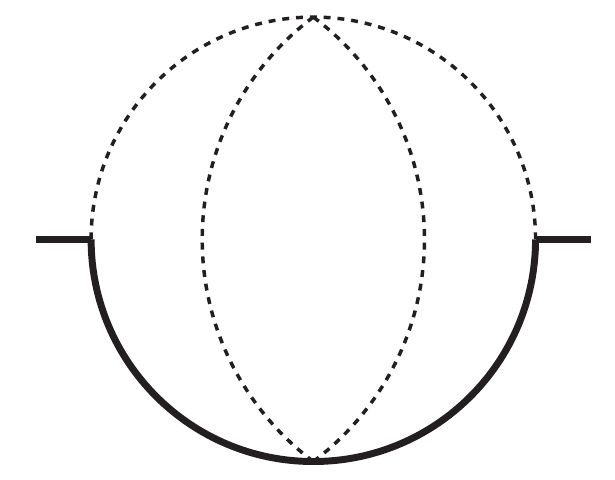}
     
     \centering $I^{3l}_5$
    \end{minipage} \\
    \bigskip
    \begin{minipage}[b]{0.19\textwidth}
     \includegraphics[width=\textwidth]{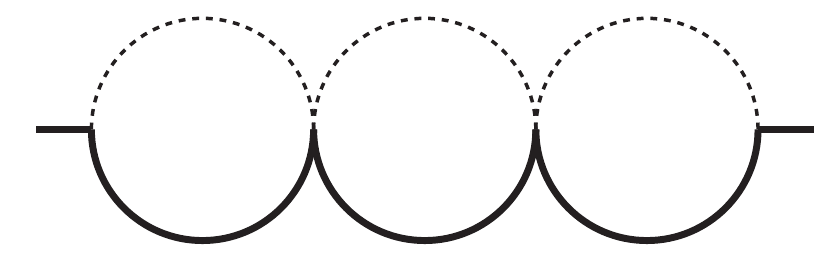}
     
     \vspace*{0.24\textwidth}
     \centering $I^{3l}_6$
    \end{minipage}
    \begin{minipage}[b]{0.19\textwidth}
     \includegraphics[width=\textwidth]{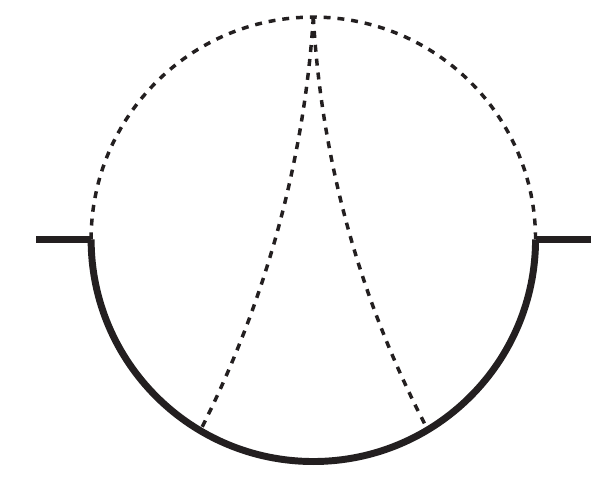}
     
     \vspace*{0.01\textwidth}
     \centering $I^{3l}_7$
    \end{minipage}
    \begin{minipage}[b]{0.19\textwidth}
     \includegraphics[width=\textwidth]{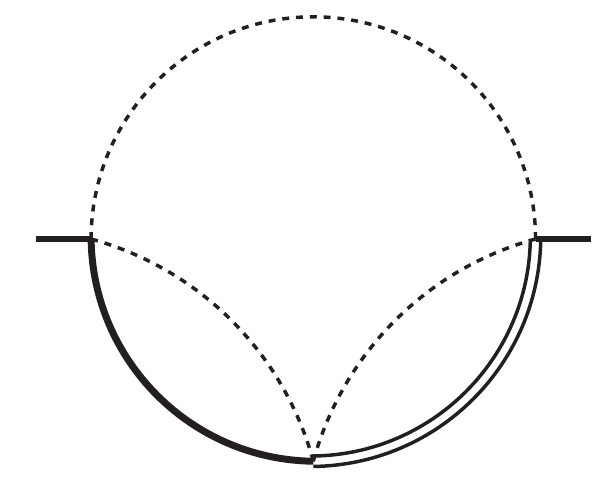}
     
     \centering $I^{3l}_8$
    \end{minipage}
    \begin{minipage}[b]{0.19\textwidth}
     \includegraphics[width=\textwidth]{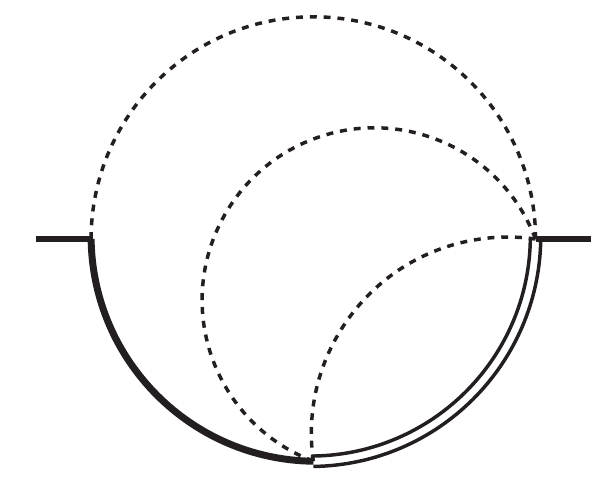}
     
     \centering $I^{3l}_9$
    \end{minipage}
    \begin{minipage}[b]{0.19\textwidth}
     \includegraphics[width=\textwidth]{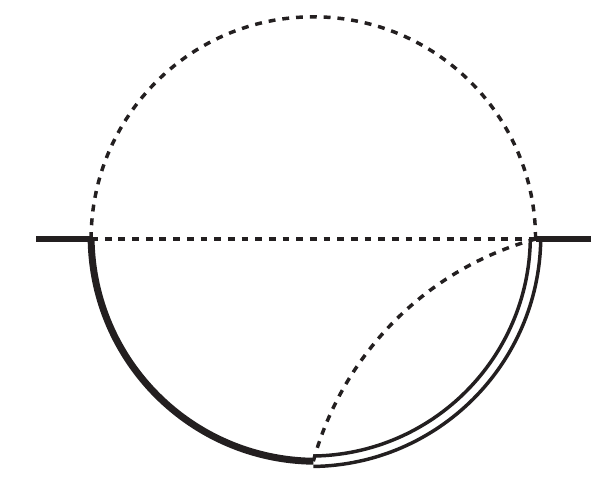}
     
     \centering $I^{3l}_{10}$
    \end{minipage}\\
    \bigskip
    \begin{minipage}[b]{0.19\textwidth}
     \includegraphics[width=\textwidth]{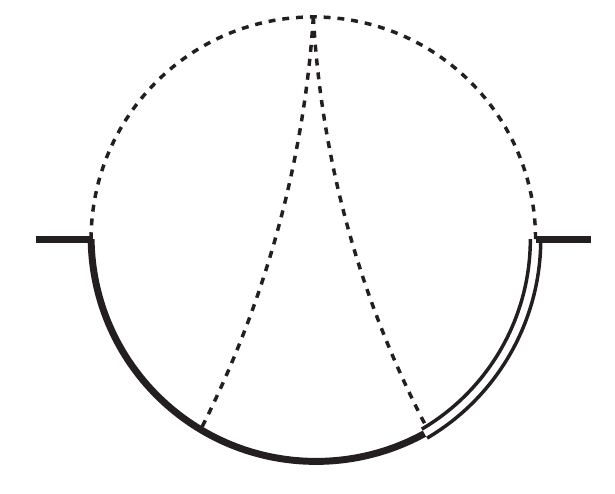}
     
     \centering $I^{3l}_{11}$
    \end{minipage}   
    \begin{minipage}[b]{0.19\textwidth}
     \includegraphics[width=\textwidth]{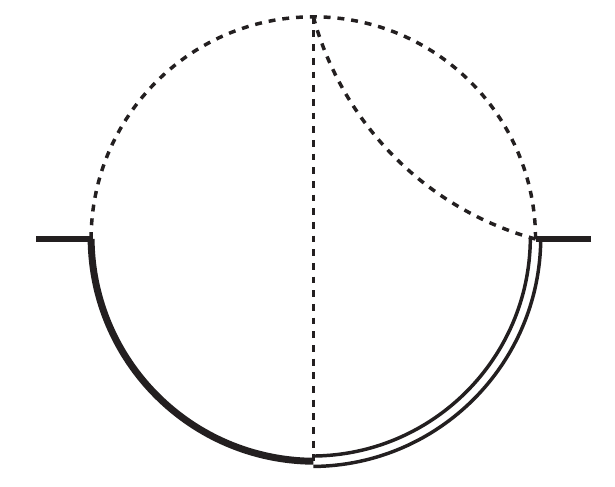}
     
     \vspace*{0.01\textwidth}
     \centering $I^{3l}_{12}$
    \end{minipage}
    \begin{minipage}[b]{0.19\textwidth}
     \includegraphics[width=\textwidth]{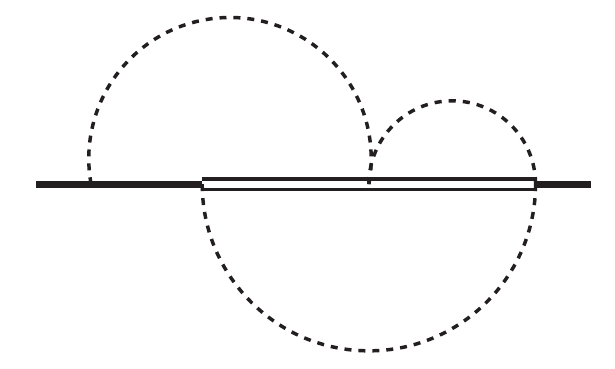}
     
     \vspace*{0.11\textwidth}
     \centering $I^{3l}_{13}$
    \end{minipage}
    \begin{minipage}[b]{0.19\textwidth}
     \includegraphics[width=\textwidth]{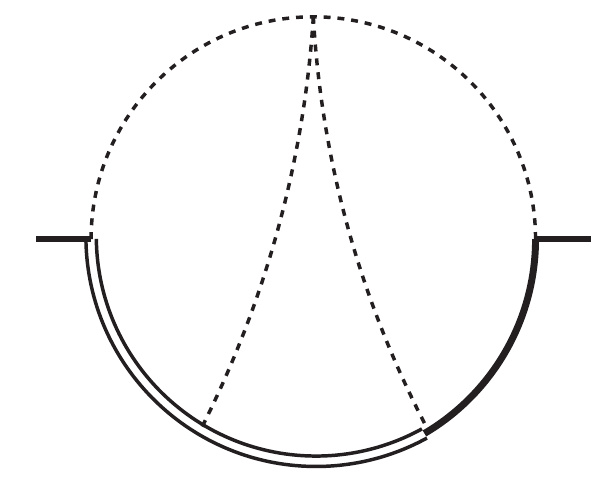}
     
     \centering $I^{3l}_{14}$
    \end{minipage}
    \begin{minipage}[b]{0.19\textwidth}
     \includegraphics[width=\textwidth]{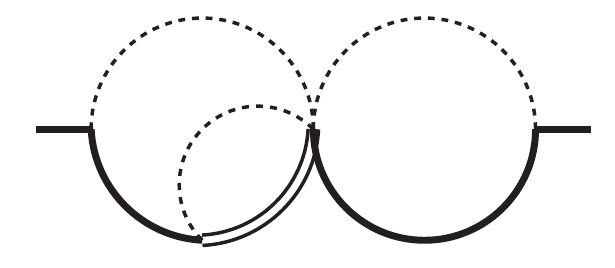}
     
     \vspace*{0.19\textwidth}
     \centering $I^{3l}_{15}$
    \end{minipage}\\
    \bigskip
    \begin{minipage}[b]{0.19\textwidth}
     \includegraphics[width=\textwidth]{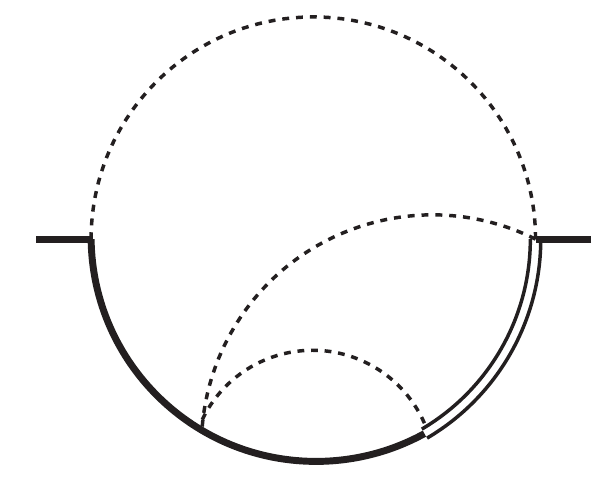}
     
     \centering $I^{3l}_{16}$
    \end{minipage}   
    \begin{minipage}[b]{0.19\textwidth}
     \includegraphics[width=\textwidth]{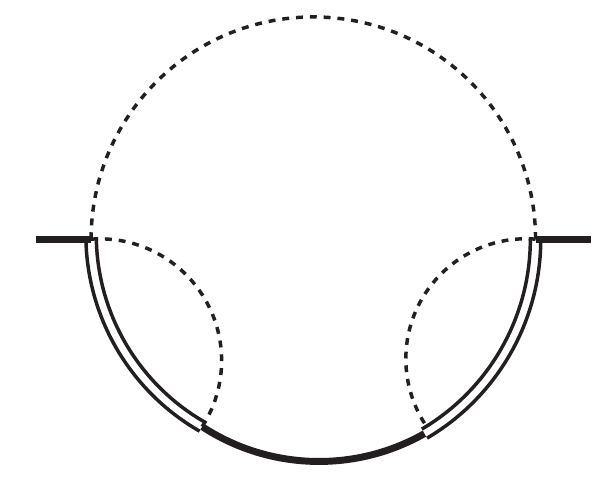}
     
     \centering $I^{3l}_{17}$
    \end{minipage}
    \begin{minipage}[b]{0.19\textwidth}
     \includegraphics[width=\textwidth]{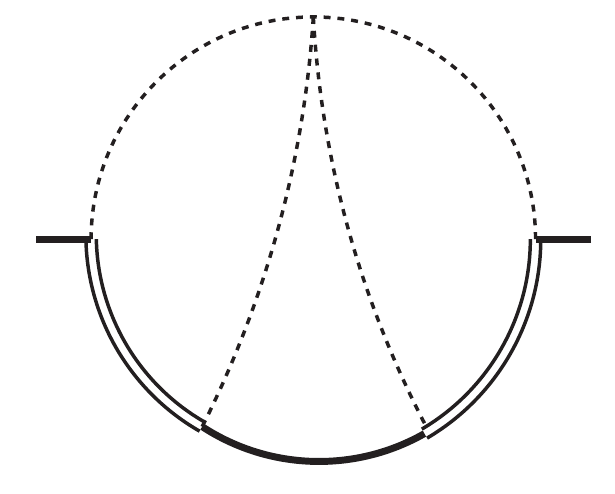}
     
     \centering $I^{3l}_{18}$
    \end{minipage}
    \begin{minipage}[b]{0.19\textwidth}
     \includegraphics[width=\textwidth]{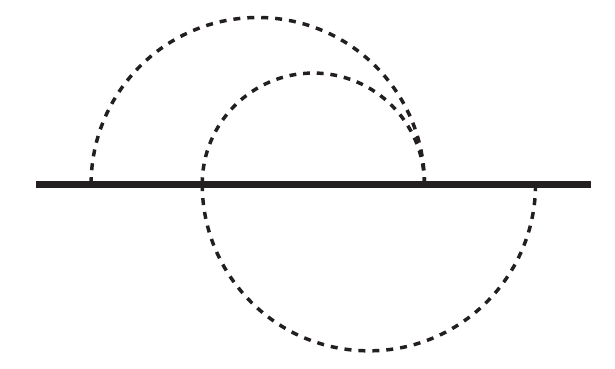}
     
     \vspace*{0.11\textwidth}
     \centering $I^{3l}_{19}$
    \end{minipage}
    \begin{minipage}[b]{0.19\textwidth}
     \includegraphics[width=\textwidth]{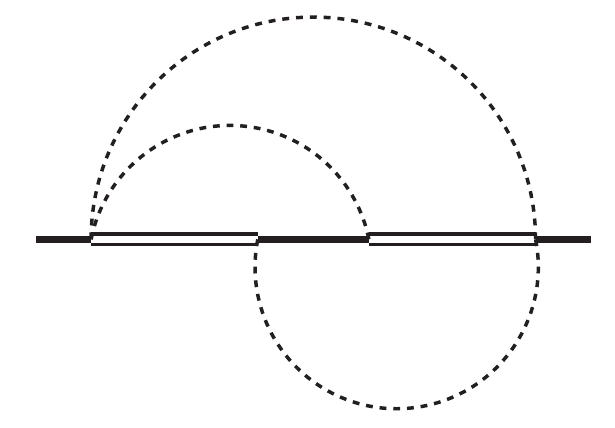}
     
     \vspace*{0.11\textwidth}     
     \centering $I^{3l}_{20}$
    \end{minipage}
    \caption{The three-loop master integrals for the triple ultra-soft region. 
      The same notation as in Fig.~\ref{fig:mi2l} is used.}
    \label{fig:MI3l}
\end{figure}


For the remaining nine integrals in the (uuu) region, we obtained
analytic results for the $\epsilon$ expansion with the help of the
Mellin-Barnes method~\cite{Smirnov:2012gma}. We managed to derive up to
four-dimensional representations.  In the case of one- and two-dimensional
Mellin-Barnes representations (which applies to 7 master integrals)
we computed the $\epsilon$ expansion by closing
the integration contour and summing up the residues analytically with the
packages \texttt{Sigma}~\cite{Schneider:2007},
\texttt{EvaluateMultiSums}~\cite{Ablinger:2010pb} together with
\texttt{HarmonicSums}~\cite{HarmonicSums}.  
For the analytic manipulation of the Mellin-Barnes integrals, the program
package {\tt MB}~\cite{Czakon:2005rk,Smirnov:2009up} was very useful.
Additionally, we managed to
obtain high-precision numerical results and use the PSLQ~\cite{PSLQ} algorithm
to reconstruct the analytic expressions.
To obtain these results for higher dimensional integrals, the
program {\tt mpmath}~\cite{mpmath} was used.
All of our analytic expressions were cross-checked using the
program {\tt FIESTA}~\cite{Smirnov:2015mct}.

For the master integral $I_{11}^{3l}$, we obtained initially
a threefold Mellin-Barnes representation, which could be reduced to 
a twofold representation by applying Barnes-Lemmas after
the $\epsilon$-expansion. Then we proceeded as described above.

The only master integral we were not able to determine with Mellin-Barnes
methods to the necessary order in $\epsilon$ was $I_{7}^{3l}$.  
We mention that we calculate $I_{7}^{3l}$
up to transcendental weight 5, i.e. one order
higher then needed for the current calculation.  The higher order terms
are necessary for the calculation in~\cite{FSS20}.  For the
calculation of $I_{7}^{3l}$ it was necessary to apply a different strategy.
For this integral, we introduced a second mass scale $x$ in the bottom-middle
and bottom-right propagator. When this mass is zero ($x=0$), the {\ms integral
reduces} to $I_{14}^{3l}$ which can be obtained by Mellin-Barnes
methods. Thus, we constructed a set of differential
equations~\cite{Kotikov:1990kg,Gehrmann:1999as,Henn:2013pwa}, applied boundary
conditions at $x=0$, and evaluated the solution at $x=1$, which provided the
desired integral. More details on the computation are given in
Appendix~\ref{app:I7}.

The analytic results for the $\epsilon$ expansion of the remaining nine master
integrals --- ordered according to complexity --- are
\begin{align}
    I_{16}^{3l} &= N^3 y^{3d-9} (m_b^2)^{3-3d/2}\,
    \frac{\Gamma^3(d/2-1)}{\Gamma(d-2)} \notag \\
    \times &\frac{1}{2 \pi i}
    \int_{-i \infty}^{+i\infty}
    dw 
    \frac{
    \Gamma(3-d+w)
    \Gamma(6+w-2d)
    \Gamma(9-3d+w)
    \Gamma(3d-8-w)
    \Gamma(-w)
    }{\Gamma(4-d+w)} \notag \\
    &= N^3
    y^{3-6\epsilon} (m_b^2)^{3\epsilon-3}
    \frac{
    \Gamma^3(d/2-1)
    \Gamma(3-d)
    \Gamma(6-2d)
    \Gamma(2d-5)
    \Gamma(9-3d)
    }
    {
    \Gamma(4-d)
    }
    \notag \\ 
    & \times {}_3 F_2 \left( \begin{matrix}6-2d,3-d,2d-5 \\ 1,4-d  \end{matrix} \ ; 1 \right)
    \notag \\
    &= \tilde{N}^3
    y^{3-6\epsilon} (m_b^2)^{3\epsilon-3}
    \Biggl[
    \frac{1}{48 \epsilon^3}
    +\frac{1}{8 \epsilon^2}
    +\left(\frac{7}{9}+\frac{29}{32} \zeta_2 \right)
    \frac{1}{\epsilon}
    +\frac{37}{6} 
    +\frac{87}{16} \zeta_2
    -\frac{119}{48} \zeta_3
    \notag \\ &
    + \left(
      \frac{101}{2}
    + \frac{163}{6} \zeta_2
    + \frac{14603}{640} \zeta_2^2
    - \frac{119}{8} \zeta_3
    \right) \epsilon
    + \biggl(
      \frac{6949}{18}
    + \frac{641}{4} \zeta_2
    + \frac{43809}{320} \zeta_2^2
    \notag \\ &
    - \frac{1659}{32} \zeta_2 \zeta_3
    - \frac{713}{9} \zeta_3
    - \frac{9951}{80} \zeta_5
    \biggr) \epsilon^2 + \mathcal{O}(\epsilon^3)
    \Biggr] , \\
    I_{19}^{3l} &= N^3y^{3d-9} (m_b^2)^{3-3d/2}\,
    \frac{
    \Gamma^3(d/2-1)
    \Gamma(3-d)
    \Gamma^2(6-2d)
    \Gamma(9-3d)
    }
    {
    \Gamma(4-d)
    \Gamma(12-4d)
    }
    \notag \\ 
    & \times {}_3 F_2 \left( \begin{matrix}6-2d,6-2d,3-d \\ 12-4d,4-d  \end{matrix} \ ; 1 \right)
    \notag \\
    &= \tilde{N}^3
    y^{3-6\epsilon} (m_b^2)^{3\epsilon-3}
    \Biggl[
    \frac{1}{24 \epsilon^3}
    +\frac{1}{4 \epsilon^2}
    +\left(\frac{13}{18}+\frac{5}{16} \zeta_2 \right)
    \frac{1}{\epsilon}
    -\frac{7}{6} 
    +\frac{15}{8} \zeta_2
    +\frac{49}{24} \zeta_3
    \notag \\ &
    + \left(
    - \frac{71}{2}
    + \frac{25}{12} \zeta_2
    - \frac{2437}{320} \zeta_2^2
    + \frac{49}{4} \zeta_3
    \right) \epsilon
    + \biggl(
    - \frac{6193}{18}
    - \frac{251}{4} \zeta_2
    - \frac{7311}{160} \zeta_2^2
    \notag \\ &
    + \frac{693}{16} \zeta_2 \zeta_3
    + \frac{1237}{18} \zeta_3
    + \frac{4929}{40} \zeta_5
    \biggr) \epsilon^2 + \mathcal{O}(\epsilon^3)
    \Biggr] , \\
    I_{12}^{3l} &= N^3y^{3d-10} (m_b^2)^{4-3d/2}\,
    \frac{
    \Gamma^2(d/2-1)
    \Gamma(10-3d)
    \Gamma(3d-9)
    }{
    \Gamma(d-2)
    \Gamma(3d/2-4)
    } 
    \times \frac{1}{2 \pi i}
    \int_{-i \infty}^{+i\infty}
    dw 
    \notag \\
    &\frac{
    \Gamma(-w)
    \Gamma(d-3-w)
    \Gamma(2d-6-w)
    \Gamma(d/2-1+w)
    \Gamma(2-d/2+w)
    \Gamma(7-2d+2w)
    }{
    \Gamma(1-w)
    \Gamma(d-2+2w)
    } \notag \\
    &=\tilde{N}^3 
    y^{2-6\epsilon} (m_b^2)^{3\epsilon-2}
    \Biggl[
    -\frac{1}{24 \epsilon^3}
    -\frac{7}{24 \epsilon^2}
    +\left(-\frac{9}{8}-\frac{103}{48} \zeta_2 \right)
    \frac{1}{\epsilon}
    -\frac{31}{24} 
    -\frac{721}{48} \zeta_2
    -\frac{41}{24} \zeta_3
    \notag \\ &
    + \left(
      \frac{187}{8}
    - \frac{1055}{16} \zeta_2
    - \frac{67073}{960} \zeta_2^2
    - \frac{287}{24} \zeta_3
    \right) \epsilon
    + \biggl(
      \frac{6989}{24}
    - \frac{9593}{48} \zeta_2
    - \frac{469511}{960} \zeta_2^2
    \notag \\ &
    - \frac{895}{48} \zeta_2 \zeta_3
    - \frac{593}{8} \zeta_3
    - \frac{12227}{120} \zeta_5
    \biggr) \epsilon^2 + \mathcal{O}(\epsilon^3)
    \Biggr]\,,
\nonumber\\
    I_{18}^{3l} &=N^3 y^{3d-11} (m_b^2)^{4-3d/2}\,
    \frac{
    \Gamma(11-3d)
    \Gamma(3d-10)
    }{
    \Gamma^2(d-3)
    }
    \times \left(\frac{1}{2 \pi i}\right)^2
    \int_{-i \infty}^{+i\infty}
    dw_1
    \int_{-i \infty}^{+i\infty}
    dw_2
    \notag \\ &
    \biggl( 
    \Gamma(-w_1)
    \Gamma(d/2-2-w_1)
    \Gamma(5-d+2w_1)
    \Gamma(d-4-w_1)
    \Gamma(1+w_1)
    \times
    \bigl\{ w_1 \leftrightarrow w_2 \bigr\}
    \biggr)
    \notag \\ &
    \times 
    \frac{
    \Gamma(d/2+w_1+w_2)
    }{
    \Gamma(-w_1-w_2)
    \Gamma(d+2w_1+2w_2)
    }
    \notag \\
    &=\tilde{N}^3 
    y^{1-6\epsilon} (m_b^2)^{3\epsilon-2}
    \Biggl[
    -\frac{\zeta_3}{\epsilon}
    - \frac{3}{5} \zeta_2^2
    - 8 \zeta_3
    + \biggl(
      \frac{9}{2} \zeta_2 \zeta_3
    - \frac{24}{5} \zeta_2^2
    - 52 \zeta_3
    - 57 \zeta_5
    \biggr) \epsilon
    + \mathcal{O}(\epsilon^2)
  \Biggr] \, ,
    \\
    I_{14}^{3l} &=N^3 y^{3d-11} (m_b^2)^{4-3d/2}\,
    \frac{
    \Gamma(11-3d)
    }{
    \Gamma(d-3)
    \Gamma(8-2d)
    }
    \times \left(\frac{1}{2 \pi i}\right)^2
    \int_{-i \infty}^{+i\infty}
    dw_1
    \int_{-i \infty}^{+i\infty}
    dw_2
    \notag \\ &
    \biggl( 
    \Gamma(-w_1)
    \Gamma(1+w_1)
    \Gamma(d/2-2-w_1)
    \Gamma(7-2d-w_1)
    \Gamma(2d-6+2w_1)
    \notag \\ &
    \Gamma(-w_2)
    \Gamma(1+w_2)
    \Gamma(d/2-2-w_2)
    \Gamma(5-d+2w_2)
    \Gamma(d-4-w_2)
    \notag \\ &
    \frac{
    \Gamma(d/2+w_1+w_2)
    }{
    \Gamma(-w_1-w_2)
    \Gamma(d+2w_1+2w_2)
    }
    \biggr)
    \notag \\
    &=\tilde{N}^3 
    y^{1-6\epsilon} (m_b^2)^{3\epsilon-2}
    \Biggl[
    -\frac{\zeta_2}{6 \, \epsilon^2}
    + \biggl(
    -\frac{4}{3} \zeta_2
    +\frac{2}{3} \zeta_3
    \biggr) \frac{1}{\epsilon}
    -\frac{26}{3} \zeta_2
    +\frac{16}{3} \zeta_3
    -\frac{391}{60} \zeta_2^2
    \notag \\ &
    + \biggl(
    -\frac{160}{3} \zeta_2
    +\frac{104}{3} \zeta_3
    -\frac{782}{15} \zeta_2^2
    +\frac{197}{6} \zeta_2 \zeta_3
    +\frac{44}{3} \zeta_5  
    \biggr) \epsilon
    + \mathcal{O}(\epsilon^2)
    \Biggr] \, ,
    \\
    I_{5}^{3l} &=N^3 y^{3d-10} (m_b^2)^{4-3d/2}\,
    \frac{
    \Gamma^2(d/2-1)
    }{
    \Gamma(d-2)
    }
    \times \left(\frac{1}{2 \pi i}\right)^2
    \int_{-i \infty}^{+i\infty}
    dw_1
    \int_{-i \infty}^{+i\infty}
    dw_2
    \notag \\ &
    \biggl( 
    \frac{
    \Gamma(-w_1)\Gamma(1+w_1)
    }{
    \Gamma(1-w_1)
    }
    \Gamma(d-3-w_1)
    \Gamma(d/2-1+w_1)
    \Gamma(-w_2)
    \Gamma(d-2+2w_1+w_2)
    \notag \\ &
    \Gamma(2-d/2+w_1+w_2)
    \frac{
    \Gamma(7-2d+2w_1+w_2)
    \Gamma(3-d-2w_1-w_2)
    }{
    \Gamma(d-2+2w_1)
    \Gamma(3-d/2+2w_1+w_2)
    }
    \Biggr)
    \notag \\
    &=\tilde{N}^3 
    y^{2-6\epsilon} (m_b^2)^{3\epsilon-2}
    \Biggl[
    \frac{1}{12 \epsilon^3}
    +\frac{7}{12 \epsilon^2}
    + \biggl(
    \frac{9}{4}
    + \frac{7}{24} \zeta_2
    \biggr) \frac{1}{\epsilon}
    + \frac{31}{12}
    + \frac{49}{24} \zeta_2
    + \frac{41}{12} \zeta_3
    \notag \\ &
    - \epsilon \biggl(
    \frac{187}{4}
    + \frac{1}{8} \zeta_2
    - \frac{287}{12} \zeta_3
    + \frac{8767}{480} \zeta_2^2
    \biggr)
    + \epsilon^2 \biggl(
    - \frac{6989}{12}
    - \frac{2983}{24} \zeta_2
    + \frac{593}{4} \zeta_3
    \notag \\ &
    - \frac{61369}{480} \zeta_2^2
    + \frac{1951}{24} \zeta_2 \zeta_3
    + \frac{12227}{60} \zeta_5  
    \biggr)
    + \mathcal{O}(\epsilon^3)
    \Biggr]\,,
\nonumber\\
    I_{20}^{3l} &=N^3 y^{3d-9} (m_b^2)^{3-3d/2}\,
    \frac{
    \Gamma(9-3d)\Gamma(d/2-1)
    }{
    \Gamma(d-2)
    }
    \times \left(\frac{1}{2 \pi i}\right)^2
    \int_{-i \infty}^{+i\infty}
    dw_1
    \int_{-i \infty}^{+i\infty}
    dw_2
    \notag \\
    & \Bigg[
    \Gamma(d-3-w_1) \Gamma(1+w_1)\Gamma(2d-5+w_1) 
    \Gamma(d/2-2-w_2)\Gamma(d-3+w_1-w_2) \notag \\
    &\times\frac{
    \Gamma(w_2+1)
    \Gamma(d/2+w_2)
    \Gamma(4-d-w_1+2w_2)
    }
    {
    \Gamma(d-2+w_1)\Gamma(d+2w_2)
    }
    \Bigg] \notag \\
    &=\tilde{N}^3 y^{3-6\epsilon} (m_b^2)^{3\epsilon-3}
    \Biggl[
    \frac{1}{\epsilon}
    \left(
    -\frac{5}{18}+\frac{\zeta_2}{6}
    \right)
    -\frac{9}{2}+\zeta_2 +\frac{7}{3}\zeta_3 \notag \\ &
    + \epsilon
    \biggl(
    -\frac{91}{2}
    -\frac{17}{12} \zeta_2
    +14 \zeta_3
    +\frac{45}{4} \zeta_2^2
    \biggr)
    + \mathcal{O}(\epsilon^2)
    \Bigg]\,,
\nonumber\\
    I_{11}^{3l} &=N^3 y^{3d-11} (m_b^2)^{4-3d/2} 
    \left(\frac{1}{2\pi i}\right)^3
    \int_{-i \infty}^{+i\infty} dw_1 \, dw_2\, dw_3 \,
    \Gamma (-w_1)\Gamma (-w_2) \Gamma (-w_3) \notag \\[5pt]
    & \times \Gamma (w1+1)  
    \Gamma (w_3+1) 
    \Gamma (d/2-w_1-2) 
    \Gamma (d-w_1-4) 
    \Gamma (-d+2 w_1+5) 
    \Gamma (3 d-w_2-10) \notag \\[5pt]
    & \times
    \Gamma (-3 d+w_2+11) \Gamma \left(d/2-w_3-2\right) \Gamma \left(d/2+w_1+w_3\right) 
    \notag \\[5pt]
    & \times 
    \frac{     \Gamma (d-w_2-w_3-4) \Gamma (-d+w_2+2 w_3+5)}{\Gamma (d-3) \Gamma (d-w_2-3) \Gamma (-w_1-w_3) \Gamma (d+2 w_1+2  w_3)}\notag \\
    &=\tilde{N}^3 y^{1-6 \epsilon } (m_b^2)^{3 \epsilon -2} 
    \Bigg[
    -\frac{\zeta_2}{6 \epsilon^2}
    -\frac{1}{\epsilon}
    \left(
    \frac{4}{3}\zeta_2
    +\frac{\zeta_3}{3}
    \right)
    -\frac{26}{3}\zeta_2
    -\frac{8}{3}\zeta_3
    -\frac{187}{60}\zeta_2^2
    \notag \\[5pt] &
    + \biggl(
    - \frac{160}{3} \zeta_2
    - \frac{52}{3} \zeta_3
    - \frac{374}{15} \zeta_2^2
    - \frac{32}{3} \zeta_2 \zeta_3
    + \frac{83}{3} \zeta_5  
    \biggr) \epsilon
    +O(\epsilon^2)
    \Bigg]\,,
\nonumber\\
    I_{7}^{3l} &=N^3 y^{3d-11} (m_b^2)^{4-3d/2} 
    \left(\frac{1}{2\pi i}\right)^4
    \int_{-i \infty}^{+i\infty} dw_1 \, dw_2\, dz_1\, dz_2     \,
    \Gamma (-w_1) 
    \Gamma (-w_2) 
    \Gamma (-z_1) 
    \Gamma (-z_2)    
    \notag \\[5pt] & 
    \times 
    \Gamma (z_1+1) 
    \Gamma (z_2+1) 
    \Gamma (d/2-z_1-2) 
    \Gamma (d/2-z_2-2) 
    \Gamma (d-w_1-z_1-4) 
    \notag \\[5pt] & 
    \times 
    \Gamma (d-w_2-z_2-4)
    \Gamma (-d+w_1+2 z_1+5) 
    \Gamma (-d+w_2+2 z_2+5) 
    \notag \\[5pt] & 
    \frac{
    \Gamma (d/2+z_1+z_2)   
    \Gamma (3 d-w_1-w_2-10) 
    \Gamma (-3 d+w_1+w_2+11)
    }{
    \Gamma (d-w_1-3) 
    \Gamma (d-w_2-3) 
    \Gamma (-z_1-z_2) 
    \Gamma (d+2 z_1+2 z_2)}
    \notag \\[5pt] & 
    = \tilde{N}^3 y^{1-6 \epsilon } (m_b^2)^{3 \epsilon -2} 
    \Bigg[
    -\frac{\zeta_2}{3 \epsilon^2}
    +\frac{1}{\epsilon}
    \left(
    \frac{\zeta_3}{3}
    -\frac{8}{3}\zeta_2
    \right)
    -\frac{52}{3}\zeta_2
    +\frac{8}{3}\zeta_3
    -\frac{49}{30}\zeta_2^2
    \notag \\ & 
    +\biggl(
      -\frac{320}{3} \zeta_2
      +\frac{52}{3} \zeta_3
      -\frac{196}{15} \zeta_2^2
      -\frac{11}{6} \zeta_2 \zeta_3
      -\frac{83}{3} \zeta_5
    \biggr) \epsilon
    +O(\epsilon^2)
    \Bigg]\,,
\end{align}
with 
\begin{eqnarray}
  \tilde{N} = N e^{-\ep \gamma_E} \,.
\end{eqnarray} 


The master integrals in the (uuh) and (uhh) regions
factorize into the products of one- and two-loop integrals.
For the (uuh) region, they are given by 
\begin{align}
    J_1^{3l} 
    &= J^{1l} I_{3}^{2l}
    = N^3y^{2d-5} (m_b^2)^{1-d/2} \Gamma^2(d/2-1) \Gamma(1-d/2) \Gamma(5-2d)\,,\nonumber
    \\
    J_2^{3l}
    &= J^{1l} I_{1}^{2l}
    = N^3y^{2d-6} (m_b^2)^{1-d/2} \Gamma^2(d/2-1) \Gamma^2(3-d) \Gamma(1-d/2)\,,\nonumber
    \\
    J_3^{3l} 
    &= J_1^{1l} I_{2}^{2l}
    = N^3y^{2d-6} (m_b^2)^{1-d/2} \Gamma^2(d/2-1) \Gamma(1-d/2) \Gamma(3-d)
    \frac{\Gamma(2d-5)\Gamma(6-2d)}{\Gamma(d-2)} \, ,
\end{align}
while for the (uhh) region we have
\begin{align}
    K_1^{3l} 
    &= I^{1l} \left( J^{1l} \right)^2
    = N^3y^{d-3} (m_b^2)^{d/2-1} \Gamma(d/2-1) \Gamma(3-d) \Gamma^2(1-d/2)\,,
    \nonumber \\
    K_2^{3l} 
    &= I^{1l} \myintB \frac{1}{[-k_1^2+m_b^2][-k_2^2+m_b^2][-(k_1+k_2+p)^2 + m_b^2]}
    \nonumber\\
    &= \tilde{N}^3y^{d-3} (m_b^2)^{d/2-2} \biggl[
    \frac{3}{4 \epsilon^3} 
    + \frac{29}{8 \epsilon^2} 
    + \frac{1}{\epsilon} \biggl(
    \frac{175}{16}
    +\frac{21}{8} \zeta_2
    \biggr)
    + \frac{765}{32}
    + \frac{267}{16} \zeta_2
    - \frac{9}{4} \zeta_3
    \notag \nonumber\\ &
    + \epsilon \biggl(
    \frac{1943}{64}
    +\frac{2313}{32}\zeta_2
    -24 \ln(2) \zeta_2
    + \frac{963}{160} \zeta_2^2
    + \frac{25}{8} \zeta_3
    \biggr)
    +\mathcal{O}(\epsilon^2)
    \biggr]\,,
  \nonumber\\
    K_3^{3l} 
    &= I^{1l} \myintB \frac{1}{[-k_1^2][-k_2^2][-(k_1+k_2+p)^2 + m_b^2]}
    \nonumber\\
    &= N^3y^{d-3} (m_b^2)^{d/2-2} \Gamma^3(d/2-1) \Gamma^2(3-d)
    \frac{\Gamma(2d-5)\Gamma(2-d/2)}{\Gamma(3d/2-3)\Gamma(d-2)}\,.
\end{align}
Higher orders in $\epsilon$ for $K_2^{3l}$ are also known,
but not needed for our calculation.  They could be obtained by employing Eq.~(30)
of Ref.~\cite{Broadhurst:1993mw} and Eq.~(27) of Ref.~\cite{Berends:1997vk}.
Analytic results for all master integrals can be found in the
ancillary file to this paper~\cite{progdata}.

For the effects of a virtual charm quark, we have additional master
integrals in the (uuh) and (uhh) regions.
In the (uuh) region, they factorize into 
$I_i^{2l}$, ($i=1,2,3$) times the one-loop charm mass tadpole.
In the second region, they factorize into $I^{1l}$ and 
two-loop on-shell integrals with two different masses which were
calculated in Ref.~\cite{Bekavac:2009gz}.
Since these master integrals will not contribute to the final result
(cf. Section~\ref{sec::charm}), we
do not give the explicit expressions here.


\section{\label{sec::charm}Charm quark mass effects}

In this Section we consider charm mass effects to the
bottom mass relations. 
Charm mass effects to the $\overline{\text{MS}}$-OS mass relation were computed
at two loops in Ref.~\cite{Broadhurst:1991fy} and at three loops in Refs.~\cite{Bekavac:2007tk,Fael:2020bgs} 
(for the two-loop expression, see also Ref.~\cite{Davydychev:1998si}).
Using these analytic results, it is straightforward to see that
the inclusion of a few expansion terms in the limit $m_c \ll m_b$ 
provide precise predictions for physical values of the quark masses.

\begin{figure}[t]
  \centering
     \includegraphics[width=.3\textwidth]{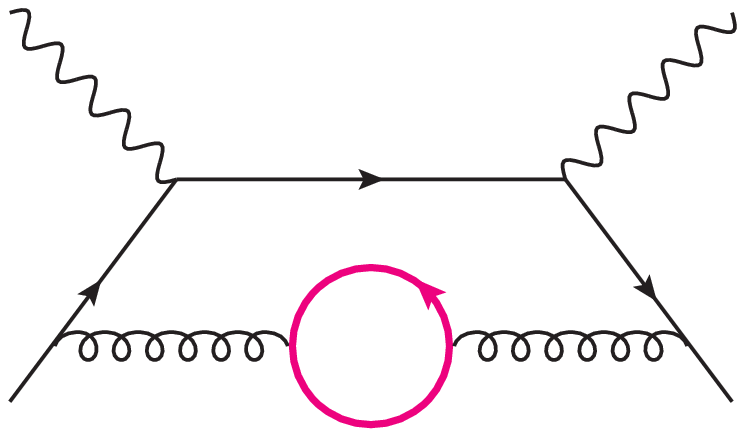}
     \hfill
     \includegraphics[width=.3\textwidth]{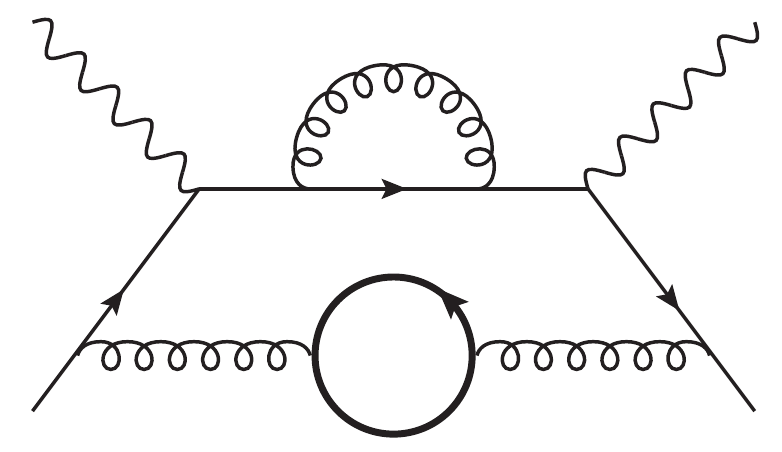}
     \hfill
     \includegraphics[width=.3\textwidth]{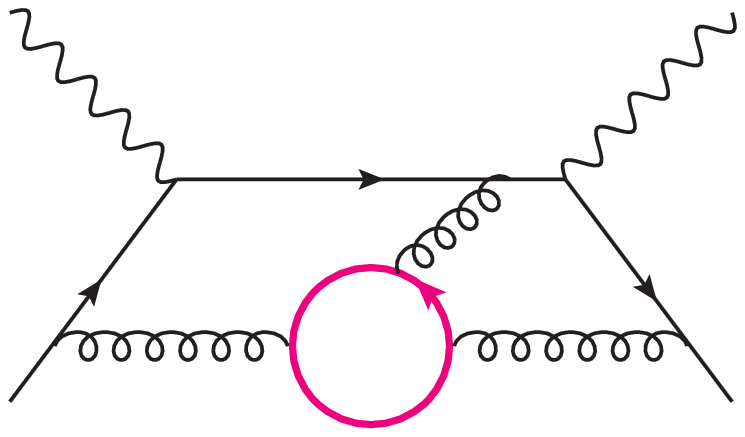}
  \caption{\label{fig::sample_diags_mc}Sample Feynman diagrams
    containing closed loops with charm quarks. The same notation
    as in Fig.~\ref{fig::sample_diags} is used.}
\end{figure}

No charm mass effects for the relation between the kinetic and the
on-shell mass for bottom are available. For their evaluation we have
to demand $|y| \ll m_c^2,m_b^2$ which means that no cuts through the charm quark
loop are possible.  At two-loop order, there are four Feynman diagrams
that contain a closed charm quark loop (cf. Fig.~\ref{fig::sample_diags_mc}).
In this case, all charm quark mass effects are generated by the well-known
one-loop decoupling relation
between $\alpha_s^{(n_l)}$ and $\alpha_s^{(n_l+1)}$.

At three-loop order, also another kind of diagrams contributes, namely
those where the charm loop is connected to the heavy quark by three
gluons (see Fig.~\ref{fig::sample_diags_mc}).  In the threshold limit,
these diagrams factorize into on-shell or {\ms tadpole} integrals, where
the mass scales are given by the charm and bottom quarks, and
integrals with ultra-soft loop momenta.  For the bare three-loop
diagrams, we obtain a non-trivial dependence on $m_c/m_b$. However,
incorporating the proper on-shell counterterms for the wave function
and heavy quark masses, only logarithmic contributions
remain. These logarithmic contributions disappear if $\alpha_s^{(3)}$
is chosen as expansion parameter. Moreover, the non-logarithmic part
of the resulting $n_l=3$ expression is identical to the
one obtained for massless charm quarks.

To summarize, all charm quark mass effects in the $n_l=4$ flavour theory are
decoupling effects. Thus, one can start the calculation in a theory where both
charm and bottom quarks are integrated out. The transition from
$\alpha_s^{(3)}$ to $\alpha_s^{(4)}$ generates $\ln(\mu_{\rm dec}^2/m_c^2)$
terms, where $\mu_{\rm dec}$ is the scale where the charm quark is
decoupled,\footnote{Note that in the formulae, which we present below, we set
  $\mu_{\rm dec}=\mu_s$, where $\mu_s$ is the renormalization scale of
  $\alpha_s$.}  and, at three-loop order, also constant contributions. The
$m^{\rm kin}$--$m^{\rm OS}$ relation one obtains this way agrees with our
explicit calculation in the four flavour theory, assuming the scaling
$|y| \ll m_c^2,m_b^2$.

Note that due to the definition of the kinetic mass in the heavy quark limit
we are forced to choose $m_c=0$ in case we assume the scaling $m_c^2\sim |y|$,
see also Ref.~\cite{Czarnecki:1997sz} for explicit results at order $\alpha_s^2$.


\section{\label{sec::ana}Analytic results}


\subsection{Renormalization}
Before presenting analytic results for the quark mass relations, we want to
discuss the renormalization of the parameters and the wave function of the external 
quarks. Note that the quantity we compute is infra-red finite.

At one-loop order there is no counterterm contribution to the
imaginary part of the forward-scattering amplitude.  In order to treat
the ultra-violet divergences at two and three loops, we have to
renormalize the strong coupling constant, the heavy quark wave
function, the heavy quark mass and the mass parameter $m$ in the
definition of the scalar current, see Eq.~(\ref{eq::J}).  
We renormalize $\alpha_s$ in the $\overline{\rm MS}$-scheme.  The
on-shell wave function renormalization constant $Z_2^{\rm OS}$ is
needed up to two-loop
order~\cite{Broadhurst:1991fy,Melnikov:2000zc,Marquard:2007uj} where
also finite charm quark mass effects are 
needed~\cite{Broadhurst:1991fy,Bekavac:2007tk}.  We choose to
renormalize the scalar current in the $\overline{\rm MS}$-scheme to
match the renormalization scheme used for the virtual corrections,
cf. Eq.~(\ref{eq:virtalS}).  The respective renormalization constant
$Z_m^{\overline{\rm MS}}$ is needed up to two-loops.  For the
renormalization of the heavy quark mass, which is present in the
virtual propagators, we introduce the corresponding counterterm in
each one- and two-loop diagram and compute the corresponding higher
order contributions together with bare contributions at the respective
loop order.  The renormalization constant $Z_m^{\rm OS}$ is again
needed up to two loops~\cite{Gray:1990yh}, including finite charm
quark mass contributions~\cite{Broadhurst:1991fi,Bekavac:2007tk}.
Note that the heavy quark mass counterterms generate gauge
depend terms which are needed in order to cancel the gauge 
dependence of the bare diagrams.

In our approach, we also generate diagrams which contain closed massive quark loops
(bottom or charm), which means that all quarks contribute to the running of $\alpha_s$.
To arrive at the theory with only $n_f-1$ light flavors, we apply
the decoupling relation (see, e.g., Ref.~\cite{Chetyrkin:1997un})
\begin{eqnarray}
  \alpha_s^{(n_f)}(\mu_{th})
  &=& \alpha_s^{(n_f-1)}(\mu_{th})
        \Biggl[
        1 
        + \frac{\alpha_s^{(n_f-1)}(\mu_{th})}{\pi} c_1\left(\frac{m}{\mu_{th}}\right) 
        + \left(\frac{\alpha_s^{(n_f-1)}(\mu_{th})}{\pi}\right)^2 c_2\left(\frac{m}{\mu_{th}}\right) 
        \Biggr]
        \,,
      \nonumber\\
  \label{eq::asdec}
\end{eqnarray}
where for later convenience we provide explicit results for the one-
and two-loop coefficients
\begin{eqnarray}
  c_1\left(\frac{m}{\mu_{th}}\right) &=& \frac{1}{3} n_h T_F \ln\left(\frac{\mu_{th}^2}{m^2} \right)  
  \,,
\label{eq:decoupling1}
\\
  c_2\left(\frac{m}{\mu_{th}}\right) &=& n_h T_F \biggl[
        \frac{1}{9} n_h T_F \ln^2\left(\frac{\mu_{th}^2}{m^2} \right)  
        + \left( \frac{5}{12} C_A - \frac{1}{4} C_F \right) \ln\left(\frac{\mu_{th}^2}{m^2} \right)  
        - \frac{2}{9} C_A + \frac{15}{16} C_F 
  \biggr]
  \,.
  \nonumber \\
  \label{eq:decoupling2}
\end{eqnarray}
In the above formulae, $m$ denotes the on-shell mass. In case
massive charm effects are considered, one first has to use
Eq.~(\ref{eq::asdec}) for the bottom and subsequently for the charm
quark.  Note that this is possible since up to two-loop order there
are no genuine $m_c/m_b$ effects (see also Ref.~\cite{Grozin:2011nk}).
Note that $c_1 \sim \ln(m^2/\mu_{th}^2)$ and thus $c_1(1)=0$.
However, the two-loop term has a finite remainder, i.e. $c_2(1)\not=0$.

The renormalization of the structure functions shows some interesting features.
At two loops the (uu) region is already finite after renormalization of the
strong coupling constant.
The (uh) region does only contribute to the $C_F^2$ and $C_F T_F n_h$ color
factors but not to $C_F C_A$ and $C_F T_F n_l$. Furthermore, it has terms that scale as
$\sim y^{-2}$.
These terms are cancelled by the (on-shell) quark mass counterterms.
After applying the wave function counterterm, the $C_F^2$ term is
exactly cancelled for the vector current. In the scalar case, there 
is a residual $C_F^2$ term which cancels against the non-vanishing virtual 
corrections when calculating the kinetic mass relation.
The remaining terms proportional to $n_h$ are eliminated by decoupling the
heavy quarks from the running of $\alpha_s$.

Very similar observations can be made at three loops.  The
all-ultra-soft region (uuu) is finite after coupling constant
renormalization.  Here, the (uuh) and (uhh) regions scale up to $\sim
y^{-3}$.  Again these terms are cancelled by the heavy quark mass
counterterms.  In case finite charm quark mass effects are considered, the
(uhh) region has a non-trivial dependence on $m_c/m_b$.
After wave function renormalization these contributions, as well as the
whole contributions to the color factors $C_F^3$ and $C_F^2 C_A$, vanish
for an external vector current.  For an external scalar
current, also the virtual corrections are needed to establish the
cancellation.  Moreover, the remaining terms proportional to $n_h$ are
again absorbed by decoupling.

The above observations can be summarized as follows:  both at two- and
three-loop order after including all relevant counterterm contributions
and after expressing the final result in terms of $\alpha_s^{(n_l)}$, 
only the pure-ultra-soft contributions survive and
all contributions proportional to $n_h$ vanish.  This means that one
could have performed the calculation from the beginning in the
effective $n_l$-flavour QCD. Furthermore, at the step where the
asymptotic expansion is applied only ultra-soft regions have to be
considered. From the physical point of view this behaviour is expected
since the kinetic mass is defined via the radiation of soft gluons
from the heavy quark.  Since in our ``full-theory'' approach the
cancellation of the $n_h$ contribution is non-trivial, we consider it
as a welcome consistency check for the correctness of our calculation.


\subsection{Quark mass relations}

In this subsection we discuss various relations between the different
definitions of the heavy quark masses.  We consider QCD with $n_f$
active flavours, where $n_f=5$ for bottom and $n_f=4$ for
charm. Furthermore, we denote by $n_l$ the number of massless quarks.
It is interesting to consider charm mass effects to
the bottom mass relations where we have $n_f=5$ and $n_l=3$.
Since one can consider different numbers of active quarks for the 
running of $\alpha_s$, we introduce $n_r$, i.e.\ the number
of active quark flavors in the running of $\alpha_s$.  
Charm effects in the $\overline{\rm MS}$-on-shell relation can be
found in Refs.~\cite{Gray:1990yh} and~\cite{Bekavac:2007tk,Fael:2020bgs} to two-
and three-loop accuracy, respectively. The charm mass effects in
the $m^{\rm kin}$--$m^{\rm OS}$ relation were discussed in
Section~\ref{sec::charm}.

Let us in a first step present results for the relation between the kinetic
and the on-shell mass, see Eq.~(\ref{eq::mkin}).
Up to three-loop order our results reads
\begin{eqnarray}
  \frac{m^{\text{kin}}}{m^{\text{OS}}} &=& 
    1
    + \frac{\alpha_s^{(n_r)}}{\pi} 
    \left[
        \frac{\mu}{m^{\text{OS}}} t^{(1,1)}
      + \frac{\mu^2}{\left(m^{\text{OS}}\right)^2} t^{(1,2)}
    \right]
    \nonumber \\ && \mbox{}    
    + \left(\frac{\alpha_s^{(n_r)}}{\pi}\right)^2 
    \left[ 
      \frac{\mu}{m^{\text{OS}}} t^{(2,1)}
      + \frac{\mu^2}{\left(m^{\text{OS}}\right)^2} t^{(2,2)}
      + \delta_{n_r,4} \, \Delta_{m_c}^{\rm kin,(2)}(m_c^{\text{OS}},m^{\text{OS}})     
    \right]
    \nonumber \\ && \mbox{}
    + \left(\frac{\alpha_s^{(n_r)}}{\pi}\right)^3 
    \left[ 
      \frac{\mu}{m^{\text{OS}}} t^{(3,1)}
      + \frac{\mu^2}{\left(m^{\text{OS}}\right)^2} t^{(3,2)}
      + \delta_{n_r,4} \, \Delta_{m_c}^{\rm kin,(3)}(m_c^{\text{OS}},m^{\text{OS}}) 
    \right]
                    \,,
  \label{eq::mkinmOS}
\end{eqnarray}
with
\begin{eqnarray}
  t^{(1,1)} &=& -\frac{4}{3} C_F
  \,,\nonumber\\
  t^{(1,2)} &=& -\frac{1}{2} C_F
  \,,\nonumber\\
  t^{(2,1)} &=& 
  C_F \Biggl[
    C_A 
    \biggl(
      - \frac{215}{27}
      + \frac{2\pi^2}{9} 
      + \frac{22}{9} l_\mu
    \biggr)
    + n_l T_F 
    \biggl(
      \frac{64}{27}
      - \frac{8}{9} l_\mu
    \biggr)
  \Biggr]
  \,,\nonumber\\
  t^{(2,2)} &=& 
  C_F
  \Biggl[
    C_A 
    \biggl(
      - \frac{91}{36}
      + \frac{\pi^2}{12}
      + \frac{11}{12} l_\mu
    \biggr)
    + n_l T_F
    \biggl(
      \frac{13}{18}
      - \frac{1}{3} l_\mu
    \biggr)
  \Biggr]
  \,,\nonumber\\
  t^{(3,1)} &=& 
  C_F 
  \Biggl[
    C_A^2
    \biggl(
      - \frac{130867}{1944}
      + \frac{511 \pi^2}{162}
      + \frac{19 \zeta_3}{2} 
      - \frac{\pi^4}{18}
      + \biggl(
        \frac{2518}{81}
        - \frac{22 \pi^2}{27}
      \biggr) l_\mu
      - \frac{121}{27} l_\mu^2
    \biggr)
    \nonumber \\ &&\mbox{}
    + C_A n_l T_F
    \biggl(
      \frac{19453}{486}
      - \frac{104 \pi^2}{81} 
      - 2 \zeta_3
      + \biggl(
        - \frac{1654}{81}
        + \frac{8\pi^2}{27}
      \biggr) l_\mu
      + \frac{88}{27} l_\mu^2
    \biggr)
    \nonumber \\ &&\mbox{}
    + C_F n_l T_F
    \biggl(
      \frac{11}{4}
      - \frac{4 \zeta_3}{3} 
      - \frac{2}{3} l_\mu
    \biggr)
    + n_l^2 T_F^2
    \biggl(
      - \frac{1292}{243}
      + \frac{8\pi^2}{81}
      + \frac{256}{81} l_\mu
      - \frac{16}{27} l_\mu^2
    \biggr)
  \Biggr]
  \,,\nonumber\\
  t^{(3,2)} &=& 
  C_F
  \Biggl[
    C_A^2
    \biggl(
      - \frac{96295}{5184}
      + \frac{445 \pi^2}{432} 
      + \frac{57 \zeta_3}{16} 
      - \frac{\pi^4}{48}
      + \biggl(
        \frac{2155}{216}
        - \frac{11 \pi^2}{36} 
      \biggr) l_\mu
      - \frac{121}{72} l_\mu^2
    \biggr)
    \nonumber \\ &&\mbox{}
    + C_A n_l T_F
    \biggl(
      \frac{13699}{1296}
      - \frac{23 \pi^2}{54}
      - \frac{3 \zeta_3}{4}
      + \biggl(
        - \frac{695}{108}
        + \frac{\pi^2}{9}
      \biggr) l_\mu
      + \frac{11}{9} l_\mu^2
    \biggr)
    \nonumber \\ &&\mbox{}
    + C_F n_l T_F
    \biggl(
      \frac{29}{32}
      - \frac{\zeta_3}{2}
      - \frac{1}{4} l_\mu
    \biggr)
    + n_l^2 T_F^2
    \biggl(
      - \frac{209}{162}
      + \frac{\pi^2}{27}
      + \frac{26}{27} l_\mu
      - \frac{2}{9} l_\mu^2
    \biggr)
  \Biggr]\,,
  \label{eq::tij}
\end{eqnarray}
where $l_\mu = \ln(2\mu/\mu_s)$ and $\mu_s$ is the renormalization scale of the
strong coupling constant $\alpha_s^{(n_r)}(\mu_s)$;
$\mu$ is the Wilsonian cutoff.
Note that $m^{\rm kin}$ on the r.h.s.\ of Eq.~(\ref{eq::mkin}) has been replaced 
by $m^{\rm OS}$ by applying the $m^{\rm kin}$--$m^{\rm OS}$ relation iteratively.
In Eqs.~(\ref{eq::mkinmOS}) and~(\ref{eq::tij}) we have 
$n_l=n_r=3$ for $m=m_c$ while for $m=m_b$ we have $n_l=3$ and $n_r=3$ or $4$.

$\Delta^{\rm kin,(2)}$ and $\Delta^{\rm kin,(3)}$ denote the two- and
three-loop finite-$m_c$ corrections, {\ms respectively,} which have to be taken into account
if the bottom quark relation is considered for $n_r=4$. The analytic
expressions are given by
\begin{eqnarray}
  \Delta_{m_c}^{\rm kin,(2)}(m_c,m) &=& 
  - c_1\left(\frac{m_c}{\mu_{s}}\right)
  \left[ 
      \frac{\mu}{m} t^{(1,1)} 
    + \frac{\mu^2}{\left(m\right)^2} t^{(1,2)} 
  \right]
  \,,\nonumber\\
  \Delta_{m_c}^{\rm kin,(3)}(m_c,m) &=& 
  \left[ 
    2 \, c_1^2\left(\frac{ m_c }{\mu_{s}}\right)
    - c_2\left(\frac{ m_c }{\mu_{s}}\right) 
    \right] 
    \left[ 
        \frac{\mu}{m} t^{(1,1)} 
      + \frac{\mu^2}{\left(m\right)^2} t^{(1,2)} 
    \right]
    \nonumber \\ &&\mbox{}
    - 2 \, c_1\left(\frac{ m_c }{\mu_{s}}\right)
    \left[ 
        \frac{\mu}{m} t^{(2,1)} 
      + \frac{\mu^2}{\left(m\right)^2} t^{(2,2)} 
    \right]
  \,,
  \label{eq::mKINomOS}
\end{eqnarray}
where we have identified the decoupling scale $\mu_{th}$ with $\mu_s$.
Note that these corrections are pure decoupling effects and are 
absent for $n_r=3$.  The functions $c_1\left(m/\mu_{th}\right)$ and
$c_2\left(m/\mu_{th}\right)$ are given in Eqs.~(\ref{eq:decoupling1})
and~(\ref{eq:decoupling2}), respectively.

For convenience of the reader, we also present the inverted relation which
allows for the computation of the on-shell mass from the kinetic mass:
\begin{eqnarray}
  \frac{m^{\text{OS}}}{m^{\text{kin}}} &=& 
    1
    - \frac{\alpha_s^{(n_r)}}{\pi} 
    \left[
      \frac{\mu}{m^{\text{kin}}} t^{(1,1)}
      + \frac{\mu^2}{\left(m^{\text{kin}}\right)^2} t^{(1,2)}
    \right]
    \nonumber \\ &&\mbox{}
    - \left(\frac{\alpha_s^{(n_r)}}{\pi}\right)^2 
    \left[
      \frac{\mu}{m^{\text{kin}}} t^{(2,1)}
      + \frac{\mu^2}{\left(m^{\text{kin}}\right)^2} t^{(2,2)}
      + \delta_{n_r,4} \, \Delta_{m_c}^{\rm kin,(2)}(m_c^{\text{OS}},m^{\text{kin}}) 
    \right]
    \nonumber \\ && \mbox{}
    - \left(\frac{\alpha_s^{(n_r)}}{\pi}\right)^3 
    \Biggl\{ 
      \frac{\mu}{m^{\text{kin}}} t^{(3,1)}
      + \frac{\mu^2}{\left(m^{\text{kin}}\right)^2} t^{(3,2)}
      + \delta_{n_r,4}\Delta_{m_c}^{\rm kin,(3)}(m_c^{\text{OS}},m^{\text{kin}}) 
    \Biggr\}
    \,.
    \label{eq::mOSomKIN}
\end{eqnarray}
Note that in the case of the charm quark we always have $n_l=n_r=3$.

Next we want to consider the relation between the kinetic
and the $\overline{\rm MS}$ mass of the bottom quark.
In order to keep the formula compact, we choose to work
with $n_r=4$ active flavors for the running of $\alpha_s$.
In section~\ref{sec::num} we will refer to this choice as 
scheme (B).
To obtain this relation, we replace the pole mass on the
r.h.s.\ of~(\ref{eq::mKINomOS}) by the $\overline{\rm MS}$ mass using
the corresponding three-loop
relation~\cite{Chetyrkin:1999ys,Chetyrkin:1999qi,Melnikov:2000qh}.  
We choose a common renormalization scale $\mu_s$ for $\alpha_s$,
$\overline{m}_b$ and $\overline{m}_c$. In the $m^{\rm OS}_b$--$\overline{m}_b$
relation, we replace $\alpha_s^{(n_f=5)}$ in favour of $\alpha_s^{(4)}$
to have the same expansion parameter as in Eqs.~(\ref{eq::mKINomOS})
and~(\ref{eq::mOSomKIN}).  
The corresponding formula reads
\begin{eqnarray}
  m_b^{\text{OS}} &=& \overline{m}_b(\mu_s) 
      \Biggl[
        1
        + \frac{\alpha_s^{(4)}}{\pi} y_m^{(1)}
        + \left(\frac{\alpha_s^{(4)}}{\pi}\right)^2
        \left(
          y_m^{(2)} 
          + \Delta_{m_c}^{(2)}
        \right)
        + \left(\frac{\alpha_s^{(4)}}{\pi}\right)^3
        \left(
          y_m^{(3)} 
          + \Delta_{m_c}^{(3)}
        \right)
      \Biggr]
                       \,,
                       \nonumber\\
  \label{eq::mOSmMS}
\end{eqnarray}
with 
\begin{eqnarray}
  y_m^{(1)} &=& C_F \left( 1 + \frac{3}{4} l_m  \right)
  \,,
  \nonumber\\
  y_m^{(2)} &=&
  C_A C_F 
  \biggl[
        \frac{1111}{384}
        -\frac{3}{8} \zeta_3
        +\frac{\pi ^2}{12}  
        \big(
              3 l_2
              -1
        \big)
        +\frac{185 l_m}{96}
        +\frac{11 l_m^2}{32}
  \biggr]
  \nonumber\\&&\mbox{}
  + C_F T_F 
  \biggl[
        -\frac{107}{48}
        +\frac{\pi ^2}{12}
        -\frac{3 l_m}{4}
        - n_l 
        \biggl(
                 \frac{71}{96}
                +\frac{\pi ^2}{12}
                +\frac{13 l_m}{24}
                +\frac{l_m^2}{8}
        \biggr)
  \biggr]
  \nonumber\\&&\mbox{}
  + C_F^2 
  \biggl[
        -\frac{71}{128}
        +\frac{3}{4} \zeta_3
        +\frac{\pi ^2}{16}  
        \big(
                5
                -8 l_2
        \big)
        -\frac{9 l_m}{32}
        +\frac{9 l_m^2}{32}
  \biggr]
  \,,
  \nonumber\\
  y_m^{(3)} &=&
    C_F^2 T_F 
    \Biggl\{
        -\frac{397}{144}
        -\frac{16 a_4}{3}
        +\frac{7 \pi ^4}{540}
        + n_l 
        \biggl[
                -\frac{109}{72}
                -\frac{8 a_4}{3}
                +\frac{119 \pi ^4}{2160}
                -\pi ^2 
                \biggl(
                         \frac{55}{72}
                        +\frac{2 l_2^2}{9}
                        \nonumber\\&&\mbox{}
                        +\frac{13 l_m}{48}
                        -\frac{1}{9} l_2 
                        \bigl(
                                11+3 l_m
                        \bigr)
                \biggr)
                -\frac{55}{24} \zeta_3
                -\frac{l_2^4}{9}
                +\biggl(
                        -\frac{91}{384}
                        +\frac{\zeta_3}{4}
                \biggr) l_m
                -\frac{3 l_m^2}{32}
                -\frac{3 l_m^3}{32}
        \biggr]
        \nonumber\\&&\mbox{}
        +\pi ^2 
        \biggl[
                \frac{23}{216}
                -\frac{l_2^2}{9}
                -\frac{7 l_m}{48}
                +\frac{1}{3} l_2 
                \bigl(
                        1+l_m
                \bigr)
        \biggr]
        -\frac{1}{12} \zeta_3
        -\frac{2 l_2^4}{9}
        +\biggl(
                -\frac{93}{64}
                +\zeta_3
        \biggr) l_m
        -\frac{9 l_m^2}{16}
    \Biggr\}
    \nonumber\\&&\mbox{}
    + C_A C_F T_F 
    \Biggl\{
        -\frac{109589}{7776}
        +\frac{8 a_4}{3}
        +\frac{67 \pi ^4}{2160}
        +\pi ^2 
        \biggl[
                \frac{293}{108}
                -\frac{1}{8} \zeta_3
                +\frac{l_2^2}{18}
                -\frac{l_2}{6}  
                \bigl(
                        25+l_m
                \bigr)
                \nonumber\\&&\mbox{}
                +\frac{5 l_m}{24}
        \biggr]
        + n_l 
        \biggl[
                -\frac{70763}{15552}
                +\frac{4 a_4}{3}
                -\frac{19 \pi ^4}{2160}
                +\pi ^2 
                \biggl(
                        -\frac{175}{432}
                        +\frac{l_2^2}{9}
                        -\frac{l_2}{18}  
                        \bigl(
                                11+3 l_m
                        \bigr)
                        -\frac{7 l_m}{72}
                \biggr)
                \nonumber\\&&\mbox{}
                -\frac{89}{144} \zeta_3
                +\frac{l_2^4}{18}
                -\biggl(
                         \frac{869}{216}
                        +\frac{\zeta_3}{2}
                \biggr) l_m
                -\frac{373 l_m^2}{288}
                -\frac{11 l_m^3}{72}
        \biggr]
        +\frac{5}{36} \zeta_3
        +\frac{5}{8} \zeta_5
        +\frac{l_2^4}{9}
        \nonumber\\&&\mbox{}
        -\biggl(
                 \frac{12515}{1728}
                +\frac{5 \zeta_3}{4}
        \biggr) l_m
        -\frac{143 l_m^2}{144}
        -\frac{11 l_m^3}{144}
    \Biggr\}
    + C_F T_F^2 
    \Biggl\{
        \frac{5917}{1944}
        + n_l 
        \biggl(
                \frac{4135}{1944}
                +\frac{13 \pi ^2}{108}
                \nonumber\\&&\mbox{}
                +\frac{5}{9} \zeta_3
                +\frac{715 l_m}{432}
                +\frac{13 l_m^2}{36}
                +\frac{l_m^3}{36}
        \biggr)
        + n_l^2 
        \biggl(
                \frac{2353}{7776}
                +\frac{7}{18} \zeta_3
                +\frac{89 l_m}{216}
                +\frac{13 l_m^2}{72}
                +\frac{l_m^3}{36}
                \nonumber\\&&\mbox{}
                +\frac{\pi^2}{108}  
                \bigl(
                        13+6 l_m
                \bigr)
        \biggr)
        -\frac{4}{9} \zeta_3
        -\frac{\pi^2}{270} 
        \bigl(
                8+15 l_m
        \bigr)
        +\frac{251 l_m}{216}
        +\frac{l_m^2}{9}
        +\frac{l_m^3}{36}
    \Biggr\}
    \nonumber\\&&\mbox{}
    + C_F^3 
    \Biggl\{
        \frac{893}{192}
        +12 a_4
        +\frac{\pi ^4}{48}
        +\pi ^2 
        \biggl(
                \frac{643}{192}
                +\frac{1}{16} \zeta_3
                -\frac{l_2^2}{2}
                +\frac{15 l_m}{64}
                -\frac{3}{8} l_2 
                \bigl(
                        20+l_m
                \bigr)
        \biggr)
        \nonumber\\&&\mbox{}
        +\frac{87}{16} \zeta_3
        -\frac{5}{8} \zeta_5
        +\frac{l_2^4}{2}
        +\biggl(
                \frac{495}{512}
                +\frac{9 \zeta_3}{16}
        \biggr) l_m
        -\frac{63 l_m^2}{128}
        +\frac{9 l_m^3}{128}
    \Biggr\}
    + C_A^2 C_F 
    \Biggl\{
        \frac{1322545}{124416}
        \nonumber\\&&\mbox{}
        -\frac{11 a_4}{3}
        +\frac{179 \pi ^4}{3456}
        +\pi ^2 
        \biggl(
                \frac{1955}{3456}
                -\frac{51}{64} \zeta_3
                -\frac{11 l_2^2}{36}
                -\frac{11 l_m}{72}
                +\frac{1}{72} l_2 
                \bigl(
                        115+33 l_m
                \bigr)
        \biggr)
        \nonumber\\&&\mbox{}
        -\frac{1343}{288} \zeta_3
        +\frac{65}{32} \zeta_5
        -\frac{11 l_2^4}{72}
        +\biggl(
                \frac{13243}{1728}
                -\frac{11 \zeta_3}{16}
        \biggr) l_m
        +\frac{2341 l_m^2}{1152}
        +\frac{121 l_m^3}{576}
    \Biggr\}
    \nonumber\\&&\mbox{}
    + C_A C_F^2 
    \Biggl\{
        -\frac{24283}{4608}
        +\frac{4 a_4}{3}
        -\frac{65 \pi ^4}{432}
        +\pi ^2 
        \biggl(
                -\frac{533}{576}
                +\frac{19}{16} \zeta_3
                +\frac{31 l_2^2}{36}
                +\frac{49 l_m}{96}
                \nonumber\\&&\mbox{}
                +\frac{5}{144} l_2 
                \bigl(
                        16-21 l_m
                \bigr)
        \biggr)
        +\frac{755}{96} \zeta_3
        -\frac{45}{16} \zeta_5
        +\frac{l_2^4}{18}
        -\biggl(
                 \frac{4219}{1536}
                -\frac{35 \zeta_3}{32}
        \biggr) l_m
        \nonumber\\&&\mbox{}
        +\frac{21 l_m^2}{64}
        +\frac{33 l_m^3}{128}
    \Biggr\}
  \,,
  \nonumber\\
  \Delta_{m_c}^{(2)} &=&
  C_F T_F 
  \biggl[
        -\frac{3 z^2}{4}
        +\frac{\pi ^2}{12} z 
        \bigl(
                 3
                +3 z^2
                -z^3
        \bigr)
        +\frac{1}{2}
        \biggl(
                 (1-z)^2 \big(1+z+z^2\big) \ln (1-z)
                 \nonumber\\&&\mbox{}
                +(1+z)^2 \big(1-z+z^2\big) \ln (1+z)
                -z^2
        \biggr) \ln (z)
        -\frac{1}{2} z^4 \ln^2(z)
        \nonumber\\&&\mbox{}
        +\frac{1}{2} (1+z)^2 \big(1-z+z^2\big) \text{Li}_2(-z)
        +\frac{1}{2} (1-z)^2 \big(1+z+z^2\big) \text{Li}_2(z)
  \biggr]
  \,,
  \nonumber\\
  \Delta_{m_c}^{(3)} &=&
  \biggl[ 
    \frac{1}{2} C_F
    + \frac{2}{3} T_F  l_m
  \biggr] \Delta_{m_c}^{(2)}
  + C_F^2 T_F l_c \biggl[
        -\frac{3 z^2}{2}
        +\frac{1}{16} \pi ^2 z \big(3+9 z^2-4 z^3\big)
        \nonumber\\&&\mbox{}
        +\frac{3}{8} z (1+z) \big(1-z+4 z^2\big) \text{Li}_2(-z)
        -\frac{3}{8} z (1-z) \big(1+z+4 z^2\big) \text{Li}_2(z)
        \nonumber\\&&\mbox{}
        +\biggl(
                -\frac{3 z^2}{4}
                -\frac{3}{8} z (1-z) \big(1+z+4 z^2\big) \ln (1-z)
                \nonumber\\&&\mbox{}
                +\frac{3}{8} z (1+z) \big(1-z+4 z^2\big) \ln (1+z)
        \biggr) \ln (z)
        -\frac{3}{2} z^4 \ln^2(z)
  \biggr]
  \nonumber\\&&\mbox{}
  + C_F^2 T_F l_m 
  \biggl[
        \frac{3 z^2}{8}
        +\frac{1}{16} \pi ^2 z \big(3-3 z^2+2 z^3\big)
        +\frac{3}{4} z^4 \ln ^2(z)
        \nonumber\\&&\mbox{}
        +\frac{3}{8} (1-z^2) \big(2-z+2 z^2\big) \biggl( \ln (1-z) \ln (z) + \text{Li}_2(z) \biggr)
        \nonumber\\&&\mbox{}
        +\frac{3}{8} (1-z^2) \big(2+z+2 z^2\big) \biggl( \ln (1+z) \ln (z) + \text{Li}_2(-z) \biggr)
  \biggr]
  - \tilde{\Delta}_{m_c}^{(3)} \, ,
\end{eqnarray}
where
\begin{align}
  z &= \frac{\overline{m}_c}{\overline{m}_b}
  \,, &
  l_m &= \ln\left( \frac{\mu_s^2}{\overline{m}_b^2} \right) 
  \,, &
  l_c &= \ln\left( \frac{\mu_s^2}{\overline{m}_c^2} \right) 
  \,, \nonumber\\[5pt]
  l_2 &= \ln\left( 2 \right) 
  \,, &
  a_4 &= \text{Li}_4\left(\frac{1}{2}\right)
  \,.
\end{align}

The contributions due to charm quark mass are taken from
Ref.~\cite{Fael:2020bgs}.  The term $\tilde{\Delta}_{m_c}^{(3)}$
denotes the genuine three loop contribution, which vanishes for $m_c
\to 0$. It can be extracted from
the ancillary file of Ref.~\cite{Fael:2020bgs}:
\begin{eqnarray}
  \tilde{\Delta}_{m_c}^{(3)}
  &=& \left.
  \texttt{zmnum}[x, \overline{m}_b, \mu_s, \mu_s]
  - \lim_{x\to0} \texttt{zmnum}[x, \overline{m}_b, \mu_s, \mu_s] \right|_{\alpha_s^3}
  \,.
\end{eqnarray}
Alternatively, also the corresponding analytic expressions
provided in~\cite{Fael:2020bgs} can be used.

After expanding in $\alpha_s^{(4)}(\mu_s)$ up to third order and
$\mu/\overline{m}$ up to second order, we obtain the relation between
the kinetic and the $\overline{\text{MS}}$  mass of the bottom quark:
\begin{eqnarray}
  \frac{m^{\rm kin}_b(\mu)}{\overline{m}_b} &=& 
    1
    + \frac{\alpha_s^{(4)}}{\pi}
    \biggl[
      y_m^{(1)} 
      + \frac{\mu}{\overline{m}_b} t^{(1,1)}
      + \frac{\mu^2}{\overline{m}_b^2} t^{(1,2)}
    \biggr]
    \nonumber\\&&\mbox{}
    + \left(\frac{\alpha_s^{(4)}}{\pi}\right)^2
    \Biggl[
      y_m^{(2)}
      + \Delta_{m_c}^{(2)}
      + \frac{\mu}{\overline{m}_b} t^{(2,1)}
      + \frac{\mu^2}{\overline{m}_b^2}
      \biggl(
        t^{(2,2)}
        - y_m^{(1)} t^{(1,2)}
      \biggr)
      \nonumber\\&&\mbox{}
      + \Delta_{m_c}^{\rm kin,(2)}(\overline{m}_c,\overline{m}_b)
    \Biggr]
    + \left(\frac{\alpha_s^{(4)}}{\pi}\right)^3
    \Biggl[
      y_m^{(3)}
      + \Delta_{m_c}^{(3)} 
      + \frac{\mu}{\overline{m}_b} t^{(3,1)}
      \nonumber\\&&\mbox{}
      + \frac{\mu^2}{\overline{m}_b^2}
      \Biggl(
        t^{(3,2)}
        - y_m^{(1)} t^{(2,2)}
        - t^{(1,2)}
        \biggl(
          y_m^{(2)} + \Delta_{m_c}^{(2)} - \left(y_m^{(1)}\right)^2
        \biggr)
      \Biggr)
      \nonumber\\&&\mbox{}
      + 
        \Delta_{m_c}^{\rm kin,(3)}(\overline{m}_c,\overline{m}_b)
        + y_m^{(1)} \Delta_{m_c}^{\rm kin,(2)}(\overline{m}_c,\overline{m}_b)
        + \frac{\mu}{\overline{m}_b} t^{(1,1)}
        \biggl(
          c_1\left(\frac{\overline{m}_c}{\mu_s}\right) y_m^{(1)}
          \nonumber\\&&\mbox{}
          + \frac{2}{3} T_F y_{m_c}^{(1)}
        \biggr)
        + \frac{\mu^2}{\overline{m}_b^2} t^{(1,2)}
        \biggl(
          2 c_1\left(\frac{\overline{m}_c}{\mu_s}\right) y_m^{(1)}
          + \frac{2}{3} T_F y_{m_c}^{(1)}
        \biggr)
      \Biggr]
  \,,
  \label{eq::mkinmMSbar}
\end{eqnarray}
with 
\begin{eqnarray}
  y_{m_c}^{(1)} &=& C_F \left( 1 + \frac{3}{4} l_c  \right)
  \,,
\end{eqnarray}
$\overline{m}_b=\overline{m}_b(\mu_s)$, $\overline{m}_c=\overline{m}_c(\mu_s)$ 
as well as $\alpha_s^{(4)}=\alpha_s^{(4)}(\mu_s)$.
The inverted relation $\overline{m}_b/m^{\rm kin}_b(\mu)$ is obtained
from Eq.~(\ref{eq::mkinmMSbar}) in a straightforward way. We refrain
from printing it in the paper but refer to the ancillary files~\cite{progdata}
where also the relations between the kinetic and on-shell mass can be found.




\section{\label{sec::BLM}BLM corrections to four loops}

In this section we present the four-loop contribution to the $m^{\rm
  OS}$--$m^{\rm kin}$ relation for the class of diagrams that contain
three insertions of massless fermion bubbles in a gluon propagator.
Such corrections are usually referred to as ``large-$\beta_0$''
(or ``BLM''~\cite{Brodsky:1982gc}) corrections, see also~\cite{Broadhurst:1994se}. Often
one performs the replacement
\begin{eqnarray}
  n_l &\to& - \frac{3}{2}\beta_0\,,
\end{eqnarray}
with $\beta_0 = 11 C_A/3 - 4 T_F n_l/3$ and uses these corrections to estimate
unknown higher order contributions.  We adopt this approach in order to get a
hint about the size of the ${ O}(\alpha_s^4)$ corrections.

Let us have a brief look to the large-$\beta_0$ corrections at two- and
three-loop order. We consider the $m_b^{\rm kin}$-$\overline{m}_b$
relation and obtain
at two and three loops the following large-$\beta_0$ terms:\footnote{The
  numbers correspond to scheme (D) defined in Section~\ref{sec::num}.}
\begin{eqnarray}
  m_b^{\rm kin}(1~\mbox{GeV}) 
  &=& 
      4163 + 248 + (-57 + 137|_{{\rm large}-\beta_0}) 
      + (15 + 15|_{{\rm large}-\beta_0})
      \,.
      \label{eq::mbBLM}
\end{eqnarray}
One observes that at ${O}(\alpha_s^2)$ the large-$\beta_0$ {\ms term}
is about twice as big as the remaining contribution, however, it has a
different sign. Thus, it overshoots the full result by a factor of two.
At three-loop order, the large-$\beta_0$ term amounts to half
of the complete result.

The leading $n_l^n$ term at order $(n+1)$ can be obtained by dressing the
gluon propagator in each of the four one-loop diagrams with $n$ closed
(massless) fermion loops.  The bubbles can be integrated out which leads to an
effective gluon propagator raised to a symbolic power. In fact, if we denote
the momentum through the gluon line by $k_1$ it is sufficient to perform the
simple replacement
\begin{eqnarray}
  \frac{-g^{\rho \sigma}}{[-k_1^2]} 
        &\to& \frac{-g^{\rho \sigma}}{[-k_1^2]}
        \left(8\Gamma(2-d/2)\frac{\Gamma^2(d/2)}{\Gamma(d)}\frac{1}{[k_1^2]^{2-d/2}}\right)^n\,,
\end{eqnarray}
in order to obtain the $n_l^n$ contribution at order $\alpha_s^{n+1}$.
It is straightforward to obtain analytic results for any given value of $n$.
Our interest is for $n=3$ which gives
\begin{eqnarray}
  \left. \frac{m^{\rm kin}}{m^{\text{OS}}} \right|_{\alpha_s^4,BLM} &=& 
    C_F \beta_0^3 \Biggl[
      \frac{\mu}{m^{\text{OS}}} \biggl(
        - \frac{4069}{648}
        + \frac{2\pi^2}{9}
        + \frac{\zeta_3}{4} 
        + \left( \frac{323}{72} - \frac{\pi^2}{12} \right) l_\mu 
        - \frac{4}{3} l_\mu^2
        + \frac{1}{6} l_\mu^3
      \biggr)
\nonumber \\ &&
        + \frac{\mu^2}{\left(m^{\text{OS}}\right)^2} \biggl(
        - \frac{4205}{3456}
        + \frac{13\pi^2}{192}
        + \frac{3\zeta_3}{32} 
        + \left( \frac{209}{192} - \frac{\pi^2}{32} \right) l_\mu 
\nonumber \\ &&
        - \frac{13}{32} l_\mu^2
        + \frac{1}{16} l_\mu^3
      \biggr)
    \Biggr] 
                \left(\frac{\alpha_s}{\pi}\right)^4
                \,,
                \label{eq::mkin-mOS-nl3}
\end{eqnarray}
with $l_\mu = \ln(2\mu/\mu_s)$, where the counterterms
for the strong coupling constant {\ms have} been obtained from the known
lower-order results.

Next we combine Eq.~(\ref{eq::mkin-mOS-nl3}) with the corresponding terms from
the $\overline{\text{MS}}$-on-shell relation~\cite{Beneke:1994qe,Lee:2013sx}
and obtain
\begin{eqnarray}
  \left. \frac{m^{\rm kin}}{\overline{m}} \right|_{\alpha_s^4,BLM} &=& 
  C_F \beta_0^3
  \Biggl[
    \frac{42979}{442368} 
    + \frac{89 \pi^2}{1536}
    + \frac{317 \zeta_3}{1024}
    + \frac{71 \pi^4}{10240}
    + \left( \frac{1301}{9216} + \frac{13\pi^2}{256} + \frac{9 \zeta_3}{64} \right) l_m
\nonumber \\ &&
    + \left( \frac{89}{1024} + \frac{3 \pi^2}{256} \right) l_m^2
    + \frac{13}{512} l_m^3
    + \frac{3}{1024} l_m^4 
    + \frac{\mu}{\overline{m}} 
    \biggl(
        - \frac{4069}{648}
        + \frac{2\pi^2}{9}
        + \frac{\zeta_3}{4} 
\nonumber \\ &&
        + \left( \frac{323}{72} - \frac{\pi^2}{12} \right) l_\mu 
        - \frac{4}{3} l_\mu^2
        + \frac{1}{6} l_\mu^3
    \biggr)
      + \frac{\mu^2}{\overline{m}^2} 
      \biggl(
        - \frac{4205}{3456}
        + \frac{13\pi^2}{192}
        + \frac{3\zeta_3}{32} 
\nonumber \\ &&
        + \left( \frac{209}{192} - \frac{\pi^2}{32} \right) l_\mu 
        - \frac{13}{32} l_\mu^2
        + \frac{1}{16} l_\mu^3
      \biggr)
  \Biggr]
                \left(\frac{\alpha_s}{\pi}\right)^4
                \,,
   \label{eq::mKIN-mMS-BLM}
\end{eqnarray}
with $l_m = \ln(\mu_s^2/\overline{m}^2)$.

We anticipate that the numerical effect is small for the bottom quark: for $\mu=1$~GeV,
$\mu_s=\overline{m}$, $\overline{m}=4.163$~GeV and $n_l=3$ we obtain a
contribution of about $-9$~MeV 
to $m^{\rm kin}$ in Eq.~(\ref{eq::mKIN-mMS-BLM}).


\section{\label{sec::num}Numerical results}

The input values for our numerical analysis are
$\alpha_s^{(5)}(M_Z)=0.1179$~\cite{Tanabashi:2018oca},
$\overline{m}_c(3~\mbox{GeV})=0.993$~GeV~\cite{Chetyrkin:2017lif} and
$\overline{m}_b(\overline{m}_b)=4.163$~GeV~\cite{Chetyrkin:2009fv}.
We use {\tt RunDec}~\cite{Herren:2017osy} for the running of the
$\overline{\rm MS}$ parameters and the decoupling of heavy
particles. For the Wilsonian cutoff we choose $\mu=1$~GeV for
bottom~\cite{Gambino:2013rza} and $\mu=0.5$~GeV or $\mu=1$~GeV for
charm~\cite{Gambino:2010jz}.

\subsection{Charm mass}

Let us start with the charm quark where we have $n_l=3$. 
Often numerical values for $\overline{m}_c(\overline{m}_c)$ are provided. 
However, this choice suffers from small renormalization scales of the order 1~GeV. 
A more appropriate choice is thus $\overline{m}_c(2~\mbox{GeV})$ or
$\overline{m}_c(3~\mbox{GeV})$. For the three choices we obtain the
following perturbative expansions for $m_c^{\rm kin}(0.5~\mbox{GeV})$:
\begin{eqnarray}
  m_c^{\rm kin}(0.5~\mbox{GeV}) &=&  \hphantom{0}993 + 191 + 100 + 52~\mbox{MeV} = 1336~\mbox{MeV}\,,
                    \nonumber\\
  m_c^{\rm kin}(0.5~\mbox{GeV}) &=&  1099 + 163 + \hphantom{0}76 + 34~\mbox{MeV} = 1372~\mbox{MeV}\,,
                    \nonumber\\
		    m_c^{\rm kin}(0.5~\mbox{GeV}) &=&  1279 + \hphantom{0}84 + \hphantom{0}30  + 11~\mbox{MeV} = 1404~\mbox{MeV}\,.
\end{eqnarray} 
For $m_c^{\rm kin}(1~\mbox{GeV})$ we obtain:
\begin{eqnarray}
  m_c^{\rm kin}(1~\mbox{GeV}) &=&  \hphantom{0}993 + 83 + 35 + 14~\mbox{MeV} = 1125~\mbox{MeV}\,,
                    \nonumber\\
  m_c^{\rm kin}(1~\mbox{GeV}) &=&  1099 + 37 + \hphantom{0}2 - \hphantom{0}3~\mbox{MeV} = 1135~\mbox{MeV}\,,
                    \nonumber\\
  m_c^{\rm kin}(1~\mbox{GeV}) &=&  1279 - 73 - 61  - 17~\mbox{MeV} = 1128~\mbox{MeV}\,,
\end{eqnarray}
where, from top to bottom, $\mu_s=3~\mbox{GeV}, 2~\mbox{GeV}$ and
$\overline{m}_c$ was chosen for $\overline{m}_c(\mu_s)$ and $\alpha_s(\mu_s)$.  
Within each equation, the four numbers
after the first equality sign refer to the tree-level results and the one-,
two- and three-loop corrections.  One observes that for each choice of $\mu_s$
the perturbative expansion behaves reasonably. It is interesting to mention
that for $\mu_s=2~\mbox{GeV}$ and $\mu=1~\mbox{GeV}$ both the two- and
three-loop corrections are particularly small and have different signs. For
this choice of $\mu$, we also observe that with $\mu_s=3~\mbox{GeV}$ the loop
corrections are positive, whereas for $\mu_s=\overline{m}_c$ they are
negative.  The three-loop terms {\ms range} from $+14$~MeV to $-17$~MeV and roughly
cover the splitting of the final numbers for $m_c^{\rm kin}(1~\mbox{GeV})$.
For $\mu=0.5~\mbox{GeV}$ the three-loop terms range from $10$~MeV to $52$~MeV
which again covers the splitting of the final numbers for
$m_c^{\rm kin}(0.5~\mbox{GeV})$.

\subsection{Bottom mass}

Let us in the following investigate the numerical effects for the bottom quark
including finite charm mass effects. 
In principle one has two choices for the treatment of $m_c$: one can either
assume that $m_c\sim m_b$ and thus we have $m_c^2\gg |y|$. This leads to the
results discussed in Section~\ref{sec::charm}, i.e., the charm mass effects in the
$m^{\rm kin}_b$--$m^{\rm OS}_b$ relation are given by the decoupling terms in
$\alpha_s$. On the other hand, if the limit $m_c^2\ll m_b^2$ is considered in
the $m^{\rm kin}_b$--$m^{\rm OS}_b$ relation, we are forced to set
$m_c=0$. This is because
due to the various limits involved in the definition of the kinetic
mass there is no energy available to produce massive charm quarks. 
This means there is no expansion in $m_c/m_b$.

In the following we consider four schemes for the treatment of
$m_c$ effects:\footnote{This extends the considerations of
  Ref.~\cite{Gambino:2011cq} since now the charm quark mass effects are
  completed up to ${O}(\alpha_s^3)$, both for the $m^{\rm kin}_b$--$m^{\rm
    OS}_b$ and the $\overline{m}_b$--$m^{\rm OS}_b$ relations.}
\begin{itemize}
\item[(A)] We parametrize the  $\overline{m}_b$--$m^{\rm kin}_b$ relation in terms
  of $\alpha_s^{(3)}$, i.e., we assume that the charm quark is decoupled and
  that there are no $m_c$ effects in the $m^{\rm kin}_b$--$m^{\rm OS}_b$ relation.
  Charm quark mass effects only come from the $\overline{m}_b$--$m^{\rm OS}_b$
  relation. They are contained in the $\Delta_{m_c}^{(k)}$ ($k=2,3$) terms, which
  vanish in the limit $m_c\to0$, and from decoupling effects
  in the transition from $\alpha_s^{(4)}$ to $\alpha_s^{(3)}$
  in the $\overline{m}_b$--$m^{\rm OS}_b$ relation.
\item[(B)] We parametrize the $\overline{m}_b$--$m^{\rm kin}_b$ relation in terms
  of $\alpha_s^{(4)}$. The corresponding expression is obtained from scheme
  (A) by using the decoupling relations for $\alpha_s$.  The charm quark mass
  effects are contained in the quantities $\Delta_{m_c}^{(k)}$ and
  $\Delta_{m_c}^{\rm kin,(k)}$ which originate from the
  $\overline{m}_b$--$m^{\rm OS}_b$ and $m^{\rm kin}_b$--$m^{\rm OS}_b$ relations,
  respectively.
\item[(C)]
  We parametrize the $\overline{m}_b$--$m^{\rm kin}_b$ relation in terms
  of $\alpha_s^{(4)}$ but assume that $m_c^2\ll |y|$.
  Note that this requires that $n_l=4$ has to be chosen in the
  $m^{\rm kin}_b$--$m^{\rm OS}_b$ relation (whereas in all other
  scheme we have $n_l=3$).
\item[(D)]
  We assume that the charm quark is formally infinitely heavy, in particular
  heavier than the bottom quark mass. In that case, we choose $n_l=3$ (similarly to
  scheme~(A)) but we do not take into account any charm quark mass effect,
  neither from the decoupling in the $\alpha_s^{(4)}$ to $\alpha_s^{(3)}$
  transition, nor from $\Delta_{m_c}^{(k)}$ in deriving the
  formulae for the mass relation.
\end{itemize}
As compared to the schemes used in Ref.~\cite{Fael:2020iea}, we include
finite-$m_c$ effects. Note that in~\cite{Fael:2020iea} the decoupling effects
in the relation between $\alpha_s^{(4)}$ and $\alpha_s^{(3)}$ for the
$\overline{m}_b$--$m^{\rm OS}_b$ have been neglected; they are relevant for scheme (A).
The analytic result for scheme (B) can be found in
Eq.~(\ref{eq::mkinmMSbar}). The formulae for the other schemes can be
obtained from Eq.~(\ref{eq::mkinmMSbar}) in a straightforward way.

First, we use $\overline{m}_b(\overline{m}_b)$ as input to compute the kinetic
mass. We fix $\mu_s=\overline{m}_b$ but organize our formulae such that
the scale of $\overline{m}_c$ is fixed to 3~GeV. For the four schemes we obtain
\begin{eqnarray}
  \mbox{(A)}\qquad m_b^{\rm kin}(1~\mbox{GeV}) 
  &=&  4163 + 248 +
      (81+7_{\Delta_{m_c}}+12_{dec}-20_{n_c})
      \nonumber\\&&\mbox{}
                    + (30+14_{\Delta_{m_c}}+16_{dec}
                    - 30_{n_c} - 1_{n_c \times dec}
                    +0.4_{\Delta_{m_c}\times dec})~\mbox{MeV}
      \nonumber\\
    &=& 4163 + 248 + 80 + 30~\mbox{MeV} = 4520~\mbox{MeV}
                    \,,                    \nonumber\\
  \mbox{(B)}\qquad m_b^{\rm kin}(1~\mbox{GeV}) 
  &=&  4163 + 259 + (88 + 7_{\Delta_{m_c}} + 5_{\Delta_{m_c}^{\rm kin}} - 22_{n_c}) 
      \nonumber\\&&\mbox{}
                    + (34 + 16_{\Delta_{m_c}} + 10_{\Delta_{m_c}^{\rm kin}} - 34_{n_c})~\mbox{MeV}
      \nonumber\\
    &=& 4163 + 259 + 78 + 26 ~\mbox{MeV}= 4526~\mbox{MeV}
                    \,,                    \nonumber\\
  \mbox{(C)}\qquad m_b^{\rm kin}(1~\mbox{GeV}) 
  &=&  4163 + 259 + (99 + 7_{\Delta_{m_c}} - 22_{n_c}) 
                    + (59 + 16_{\Delta_{m_c}} - 34_{n_c})~\mbox{MeV}
      \nonumber\\
    &=& 4163 + 259 + 84 + 41 ~\mbox{MeV}= 4547~\mbox{MeV}
                    \,,                    \nonumber\\
  \mbox{(D)}\qquad m_b^{\rm kin}(1~\mbox{GeV}) 
  &=&  4163 + 248 + 81 + 30~\mbox{MeV}= 4521~\mbox{MeV}
      \,,
                    \label{eq::mbMS2KIN}
\end{eqnarray}
where the {\ms origins} of the charm quark mass effects have been marked with the following labels.
\begin{itemize}
 \item $dec$: it is present in scheme (A) and marks the terms in the $\overline{\rm
    MS}$-on-shell relation which originate from the decoupling of the charm
  quark in $\alpha_s$.
  
\item $n_c$: the contribution from closed charm loops (which survive even for $m_c=0$).
\item $\Delta_{m_c}$: finite $m_c$ terms from the $\overline{\rm MS}$-on-shell
  relation which vanish at $m_c=0$.
\item  $\Delta_{m_c}^{\rm kin}$: finite $m_c$ terms from the kinetic-on-shell relation
\end{itemize}

Overall, we observe that the charm mass effects from the $\overline{m}_b$--$m^{\rm OS}_b$
relation ($\Delta_{m_c}$) are sizeable and the three-loop term is about a factor of two
  bigger than the two-loop contribution.
The same is true for the charm mass effects from the $m^{\rm kin}_b$--$m^{\rm OS}_b$ relation
($\Delta_{m_c}^{\rm kin}$ in scheme (B)).
For scheme (A), the decoupling terms of $\alpha_s$ in the  $\overline{m}_b$--$m^{\rm OS}_b$
relation also provide relatively large correction.
However, for scheme (A) and (B), we see an important cancellation of these charm mass effects
against the remaining $n_c$ terms that have the opposite sign as the previous three contributions.
In scheme (C) the cancellation between $\Delta_{m_s}$ and $n_c$ is less efficient.

In the schemes (A), (B) and (D), one observes a good convergence of the perturbative series: 
the coefficients reduce by a  factor two to three when going to higher orders.
Scheme (C) behaves slightly worse. 
In our opinion, schemes (A) and (B) are the preferable choices for the phenomenological applications. 
The difference in the final results for the  kinetic mass is only 6~MeV. 
The comparison between schemes (A), (B) and (D) demonstrates that the charm quark ``wants'' to be treated as a heavy particle.
Note that the final result for scheme (D) differs from scheme (A) by only 1~MeV.

On the contrary, treating the charm as a light quark as in scheme (C) leads 
to a kinetic mass which is about 20~MeV larger than in the other schemes
which is mainly due to the finite charm quark mass effects 
in the $\overline{\rm MS}$-on-shell relation.
In this case, there is no room for a finite charm quark mass in the kinetic-on-shell relation.
Note, however, that $m_c^2\ll |y|$ is not consistent with the physical value 
of the charm quark mass and thus scheme (C) should not be used 
for practical application. On the other hand, if in nature we would have
$m_c = {\cal O}(100~\mbox{MeV})$, scheme (C) would be a perfectly
viable scheme (of course $\Delta_{m_c}$ would be much smaller in this case).

Let us compare our numerical results with our previous ones in Ref.~\cite{Fael:2020iea}.
For scheme (A) and (D), the numerical values that are not tagged by ``$dec$'', ``$\Delta_{m_c}^{\rm kin}$'' 
or ``$\Delta_{m_c}$'' agree with the first line of Eq.~(8) in~\cite{Fael:2020iea}, where we set $n_l=3$ and $\alpha_s^{(3)}$.

Scheme (C) corresponds to the scheme used in the second line in Eq.~(8) of~\cite{Fael:2020iea},
i.e., $n_l=4$ and $\alpha_s^{(4)}$ is used in the kinetic-on-shell relation.
Note that here for scheme (C), the $\overline{\rm MS}$-on-shell relation uses
$n_l=3$ instead and we mark the different charm contributions separately. 
In the limit $m_c\to0$ however, we recover the results in the second line 
in Eq.~(8) of~\cite{Fael:2020iea}.

Next we discuss the computation of the bottom quark mass in the
$\overline{\rm  MS}$ scheme using the kinetic mass $m_b^{\rm kin}=4.550$~GeV
as input. We furthermore use $\mu_s=m_b^{\rm kin}$
and obtain for the schemes (A), (B), (C) and (D)
\begin{eqnarray}
  \mbox{(A)}\qquad \overline{m}_b(m_b^{\rm kin}) 
  &=&  
      4550  - 275 - (102+6_{\Delta_{m_c}}+14_{dec}-21_{n_c}) 
      \nonumber\\&&\mbox{} 
                    - (39
                    + 13_{\Delta_{m_c}}
                    + 18_{dec}
                    + 0.4_{\Delta_{m_c}\times dec} 
                    - 30_{n_c} 
                    - 1_{dec\times n_c}) ~\mbox{MeV}
      \nonumber\\&=&
                           4550 - 275 - 101 - 40~\mbox{MeV} = 4134~\mbox{MeV}
                           \,,
                    \nonumber\\
  \mbox{(B)}\qquad \overline{m}_b(m_b^{\rm kin}) 
  &=&  4550 -  288 
      - (111 + 7_{\Delta_{m_c}}  + 5_{\Delta_{m_c}^{\rm kin}} - 23_{n_c}) 
      \nonumber\\&&\mbox{} 
      - (44 + 15_{\Delta_{m_c}} + 10_{\Delta_{m_c}^{\rm kin}} - 34_{n_c} )~\mbox{MeV}
      \nonumber\\&=&
                           4550 - 288 - 100 - 36~\mbox{MeV} = 4126~\mbox{MeV}
\,,
                    \nonumber\\
  \mbox{(C)}\qquad \overline{m}_b(m_b^{\rm kin}) 
  &=&  4550 -  288 
      - (122 +7_{\Delta_{m_c}} -23_{n_c}) 
      - (69 + 15_{\Delta_{m_c}} - 34_{n_c})~\mbox{MeV}
      \nonumber\\&=&
                           4550 - 288 - 106 - 50~\mbox{MeV} = 4106~\mbox{MeV}
\,,
                    \nonumber\\
  \mbox{(D)}\qquad \overline{m}_b(m_b^{\rm kin}) 
  &=&  4550 -  275 - 102 - 39~\mbox{MeV} = 4134~\mbox{MeV}
\,.
                                                     \label{eq::mbKIN2mbMS}
\end{eqnarray}
The convergence properties in these equations are similar to
Eq.~(\ref{eq::mbMS2KIN}).  
In a second step, we can obtain the scale-invariant $\overline{\rm MS}$ mass $\overline{m}_b(\overline{m}_b)$
 with the help of the QCD renormalization group
equations up to five-loop
accuracy~\cite{Baikov:2014qja,Luthe:2016xec,Baikov:2017ujl,Baikov:2016tgj,Herzog:2017ohr,Luthe:2017ttg,Chetyrkin:2017bjc}
as implemented in {\tt RunDec}~\cite{Herren:2017osy}.
For the four schemes we obtain
\begin{eqnarray}
  \mbox{(A)}\qquad \overline{m}_b(\overline{m}_b) &=& 4195~\mbox{MeV}\,, \nonumber\\
  \mbox{(B)}\qquad \overline{m}_b(\overline{m}_b) &=& 4189~\mbox{MeV}\,, \nonumber\\
  \mbox{(C)}\qquad \overline{m}_b(\overline{m}_b) &=& 4171~\mbox{MeV}\,, \nonumber\\
  \mbox{(D)}\qquad \overline{m}_b(\overline{m}_b) &=& 4195~\mbox{MeV}\,.
\end{eqnarray}
Excluding scheme (C) we observe a splitting of about $6$~MeV which is
more than a factor two smaller as the current uncertainty of the $\overline{\rm MS}$
bottom quark mass as extracted from experimental data or lattice calculations
(see, e.g., Ref.~\cite{Tanabashi:2018oca}).

Next, we consider the variation of the renormalization scale $\mu_s$, which is
present in the mass conversion formulae. After the inclusion of higher order
perturbative corrections, the dependence on $\mu_s$ should decrease. In fact, the
dependence on $\mu_s$ can also be used as a measure to estimate the unknown
higher order terms, i.e., four-loop corrections.
Note also that contributions from higher dimensional operators would scale as
$\alpha_s \mu^3/m_b^3$ which is numerically close to $\alpha_s^4$ if we assume
$\alpha_s\sim 0.2$, $\mu\sim1$~GeV and $m_b\sim5$~GeV.\footnote{From the discussion below
  Eq.~(40) in Ref.~\cite{Bigi:1994em}, one might draw the conclusion that the correction
  to the kinetic mass of order $\alpha_s \mu^3$ is zero. 
  This fact is reported also in Appendix A.1 of Ref.~\cite{Benson:2003kp}, 
  however a proof was never given to our knowledge.}

\begin{figure}[t]
  \includegraphics[width=0.9\textwidth]{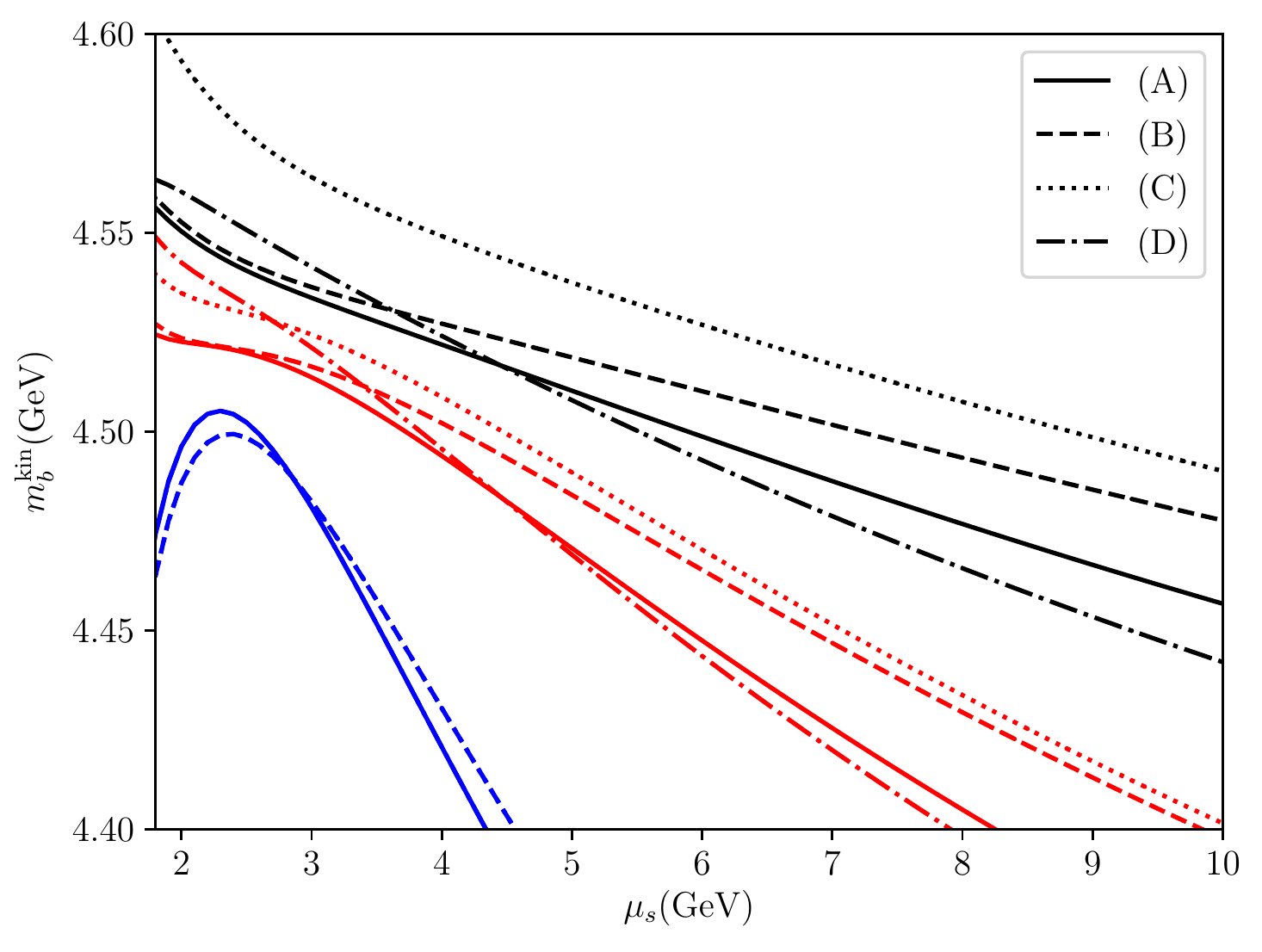}
  \caption{\label{fig::mbkin}$\overline{m}_b^{\rm kin}$ computed from
    $\overline{m}_b(\mu_s)$ using one-, two- and three-loop (blue,
    red, black) accuracy as a function of $\mu_s$. At each loop-order,
    four lines are shown, one for each of schemes (A) -- (D). 
   }
\end{figure}

\begin{figure}[t]
  \includegraphics[width=0.9\textwidth]{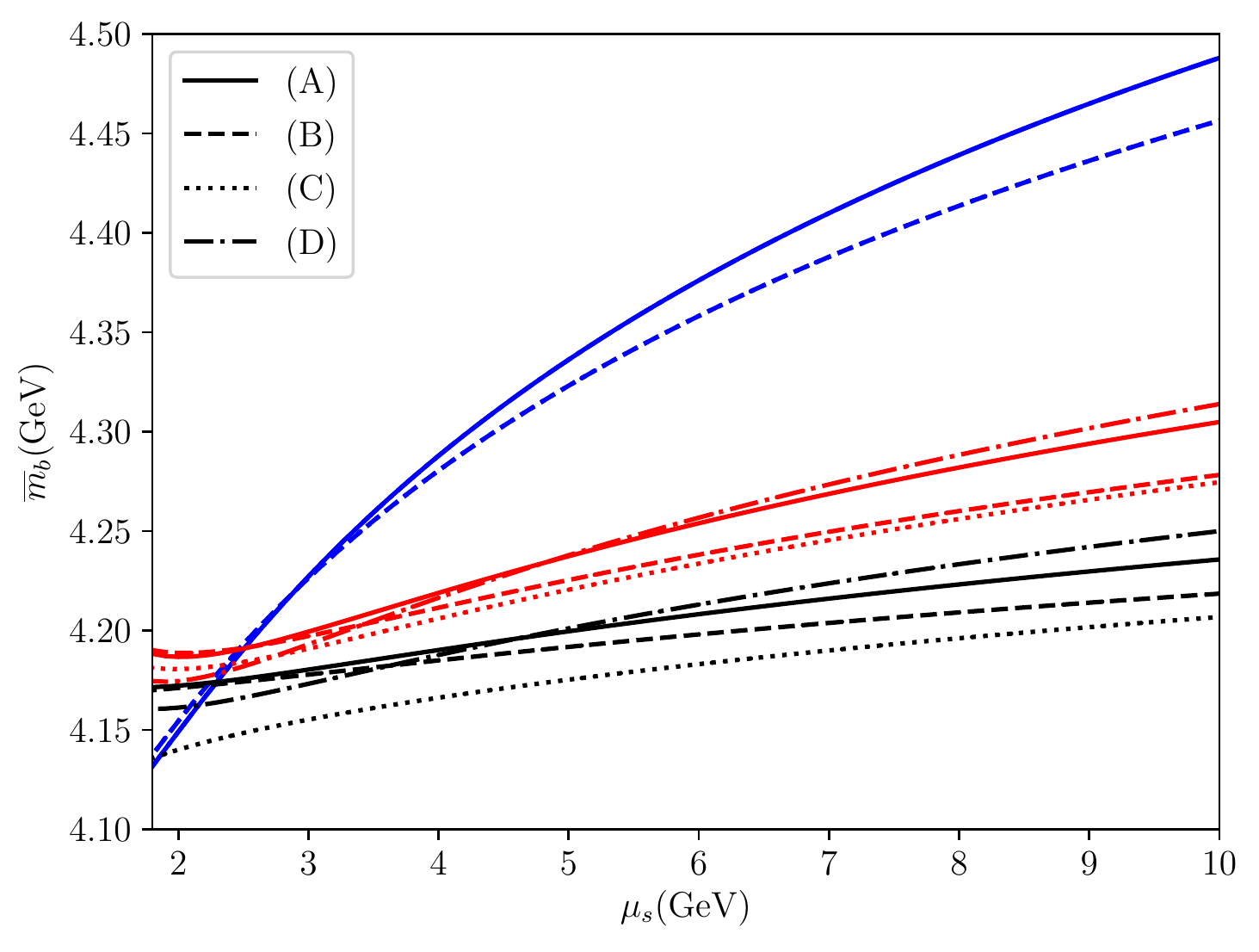}
  \caption{\label{fig::mbmb}$\overline{m}_b(\overline{m}_b)$ computed
    from $m^{\rm kin}$ using one-, two- and three-loop (blue, red,
    black) accuracy as a function of $\mu_s$. At each loop-order,
    four lines are shown, one for each of the scheme (A) -- (D).}
\end{figure}

In Fig.~\ref{fig::mbkin} we show $m_b^{\rm kin}$ obtained from
$\overline{m}_b(\mu_s)$ with initial value
$\overline{m}_b(\overline{m}_b)=4.163$~GeV as a function of
$\mu_s$. Results based on one-, two- and three-loop conversion formulae are
shown. On the horizontal axis we vary the intermediate scale $\mu_s$ between
1.5~GeV and 10~GeV.  Note that a similar plot can be found in Fig.~9 of
Ref.~\cite{Gambino:2011cq}.  However, there the scale of $\overline{m}_b$ was
fixed and only the scale of $\alpha_s$ ($\mu_s$) varied.
Figure~\ref{fig::mbmb} shows the corresponding results where
$\overline{m}_b(\overline{m}_b)$ is computed from $m_b^{\rm kin}=4.550$~GeV.

Both in Fig.~\ref{fig::mbkin} and Fig.~\ref{fig::mbmb}, we observe a flattening of the curves after including
higher order corrections. If we restrict ourselves to values of $\mu_s\ge 3$~GeV,
the three-loop curves varies by about 25~MeV and 50~MeV, respectively,
which suggests an uncertainty of $\pm13$~MeV and $\pm25$~MeV.
Note, however, that a stronger $\mu_s$ dependence is observed below 2 GeV.

There are several options to estimate of the theoretical uncertainty associated to
the $\overline{m}_b$--$m_b^{\rm kin}$ conversion formula.
\begin{itemize}
\item[1.] We proceed as in Ref.~\cite{Fael:2020iea} and use
  half of the three-loop correction as an estimate on the 
  size of the unknown higher orders. This leads to 
  an uncertainty of about 15~MeV (excluding scheme (C)).
  Note that the same criterion applied to the two-loop
  mass relation leads to an uncertainty of about 40~MeV.
  Thus, the three-loop term leads to a reduction of the uncertainty by about a factor of two.
\item[2.] An estimate of higher order effects is also obtained by varying
  $\mu_s$.
  If we choose $3~\mbox{GeV}\le \mu_s\le 9$~GeV we obtain an uncertainty of $\pm \{17,13,18,24\}$~MeV
  for the four schemes (see also Fig.~\ref{fig::mbkin}).
  The same prescription at order $\alpha_s^2$ leads to 
  an uncertainty of $\pm \{33,25,27,39\}$~MeV.
\item[3.] An optimistic uncertainty estimate could be based on the four-loop large-$\beta_0$ 
  approximation computed in Section~\ref{sec::BLM}. In that case we obtain
  9~MeV for the missing four-loop term. 
\end{itemize}
We recommend to use option 1.

Finally, we present simple formulae which can be used to convert the
scale-invariant bottom quark mass to the kinetic scheme or vice versa
using the preferred input values for the mass and strong coupling constant.
We find
\begin{align}
  \frac{\overline{m}_b(\overline{m}_b)}{\mbox{MeV}} =& \{4195,4189,4171,4195\} +
                                                      \Delta_{\rm kin} \{18,18,18,18\} 
                                                      \nonumber\\&
  - \Delta_{\alpha_s} \{7,7,8,7\} 
  \pm \{20,18,25,19\}
                                     \,, 
                                     \nonumber\\
  \frac{m_b^{\rm kin}}{\mbox{MeV}} =& \{4520,4526,4547,4521\}
                                     + \Delta_{\overline{\rm MS}} \{18,18,18,18\} 
                                     \nonumber\\&
  +  \Delta_{\alpha_s} \{8,8,10,8\}
  \pm \{15,13,20,15\}
  \,,
\end{align}
where the three numbers in the curly brackets refer to schemes (A),
(B), (C) and (D), respectively.  For the quantities $\Delta_X$ we have
\begin{eqnarray}
  \Delta_{\rm kin} &=& (m_b^{\rm kin}/\mbox{MeV}-4550)/20\,,\nonumber\\
  \Delta_{\overline{\rm MS}} &=& (\overline{m}_b(\overline{m}_b)/\mbox{MeV}-4163)/16\,,\nonumber\\ 
  \Delta_{\alpha_s} &=& (\alpha_s-0.1179)/0.001\,.
\end{eqnarray}

All numerical results presented in this Section can be reproduced using
the implementation in {\tt RunDec} and {\tt
  CRunDec}~\cite{Herren:2017osy,rundec} in the functions \verb|mKIN2mMS[...]|
and \verb|mMS2mKIN[...]|.  For more details concerning the arguments, we refer
to the latest version which can be downloaded from~\cite{rundec}.


\section{\label{sec::concl}Conclusions}

The aim of this paper has been the computation of the three-loop
corrections to the relation between the heavy quark masses defined in
the kinetic and the $\overline{\rm MS}$ schemes. 

We described in detail the methods employed for the calculation
of the mass relation.  In particular the application of
the asymptotic expansion in the threshold limit and the computation of
the master integrals. For the latter we provided explicit analytic results.
Our strategy is in principle extendable to $\alpha_s^4$, 
if such precision will ever become necessary in the future.
We furthermore discussed in detail finite charm quark mass
effects for the bottom mass relations.

The numerical analysis of the $m^{\rm kin}$--$m^{\overline{\rm MS}}$ relation
shows a good convergence of the perturbative series, both for charm and
bottom quark.  Altogether, charm quark mass effects to the bottom mass are
small and do not destabilize the convergence property of the series.  The mass
relation is mainly sensitive to the number of massless quarks, while it is
rather insensitive to charm which behaves as a heavy degree of freedom.  The
new correction terms at three loops reduce the uncertainty due to scheme
conversion by about a factor two.

The extraction of $|V_{cb}|$ from inclusive semileptonic $B$ decays is founded
upon the kinetic scheme for the heavy quark masses and the HQE
parameters. Therefore, the work presented in this paper is pivotal for future
precision determinations of $|V_{cb}|$ at Belle II.



\section*{Acknowledgements}  

We thank Andrzej Czarnecki, Paolo Gambino and Miko{\l}aj Misiak for useful
discussions and communications and Alexander Smirnov for help in the
use of {\tt asy}~\cite{Pak:2010pt}.  We furthermore
thank Go Mishima for useful hints in connection to the application of
the Mellin-Barnes method. We are grateful to Florian Herren
for providing us with his program \texttt{LIMIT}~\cite{Herren:2020ccq}  which automates the
partial fraction decomposition in case of linearly dependent
denominators.  This research was supported by the Deutsche
Forschungsgemeinschaft (DFG, German Research Foundation) under grant
396021762 --- TRR 257 ``Particle Physics Phenomenology after the Higgs
Discovery''.


\begin{appendix}


\section{Perturbative contributions to HQET parameters}

In this section we give the perturbative contributions  
to HQET parameters $\overline{\Lambda}, \mu_\pi^2, \rho_D$ and $r_E$ up to order~$\alpha_s^3$.
These expressions can be employed to renormalize their non-perturbative versions in the so-called kinetic scheme. 
The definitions of $\overline{\Lambda}|_{\rm pert}$  and $\mu_\pi^2|_{\rm pert}$ were given in Eq.~\eqref{eq::Lam_mupi}. The HQET parameters $\rho_D$ and $r_E$ are defined as:
\begin{align}
  2 M_B \rho_D^3 &= \bra{H_\infty} \bar h_v iD_{\perp \mu} (i v \cdot D) iD_\perp^\mu h_v \ket{H_\infty},\notag  \\
  2 M_B r_E^4 &= - \bra{H_\infty} \bar h_v iD_{\perp \mu} (i v \cdot D)^2 iD_\perp^\mu h_v \ket{H_\infty}.
\end{align}
Their perturbative versions are given by the SV sum rules:
\begin{align}
  [\rho_D^3(\mu)]_{\rm pert}
  &= 
      \lim_{\vec{v}\to0}\lim_{m_b\to\infty} 
      \frac{3}{\vec{v}\,^2} 
      \frac{
      \displaystyle
      \int_0^\mu \omega^3 \, W(\omega,\vec{v}\,) \, {\rm d}\omega}
      {
      \displaystyle
      \int_0^\mu W(\omega,\vec{v}\,) {\rm d}\omega}
      \,,\notag\\[5pt]
      [r_E^4(\mu)]_{\rm pert} &= 
      \lim_{\vec{v}\to0}\lim_{m_b\to\infty} 
      \frac{3}{\vec{v}\,^2} 
      \frac{\displaystyle
      \int_0^\mu \omega^4 \, W(\omega,\vec{v}\,) \, {\rm d}\omega}
      {\displaystyle
      \int_0^\mu W(\omega,\vec{v}\,) {\rm d}\omega}
      \,.
\end{align}
Up to $O(\alpha_s^3)$, the HQET parameters are given by the following expressions:
\begin{eqnarray}
  \left[ \overline{\Lambda}(\mu) \right]_{\text{pert}} &=&
  \frac{ \alpha_s^{(n_l)} }{ \pi } C_F \, \mu 
  \Biggl\{
  \frac{4}{3}   
  + \frac{ \alpha_s^{(n_l)} }{ \pi } 
  \biggl[
        \frac{1}{27} C_A 
        \bigl(
                215
                -6 \pi ^2
                -66 l_{\mu }
        \bigr)
        -\frac{8}{27} n_l T_F
        \bigl(
                8
                -3 l_{\mu }
        \bigr) 
  \biggr]
  \nonumber \\ &&
  +\left( \frac{ \alpha_s^{(n_l)} }{ \pi } \right) ^2  
  \biggl[
        C_A n_l T_F 
        \biggl(
                -\frac{19453}{486}
                +2 \zeta_3
                +\frac{1654 l_{\mu }}{81}
                -\frac{88 l_{\mu }^2}{27}
                +\frac{8 \pi ^2 }{81} 
                \bigl(
                         13
                        -3 l_{\mu }
                \bigr)
        \biggr)
        \nonumber \\ &&
        +C_A^2 
        \biggl(
                \frac{130867}{1944}
                +\frac{\pi ^4}{18}
                -\frac{19}{2} \zeta_3
                -\frac{2518 l_{\mu }}{81}
                +\frac{121 l_{\mu }^2}{27}
                -\frac{\pi ^2}{162}  
                \bigl(
                         511
                        -132 l_{\mu }
                \bigr)
        \biggr)
        \nonumber \\ && 
        + n_l^2 T_F^2
        \biggl(
                \frac{1292}{243}
                -\frac{8 \pi ^2}{81}
                -\frac{256 l_{\mu }}{81}
                +\frac{16 l_{\mu }^2}{27}
        \biggr)
        \nonumber \\ &&
        - C_F n_l T_F
        \biggl(
                 \frac{11}{4}
                -\frac{4 \zeta_3}{3}
                -\frac{2 l_{\mu }}{3}
        \biggr)
  \biggr]
  \Biggr\}
  \,, \\
  \left[ \mu_\pi^2(\mu) \right]_{\text{pert}} &=&
  \frac{ \alpha_s^{(n_l)} }{ \pi } C_F \, \mu^2 
  \Biggl\{
        1
        + \frac{ \alpha_s^{(n_l)} }{ \pi } 
        \biggl[
          \frac{1}{18} C_A 
          \bigl(
                91
                -3 \pi ^2
                -33 l_{\mu }
          \bigr)
          -\frac{1}{9} n_l T_F 
          \bigl(
                13
                -6 l_{\mu }
          \bigr) 
        \biggr]
        \nonumber \\ &&
        +\left( \frac{ \alpha_s^{(n_l)} }{ \pi } \right) ^2
        \biggl[
        C_A n_l T_F
        \biggl(
                \frac{3}{2} \zeta_3
                -\frac{13699}{648}
                +\frac{1}{27} \pi ^2 
                \bigl(
                        23
                        -6 l_{\mu }
                \bigr)
                +\frac{695 l_{\mu }}{54}
                -\frac{22 l_{\mu }^2}{9}
        \biggr)
        \nonumber \\ &&
        +C_A^2 
        \biggl(
                \frac{96295}{2592}
                +\frac{\pi ^4}{24}
                -\frac{57}{8} \zeta_3
                -\frac{2155 l_{\mu }}{108}
                +\frac{121 l_{\mu }^2}{36}
                -\frac{1}{216} \pi ^2 
                \bigl(
                         445
                         -132 l_{\mu }
                \bigr)
        \biggr)
        \nonumber \\ &&
        + C_F n_l T_F
        \biggl(
                -\frac{29}{16}
                +\zeta_3
                +\frac{l_{\mu }}{2}
        \biggr) 
        + n_l^2 T_F^2
        \biggl(
                \frac{209}{81}
                -\frac{2 \pi ^2}{27}
                -\frac{52 l_{\mu }}{27}
                +\frac{4 l_{\mu }^2}{9}
        \biggr) 
    \biggr]
  \Biggr\}
  \,, \\
  \left[ \rho_D^3(\mu) \right]_{\text{pert}} &=&
  \frac{ \alpha_s^{(n_l)} }{ \pi } C_F \mu^3
  \Biggl\{
    \frac{2}{3} 
    + \frac{ \alpha_s^{(n_l)} }{ \pi } 
    \biggl[
        \frac{1}{18} C_A 
        \bigl(
                57-2 \pi ^2-22 l_{\mu }
        \bigr)
        -\frac{4}{9} n_l T_F
        \bigl(
                2-l_{\mu }
        \bigr) 
    \biggr]
    \nonumber \\ &&
    +\left( \frac{ \alpha_s^{(n_l)} }{ \pi } \right) ^2 
    \biggl[
        C_A n_l T_F 
        \biggl(
                 \zeta_3
                -\frac{12133}{972}
                +\frac{217 l_{\mu }}{27}
                -\frac{44 l_{\mu }^2}{27}
                +\frac{4}{81} \pi ^2 
                \bigl(
                        11-3 l_{\mu }
                \bigr)
        \biggr)
        \nonumber \\ &&
        +C_A^2
        \biggl(
                \frac{86707}{3888}
                +\frac{\pi ^4}{36}
                -\frac{19}{4} \zeta_3
                -\frac{113 l_{\mu }}{9}
                +\frac{121 l_{\mu }^2}{54}
                -\frac{1}{108} \pi ^2 
                \bigl(
                        141-44 l_{\mu }
                \bigr)
        \biggr)
        \nonumber \\ &&
        + C_F n_l T_F
        \biggl(
                \frac{1}{72} 
                \bigl(
                         48 \zeta_3 
                         - 83
                \bigr)
                +\frac{l_{\mu }}{3}
        \biggr) 
        + n_l^2 T_F^2 
        \biggl(
                \frac{358}{243}
                -\frac{4 \pi ^2}{81}
                -\frac{32 l_{\mu }}{27}
                +\frac{8 l_{\mu }^2}{27}
        \biggr) 
    \biggr]
  \Biggr\}
  \,, \nonumber \\
  \\
  \left[ r_E^4(\mu) \right]_{\text{pert}} &=&
  \frac{ \alpha_s^{(n_l)} }{ \pi } C_F \mu^4
  \Biggl\{
    \frac{1}{2}
    +
    \frac{ \alpha_s^{(n_l)} }{ \pi } 
    \biggl[ 
        \frac{C_A}{144} 
        \bigl(
                331
                -12 \pi ^2
                -132 l_{\mu }
        \bigr)
        -\frac{1}{36} n_l T_F
        \bigl(
                 23
                -12 l_{\mu }
        \bigr) 
    \biggr]
    \nonumber \\ &&
    + \left(\frac{ \alpha_s^{(n_l)} }{ \pi }\right)^2 
    \biggl[
        C_A n_l T_F 
        \biggl(
                 \frac{3}{4} \zeta_3
                -\frac{1427}{162}
                +\frac{1}{108} \pi ^2 
                \big(
                        43
                        -12 l_{\mu }
                \big)
                +\frac{629 l_{\mu }}{108}
                -\frac{11 l_{\mu }^2}{9}
        \biggr)
        \nonumber \\ &&
        +C_A^2 
        \biggl(
                \frac{20569}{1296}
                +\frac{\pi ^4}{48}
                -\frac{57}{16} \zeta_3
                -\frac{3947 l_{\mu }}{432}
                +\frac{121 l_{\mu }^2}{72}
                -\frac{1}{108} \pi ^2 
                \bigl(
                        103
                        -33 l_{\mu }
                \bigr)
        \biggr)
        \nonumber \\ &&
        +C_F n_l T_F 
        \biggl(
                -\frac{27}{32}
                +\frac{\zeta_3}{2}
                +\frac{l_{\mu }}{4}
        \biggr) 
        +n_l^2 T_F^2
        \biggl(
                \frac{331}{324}
                -\frac{\pi ^2}{27}
                -\frac{23 l_{\mu }}{27}
                +\frac{2 l_{\mu }^2}{9}
        \biggr) 
    \biggr]
  \Biggr\}
  \,,
\end{eqnarray}  
with $l_\mu = \ln(2 \mu/\mu_s)$.
Note that we chose to parametrize the relations in terms
of $\alpha_s^{(n_l)}(\mu_s)$ since all dependence on heavy quark
masses decouples. The analytic expressions from this appendix
can be downloaded from~\cite{progdata}.


\section{Calculation of $I_7^{3l}$}
\label{app:I7}

In this Appendix we describe the calculation of the master integral
\begin{align}
    {I_7^{3l}} = \int \frac{{\rm d}^d \vec{v}}{(2\pi)^{3d}}
  \frac{1}{[-k_1^2][-k_2^2][-(k_1-k_3)^2][-(k_2-k_3)^2][-2 k_1.p_1+y][-2
  k_2.p_1+y]}
\nonumber\\                                                            
  \times \frac{1}{[-2k_3.p_1+y]} \,,
\end{align} 
with $p_1^2 = m^2$. We compute $I_7^{3l}$ including the terms of order
$\epsilon$.  Since we know that the integral scales uniformly as
$y^{3d-11} m^{8-3d}$ we can set $y=m=1$.

The direct evaluation through Mellin-Barnes integrals and subsequent summation
was not successful, since we encounter quite {\ms complicated} threefold
infinite sums already for the $\epsilon^0$ part.  We thus follow the idea
to introduce another scale $x$ into the problem and solve the associated
differential equations.  The boundary conditions may be fixed in the limit
$x \to 0$ and $I_7$ can be extracted from the limit $x \to 1$.

We look at the auxiliary integral
\begin{align}
    {\hat{I}_{12}} = \int \frac{{\rm d}^d \vec{v}}{(2\pi)^{3d}}
  \frac{1}{[-k_1^2][-k_2^2][-(k_1-k_3)^2][-(k_2-k_3)^2][-2
  k_1.p_1+1][{-2k_2.p_1}+x]}
  \nonumber\\
  \times \frac{1}{[-2 {k_3.p_1}+x]}\,,
\end{align} 
with $p_1^2=1$.
Using \texttt{LiteRed} it is straightforward to find a closed system of differential equations for this family.
The master integrals we encounter are
\begin{align}
    \hat{I}_1    &= \int \frac{{\rm d}^d \vec{v}}{(2\pi)^{3d}}
    \frac{1}{[-k_2^2][-(k_1-k_3)^2][-(k_2-k_3)^2][-2 k_1.p_1+1]}
    ,
\nonumber\\
    \hat{I}_2    &= \int \frac{{\rm d}^d \vec{v}}{(2\pi)^{3d}}
    \frac{1}{[-k_1^2][-(k_1-k_3)^2][-(k_2-k_3)^2][-2k_2.p_1+x]}
    ,
\nonumber\\
    \hat{I}_3    &= \int \frac{{\rm d}^d \vec{v}}{(2\pi)^{3d}}
    \frac{1}{[-k_2^2][-(k_1-k_3)^2][-(k_2-k_3)^2][-2 k_1.p_1+1][-2k_3.p_1+x]}
    ,
\nonumber\\
    \hat{I}_4    &= \int \frac{{\rm d}^d \vec{v}}{(2\pi)^{3d}}
    \frac{1}{[-k_2^2][-(k_1-k_3)^2][-(k_2-k_3)^2][-2 k_1.p_1+1][-2k_2.p_1+x]}
    ,
\nonumber\\
    \hat{I}_5    &= \int \frac{{\rm d}^d \vec{v}}{(2\pi)^{3d}}
    \frac{1}{[-k_1^2][-(k_1-k_3)^2][-(k_2-k_3)^2][-2 k_1.p_1+1][-2k_2.p_1+x]}
    ,
\nonumber\\
    \hat{I}_6    &= \int \frac{{\rm d}^d \vec{v}}{(2\pi)^{3d}}
    \frac{1}{[-k_1^2][-k_2^2][-(k_2-k_3)^2][-2 k_1.p_1+1][-2k_3.p_1+x]}
    ,
\nonumber\\
    \hat{I}_7    &= \int \frac{{\rm d}^d \vec{v}}{(2\pi)^{3d}}
    \frac{1}{[-k_1^2][-k_2^2][-(k_1-k_3)^2][-2k_2.p_1+x][-2 k_3.p_1+x]}
    ,
\nonumber\\
    \hat{I}_8    &= \int \frac{{\rm d}^d \vec{v}}{(2\pi)^{3d}}
    \frac{1}{[-k_1^2][-k_2^2][-(k_1-k_3)^2][-(k_2-k_3)^2][-2 k_3.p_1+x]}
    ,
\nonumber\\
    \hat{I}_9    &= \int \frac{{\rm d}^d \vec{v}}{(2\pi)^{3d}}
    \frac{1}{[-k_1^2][-k_2^2][-(k_1-k_3)^2][-2 k_1.p_1+1][-2k_2.p_1+x][-2 k_3.p_1+x]}
    ,
\nonumber\\
    \hat{I}_{10} &= \int \frac{{\rm d}^d \vec{v}}{(2\pi)^{3d}}
    \frac{1}{[-k_1^2][-k_2^2][-(k_1-k_3)^2][-(k_2-k_3)^2][-2 k_1.p_1+1][-2k_3.p_1+x]}
    ,
\nonumber\\
    \hat{I}_{11} &= \int \frac{{\rm d}^d \vec{v}}{(2\pi)^{3d}}
    \frac{1}{[-k_1^2][-k_2^2][-(k_1-k_3)^2][-(k_2-k_3)^2][-2 k_1.p_1+1][-2k_2.p_1+x]}
        ,
\nonumber\\
    \hat{I}_{12} &= \int \frac{{\rm d}^d \vec{v}}{(2\pi)^{3d}}
    \frac{1}{[-k_1^2][-k_2^2][-(k_1-k_3)^2][-(k_2-k_3)^2][-2
                   k_1.p_1+1][-2k_2.p_1+x]}
  \nonumber\\&
  \mbox{}\hspace*{25em}\times \frac{1}{[-2k_3.p_1+x]}
        ,
\nonumber\\
    \hat{I}_{13} &= \int \frac{{\rm d}^d \vec{v}}{(2\pi)^{3d}}
    \frac{1}{[-k_1^2]^2[-k_2^2][-(k_1-k_3)^2][-(k_2-k_3)^2][-2
                   k_1.p_1+1][-2k_2.p_1+x]}
  \nonumber\\&
  \mbox{}\hspace*{25em}\times \frac{1}{[-2k_3.p_1+x]}\,.
\end{align}
Note that $I_7^{3l}=\hat{I}_{12}|_{x=1}$.

Most of them can be easily computed for general $d$ and $x$ and we find
\begin{align}
    \hat{I}_1    &= \Gamma^3(d/2-1) \Gamma(7-3d),
\nonumber\\
    \hat{I}_2    &= x^{3d-7} \Gamma^3(d/2-1) \Gamma(7-3d),
\nonumber\\
    \hat{I}_3    &= \frac{\Gamma^3(d/2-1) \Gamma(8-3d)}{3-d} {}_2 F_1 \left(1,8-3d,4-d;1-x\right),
\nonumber\\
    \hat{I}_4    &= \frac{\Gamma^3(d/2-1) \Gamma(8-3d)}{5-2d} {}_2 F_1 \left(1,8-3d,6-2d;1-x\right),
\nonumber\\
    \hat{I}_5    &= \frac{\Gamma^3(d/2-1) \Gamma(8-3d)}{5-2d} {}_2 F_1 \left(1,d-2,6-2d;1-x\right),
\nonumber\\
    \hat{I}_6    &= x^{2d-5} \Gamma^3(d/2-1) \Gamma(3-d) \Gamma(5-2d),
\nonumber\\
    \hat{I}_7    &= x^{3d-8} \Gamma^3(d/2-1) \Gamma(3-d) \Gamma(5-2d),
\nonumber\\
    \hat{I}_8    &= x^{3d-9} \frac{\Gamma^2(2-d/2) \Gamma^4(d/2-1) \Gamma(3d/2-4) \Gamma(9-3d)}{\Gamma(4-d) \Gamma^2(d-2)},
\nonumber\\
    \hat{I}_9    &= \frac{\Gamma^3(d/2-1) \Gamma^2(3-d) \Gamma(6-2d)}{\Gamma(4-d)} {}_2 F_1 \left(6-2d,3-d,4-d;1-x\right).
\end{align}
The $\epsilon$-expansion of the hypergeometric functions can be
obtained using \texttt{HypExp} \cite{Huber:2005yg,Huber:2007dx} or \texttt{EvaluateMultiSums} \cite{Ablinger:2010pb}.  
The integrals $\hat{I}_{10}$ to $\hat{I}_{13}$ can be determined through differential
equations with the following boundary conditions
\begin{align}
    \left. \hat{I}_{10} \right|_{x=0} &= \left. I_{12}^{3l} \right|_{y=1,m=1} ,
    \\
    \left. \hat{I}_{11} \right|_{x\to0} &= \frac{x}{12 \epsilon^3}
        + \frac{1}{\epsilon^2} 
        \Bigl[
            -\frac{1}{24}
            + x
            \Bigl(
              \frac{2}{3}
              - \frac{1}{4} \ln(x)
            \Bigr) 
        \Bigr]
        + \frac{1}{\epsilon}
        \Bigl[
          -\frac{13}{24}
          + x 
          \Bigl(
            \frac{10}{3}
            + \frac{7}{24} \zeta_2
            - 2 \ln(x)
            \nonumber \\ &
            + \frac{1}{4} \ln^2(x)
          \Bigr)
        \Bigr]
        - \frac{109}{24}
        - \frac{29}{16} \zeta_2
        +\epsilon
        \Bigl[
          -\frac{757}{24}
          -\frac{345}{16} \zeta_2
          +\frac{23}{24} \zeta_3
          + x
          \Bigl(
            \frac{100}{3}
            + 17 \zeta_2
            - 2 \zeta_3
            \nonumber \\ &
            - \frac{6847}{480} \zeta_2^2
            - \bigl( 52 +23 \zeta_2 - \frac{23}{4} \zeta_3 \bigr) \ln(x)
            + \bigl( 11 + \frac{23}{8} \zeta_2 \bigr) \ln^2(x)
            - \frac{4}{3} \ln^3(x)
            \nonumber \\ &
            + \frac{1}{12} \ln^4(x)
          \Bigr)
        \Bigr]
        +\epsilon^2
        \Bigl[
          - \frac{4777}{24}
          - \frac{2649}{16} \zeta_2
          + \frac{395}{24} \zeta_3
          - \frac{3279}{64} \zeta_2^2
          + x 
          \Bigl(
            - \frac{16}{3}
            + \frac{85}{3} \zeta_2
            \nonumber \\ &
            + \frac{58}{3} \zeta_3
            - \frac{763}{12} \zeta_2^2
            + \frac{463}{24} \zeta_2 \zeta_3
            + \frac{1187}{60} \zeta_5
            - \bigl( 228 + \frac{253}{2} \zeta_2 - 46 \zeta_3 + \frac{749}{32} \zeta_2^2 \bigr) \ln(x)
            \nonumber \\ &
            + \bigl( 52 + 23 \zeta_2 - \frac{23}{4} \zeta_3 \bigr) \ln^2(x)
            - \bigl( \frac{22}{3} + \frac{23}{12} \zeta_2 \bigr) \ln^3(x)
            + \frac{2}{3} \ln^4(x)
            - \frac{1}{30} \ln^5(x)
          \Bigr)
        \Bigr],
    \\
    \left. \hat{I}_{12} \right|_{x=0} &= \left. I_{14}^{3l} \right|_{y=1,m=1} .
\end{align}
Note that we need the initial value of $\hat{I}_{11}$ up to $\mathcal{O}(x)$,
since the homogeneous solution of its associated
differential equation vanishes at $x=0$.
We have used a Mellin-Barnes representation to obtain the 
expansion in $x$.  
We do not need the limit $x = 0$
of $\hat{I}_{13}$ since
the master integrals $\hat{I}_{12}$ 
and $\hat{I}_{13}$ are not linearly independent
in the limit $x \to 1$.  
We use this to express the integration constants of
$\hat{I}_{13}$ through the ones for $\hat{I}_{12}$
in a later step.  
The differential equations have singular 
behaviour at $x=0$ and $x=1$, 
leading to harmonic polylogarithms.

Solving the differential equations for $\hat{I}_{10}$ and $\hat{I}_{11}$ is
simple, since they only depend on already known master integrals.  The
differential equations for $\hat{I}_{12}$ and $\hat{I}_{13}$ form a coupled
$2 \times 2$-system, which we decouple into a second order differential
equation for $\hat{I}_{12}$ using the \texttt{Mathematica} package
\texttt{OreSys} \cite{ORESYS}.  The differential equations are solved using
\texttt{HarmonicSums} \cite{HarmonicSums}.  The solution of $\hat{I}_{13}$ can
be constructed from the solution of $\hat{I}_{12}$, its derivatives and
already known master integrals.  We then use the fact that $\hat{I}_{12}$ and
$\hat{I}_{13}$ are not linearly independent at $x=1$ to fix one half of the
integration constants introduced by solving the differential equation.  The
other half can be fixed from the $x=0$ limit of $\hat{I}_{12}$.  Fixing the
boundary values in this way and expanding the general solution for $x \to 1$
we finally obtain
\begin{align}
    I_7^{3l} &= y^{3d-11} (m^2)^{4-3d/2} \left. \hat{I}_{12} \right|_{x=1}
    \nonumber \\ 
    &= y^{3d-11} (m^2)^{4-3d/2} 
    \Biggl[
        -\frac{\zeta_2}{3 \epsilon^2}
        + \frac{1}{\epsilon} \Bigl( -\frac{8}{3} \zeta_2 + \frac{1}{3} \zeta_3 \Bigr)
        - \frac{52}{3} \zeta_2
        + \frac{8}{3} \zeta_3
        - \frac{49}{30} \zeta_2^2 
        \nonumber \\ &
        + \epsilon
        \Bigl(
          - \frac{320}{3} \zeta_2
          + \frac{52}{3} \zeta_3
          - \frac{196}{15} \zeta_2^2
          - \frac{11}{6} \zeta_2 \zeta_3
          - \frac{83}{3} \zeta_5
        \Bigr)
    \Biggr] .
\end{align}


\section{\label{app:aux}Auxiliary integrals}

In this Section we present the formulae for auxiliary integrals useful for the direct integration
of the three-loop master integrals. They are given by
\begin{eqnarray}
  \lefteqn{J_0(y,n_1,n_2,n_3) =}
  \nonumber\\&&
    \int \frac{{\rm d}^d v}{(2 \pi)^d}
    \frac{1}{(-v^2)^{n_1} (-2p\cdot v)^{n_2} (-2p\cdot v + y)^{n_3}}
    \nonumber \\&&
    = (m^2)^{n_1-d/2} y^{d-2n_1-n_2-n_3}
    \frac{
    \Gamma(d-2n_1-n_2)\Gamma(d/2-n_1)\Gamma(2n_1+n_2+n_3-d)
    }{
    \Gamma(n_1)\Gamma(n_3)\Gamma(d-2n_1)
    }\,,
    \nonumber\\
    \lefteqn{J_1(y_2,y_3;n_1,n_2,n_3) =}
    \nonumber\\&&
    \int \frac{{\rm d}^d v}{(2 \pi)^d}
    \frac{1}{(-v^2)^{n_1} (-2p\cdot v+y_2)^{n_2} (-2p\cdot v + y_3)^{n_3}} \notag \\
    &=&  
    (m^2)^{n_1-d/2} \frac{
    \Gamma(d/2-n_1)
    }{
    \Gamma(n_1)\Gamma(n_2)\Gamma(n_3)\Gamma(d-2n_1)
    } 
     \frac{1}{2\pi i} \int_{-i \infty}^{+i\infty}
    dw \, y_2^w \, y_3^{d-2n_1-n_2-n_3-w} \notag\\
    &&\times \Gamma(n_2+w) \Gamma(-w) \Gamma(d-2n_1-n_2-w)\Gamma(2n_1+n_2+n_3+w-d)\,,\nonumber\\
    \lefteqn{J_2(p,q;n_1,n_2,n_3) =}
    \nonumber\\&&
    \int \frac{{\rm d}^d v}{(2 \pi)^d}
    \frac{1}{(-v^2)^{n_1} [-(v-q)^2]^{n_2} (-2 p\cdot v)^{n_3}} \notag \\
    &=&\frac{1}{\Gamma(n_1)\Gamma(n_2)\Gamma(n_3)\Gamma(d-n_1-n_2-n_3)}\notag \\
    &&\times\frac{1}{2\pi i} \int_{-i \infty}^{+i\infty}
    dw \,
    (-q^2)^{w}
    (-2p\cdot q)^{d-2n_1-2n_2-n_3-2w}
    (m^2)^{w+n_1+n_2-d/2} \notag \\
    &&\times
    \Gamma(-w)
    \Gamma(d-2n_1-n_2-n_3-w)
    \Gamma(n_1+w) \notag \\
    &&\times \Gamma(d/2-n_1-n_2-w)
    \Gamma(2n_1+2n_2+n_3+2w-d)\,.
\end{eqnarray}


\end{appendix}



\end{document}